\newif\ifanon
\anonfalse

\documentclass{article}

\pdfoutput=1

\usepackage[dvipsnames]{xcolor}
\usepackage{tikz}
\usetikzlibrary{calc,patterns, patterns.meta, 3d, arrows.meta}
\usetikzlibrary{decorations.pathreplacing,fit,shapes.misc}
\usepackage{xparse}
\usepackage{woo}
\usepackage{authblk}
\usepackage{todonotes}
\usepackage{caption}
\usepackage{float}

\newcommand{\twodclhgrid}{2D-CLH-Grid}
\newcommand{\twodclhrgrid}[1]{2D-$\text{CLH}^{(#1)}$-Grid}
\newcommand{\threedclh}{3D*-CLH}
\newcommand{\threedclhr}[1]{3D*-$\text{CLH}^{(#1)}$}
\newcommand{\threedclhgrid}{3D*-CLH-Grid}
\newcommand{\threedclhrgrid}[1]{3D*-$\text{CLH}^{(#1)}$-Grid}
\newcommand{\twodclh}{2D*-CLH}
\newcommand{\twodclhr}[1]{2D*-$\text{CLH}^{(#1)}$}

\pgfdeclarelayer{bg}    
\pgfdeclarelayer{fg}
\pgfsetlayers{bg,main,fg}

\usepackage[style=alphabetic]{biblatex}
\makeatletter
\tikzset{
  fitting node/.style={
    inner sep=0pt,
    fill=none,
    draw=none,
    reset transform,
    fit={(\pgf@pathminx,\pgf@pathminy) (\pgf@pathmaxx,\pgf@pathmaxy)}
  },
  reset transform/.code={\pgftransformreset}
}

\makeatother

\usepackage{datetime}

\newdateformat{monthyeardate}{%
  \monthname[\THEMONTH], \THEYEAR}

\title{Commuting Local Hamiltonians Beyond 2D}

\ifanon
\author{}
\date{}
\else
\author[1]{John Bostanci\thanks{Email: johnb@cs.columbia.edu}}
\affil[1]{Columbia University}
\author[2]{Yeongwoo Hwang\thanks{Email: yeongwoohwang@g.harvard.edu}}
\affil[2]{Harvard University}
\date{April 2025}
\fi

\begin{document}

\maketitle
\begin{abstract}
    Local Hamiltonians are a quantum analogue to SAT in classical complexity. As a model of ``intermediate'' complexity,
    \emph{commuting} local Hamiltonians provide a testing ground for studying many of the most interesting open
    questions in quantum information theory, including the quantum PCP conjecture and the nature of entanglement, while
    possibly being more amenable to analysis. Despite its simpler nature, the exact complexity of the commuting local
    Hamiltonian problem (CLH) remains elusive.  A number of works
    \cite{Bravyi_BV2004_CommutativeVersionKlocal,Schuch_Sch2011_ComplexityCommutingHamiltonians,Aharonov_AE2011_ComplexityCommutingLocal,
    Aharonov_AE2015_CommutingLocalHamiltonian,Irani_IJ2023_CommutingLocalHamiltonian} have shown that increasingly
    expressive families of commuting local Hamiltonians admit completely classical verifiers.  Despite intense work,
    proofs placing CLH in $\mathsf{NP}$ rely heavily on an underlying 2D lattice structure, or
    a very constrained local dimension and locality.  

    In this work, we present a new technique to analyze the complexity of various families of commuting local
    Hamiltonians: \emph{guided reductions}. Intuitively, these are a generalization of typical reduction where the
    prover provides a guide so that the verifier can construct a simpler Hamiltonian. The core of our reduction is a new
    rounding technique based on a combination of Jordan's Lemma and the Structure Lemma.  Our rounding technique is much
    more flexible than previous work and allows us to remove constraints on local dimension in exchange for a rank-$1$
    assumption. Specifically, we prove the following two results:
    \begin{enumerate}
        \item 2D-CLH for rank-1 instances are contained in $\mathsf{NP}$, independent of the qudit dimension.  It is
            notable that this family of commuting local Hamiltonians has no restriction on the local dimension \emph{or}
            the locality of the Hamiltonian terms.
        \item 3D-CLH for rank-1 instances are in $\mathsf{NP}$.  To our knowledge this is the first time a family of
            {3D} commuting local Hamiltonians has been contained in $\mathsf{NP}$.  
    \end{enumerate}
    Our results apply to Hamiltonians with large qudit degree and remain non-trivial despite the quantum Lov\'asz Local
    Lemma \cite{Ambainis_AKS2012_QuantumLovaszLocal}.
\end{abstract}
\newpage
\tableofcontents
\newpage

\section{Introduction}

The local Hamiltonian problem~\cite{Kitaev_KSV2002_ClassicalQuantumComputation} is one of the most important problems in quantum complexity theory, and the central problem of Hamiltonian complexity.  
Understanding the properties of local Hamiltonians, including the properties of their spectrum, ground states, and Gibbs states, has been a central question of study for the past $20$ years, and has greatly improved our understanding of both quantum mechanics and quantum computation.  

For those aiming to understand the properties of local Hamiltonians, commuting local Hamiltonians (CLHs) represent a kind of
half-way point between classical constraint satisfaction problems and general local Hamiltonians.  On one hand, the
ground states of commuting Hamiltonians can still exhibit high entanglement and can be non-trivial, as exhibited by the
Toric code and the recent proof of the no low-energy trivial states (NLTS) theorem~\cite{anshu2023nlts}.  On the other
hand, commutation often simplifies the analysis of quantum algorithms, as measurements in a shared basis do not experience the
uncertainty principle when multiple measurements are applied.  In addition to this, commuting local \emph{projectors}
always have an integral spectrum, and satisfy perfect completeness, making them robust to small perturbations and thus
making them prime candidates for studying constructions of quantum PCPs.  These properties make commuting local
Hamiltonians a popular test bed for physicists and computer scientists alike; many advances in Gibbs state sampling
come from analyzing the Gibbs states of commuting local
Hamiltonians~\cite{kastoryano2016quantum,bardet2023rapid,ding2024polynomialtimepreparationlowtemperaturegibbs,hwang2024gibbsstatepreparationcommuting}, and
work towards the quantum PCP conjecture has largely focused on quantum LDPC codes, and related code
Hamiltonians~\cite{anshu2023nlts, coble2023local}.  

Thus, by sharing some properties of classical constraint satisfaction problems and some properties of local
Hamiltonians, the commuting local Hamiltonian problem represents an interesting lens through which to study the
following problem: what properties of quantum computation make it more powerful than classical computation? The problem
of determining the complexity of commuting local Hamiltonians has been studied explicitly by a number of authors,
starting with the work of \cite{Bravyi_BV2004_CommutativeVersionKlocal}, who shows that the ground energy of $2$-local
commuting Hamiltonians are classically verifiable.
Since then, a long line of works have shown that determining the ground energy of commuting Hamiltonians that are
qubit 3-local or almost Euclidean~\cite{Aharonov_AE2015_CommutingLocalHamiltonian}, on locally expanding
graphs~\cite{Aharonov_AE2011_ComplexityCommutingLocal}, 2D over qubits~\cite{Aharonov_AKV2018_ComplexityTwoDimensional},
and on a square lattice over qutrits~\cite{Irani_IJ2023_CommutingLocalHamiltonian} are also problems contained in
$\mathsf{NP}$. In our work, we continue this line of inquiry and study the complexity of commuting local Hamiltonians.
However, whereas previous work largely relied either on \textbf{(a)} Hamiltonian terms being extremely local, or
\textbf{(b)} the local qudit dimension being extremely small, we study a class of commuting Hamiltonians with local
terms of arbitrarily high locality, and unrestricted qudit dimension. The catch is that we impose strong restrictions on
the rank of each term; namely, all Hamiltonian terms should be rank-1 projections.

A central theme of our work is the construction of \emph{reductions} between CLH instances. In the classical world,
reductions have provided an incredibly general framework for relating the computational complexity of two families of
problems are through reductions. Starting with the seminal Cook-Levin Theorem \cite{cook1971} which placed Boolean
satisfiability in $\cNP$, reductions have paved the way for a rich set of tools for placing increasingly exotic decision
tasks in $\cNP$. For $\cQMA$, the quantum analogue of $\cNP$, similar efforts have been made, the most successful of
which are \emph{perturbation gadgets}, which were original used to reduce the locality of $\cQMA$-hard local
Hamiltonians from 3 to 2 \cite{Kempe_KKR2005_ComplexityLocalHamiltonian}. Since then, perturbation gadgets have been
used extensively to show many other, often much more constrained, Hamiltonians are also $\cQMA$-complete
\cite{Cubitt_CM2016_ComplexityClassificationLocal}. Still, perturbation gadgets are not universally applicable and, in
particular, cannot be used to construct \emph{commuting} Hamiltonians. To address this deficit, we introduce a framework
of \emph{guided reductions} which preserve commutativity.

\subsection{On the rank-1 constraint} 
All of the results in the paper apply to a restriction of the commuting local Hamiltonian problem where all the Hamiltonian terms are rank-1 projectors.  It is already known~\cite{Irani_IJ2023_CommutingLocalHamiltonian} that restricting to projectors does to change the complexity of the commuting local Hamiltonian problem, but the rank-1 version of this problem has not been studied in detail before.  We believe this to be a worthwhile model to consider for a number of reasons.  First, we note that in related
complexity classes like $\mathsf{NP}$, $\mathsf{QMA}$, and $\mathsf{QCMA}$, the
rank-1 versions of canonical complete problems retain their hardness.  Taking circuit SAT~\cite{cook1971,levin1973universal} as an example, $k$-SAT can be viewed as a rank-1 local Hamiltonian version of SAT, where the Hamiltonian is a sum over clauses of the assignment that makes the clause false.  For $\mathsf{QMA}$, \cite{Cubitt_CM2016_ComplexityClassificationLocal} showed that the Heisenberg local Hamiltonian problem is $\mathsf{QMA}$ complete.  The Heisenberg local Hamiltonian problem is the local Hamiltonian problem where each local term is (a constant times) the projector onto the singlet state, $\frac{1}{\sqrt{2}} (\ket{01} - \ket{10})$. One important reason for studying the complexity of commuting local
Hamiltonians is that they provide a simplified setting in which one can study the complexity of the (non-commuting) local Hamiltonian
problem.  Since rank-1 local Hamiltonians are $\mathsf{QMA}$ complete, we believe that studying the rank-1 version of
commuting local Hamiltonians can provide insights into the nature of the local Hamiltonian problem, and provides a useful
test-bed for studying reductions between local Hamiltonian problems.

On the other hand, one should be careful that this problem remains non-trivial. Specifically, the quantum Lov\'azs Local
Lemma \cite{Ambainis_AKS2012_QuantumLovaszLocal} shows that if have a Hamiltonian $H$ on $d$-dimensional qudits with a
uniform bound $r$ on the rank, $k$ on the locality, and $g$ on the number of terms acting non-trivially on any qudit,
then the Hamiltonian $H$ is always satisfiable, as long as
\[
    g \leq \frac{d^k}{r e}\,,
\]
where $e$ is Euler's number. This implies that if we restrict to a rank-1 qubit Hamiltonian on the 2D square lattice (so
$g = 4$ and $d^k/re > 5.9$), the problem becomes trivial! Therefore, we necessarily need to consider more complex
geometries, such as the 2D quasi-Euclidean complexes studied by \cite{Aharonov_AKV2018_ComplexityTwoDimensional}, where
$g$ can be any $c \in \O(1)$.

\subsection{Related results}
Our results follow a series of works studying the complexity of the commuting local Hamiltonian problem, under the restriction that all terms are commuting. The line of study was initiated by \cite{Bravyi_BV2004_CommutativeVersionKlocal} who showed that the 2-local CLH problem was contained in $\cNP$. They also considered the specialized case of \emph{factorized} commuting local Hamiltonians, where each Hamiltonian term $h$ can be written as a tensor product of single-qudit operators. In this specialized setting, they showed that qubit CLH is contained in $\cNP$, regardless of geometry and locality. A key contribution of their work was the introduction of the \emph{Structure Lemma}. This lemma characterizes the local algebra of commuting operators and has been an integral part of nearly all works on CLH since then. In fact, the lack of tools beyond the Structure Lemma has been a stumbling block on extending results to higher localities and qudit dimension, as the ``structure'' induced by the lemma becomes increasing complex as the qudit dimension and locality increase.

The next major step along this line was due to \cite{Aharonov_AE2011_ComplexityCommutingLocal} who extended the result
of \cite{Bravyi_BV2004_CommutativeVersionKlocal} to showed that 3-local qubit CLHs are in $\cNP$. For qutrits, the
assumption that the underlying interaction graph is ``nearly Euclidean'' also yields containment in $\cNP$. Following this work, other authors moved onto stronger geometric restrictions. \cite{Schuch_Sch2011_ComplexityCommutingHamiltonians} considered the 2D square lattice and showed qubit Hamiltonians defined in the lattice can be placed in $\cNP$. A follow up work by \cite{Aharonov_AKV2018_ComplexityTwoDimensional} showed that Schuch's work can even be made \emph{constructive}, i.e. the verifier is able to efficiently prepare a ground state of the local Hamiltonian, rather than just be convinced about its existence. Their work can be viewed as reducing 2D lattice CLHs to a generalization of the Toric code, whose ground state can be prepared efficiently in classical polynomial time. Finally, \cite{Irani_IJ2023_CommutingLocalHamiltonian} extended the ideas contained in \cite{Schuch_Sch2011_ComplexityCommutingHamiltonians} to show that CLHs with qutrits on the 2D square lattice are also in $\cNP$. Like the proof Schuch, this result is non-constructive.

A natural question is whether regardless of qudit dimension or geometric structure, CLHs are generically in $\cNP$. One evidence that this might not be the case comes from \cite{Gosset_GMV2016_QCMAHardnessGround}, who show that the Ground State Connectivity (GSC) problem for commuting local Hamiltonians is $\cQCMA\textsf{-hard}$, which matches the complexity of GSC for general local Hamiltonians.

\subsection{Our contributions}

Our main contribution is to demonstrate a family of commuting local Hamiltonians on which a guided reduction can be performed to the 2-local CLH problem; rank-1 commuting Local Hamiltonians in 2D and 3D. 
We define a guided reduction as follows.

\begin{definition}[Guided reduction]
    \label{def:guided_reduction}
    We say that the CLH family $\mc H$ has an $(\epsilon, \delta)$ \emph{guided reduction} to another CLH family $\mc H'$ if there is a mapping
    $\mc M : \mc H \times \{0,1\}^* \rightarrow \mc H'$ such that for a $n$-qudit Hamiltonian $H \in \mc H$, 
    \begin{itemize}
        \item If $\lambda_0(H) = 0$ then there exists a string $s$ with $|s| \in \cpoly(n)$ such that $\lambda_0(\mc
            M(H, s)) = 0$.
        \item If $\lambda_0(H) > \eps$, then for all strings $s$, $\lambda_0(\mc M(H, s)) > \delta$.
    \end{itemize}
    Here $\lambda_0(H)$ is the minimum eigenvalue of $H$. When both $\epsilon$ and $\delta$ are constants in $(0,1]$, we simply call this a guided reduction.
\end{definition}
This definition is reminiscent of the ``$\mathsf{NP}$''-reductions referred to in
\cite{Irani_IJ2023_CommutingLocalHamiltonian}, where the reduction can not be performed in polynomial time, but the
prover can attach an additional proof that allows the verifier to efficiently transform the instance to an equivalent
one. The important property of this type of reduction is that such a valid guiding string only exists for ``yes''
instances of the problem. The result of \cite{Aharonov_AKV2018_ComplexityTwoDimensional} can also be interpreted as providing a guided
reduction from 2D qubit CLHs to 2-local CLHs (see \Cref{sec:techniques}).

Our main results are the following two theorems.
\begin{theorem}[Guided reduction from rank-1 2D CLH to 2-local CLH (informal)]
    There is a guided reduction from the family of rank-1 2D CLHs to the family of 2-local CLHs.
\end{theorem}

\begin{theorem}[Guided reduction from rank-1 3D* CLH to 2-local CLH (informal)]
    There is a guided reduction from the family of rank-1 3D* CLHs to the family of 2-local CLHs.
\end{theorem}

In the above theorem, the ``*'' refers to some constraints on the types of 3D Hamiltonians for which our reduction applies; we cover these details in \Cref{sec:geometrically_contrained}.

\paragraph{New techniques for rounding commuting local Hamiltonians}

Our guided reductions will be constructed using the building block of ``rounding schemes'' (\Cref{def:rounding_scheme})
in which a Hamiltonian $H$ over a set of registers $\reg R = \reg R_1 \otimes \dots \otimes \reg R_n$ is ``rounded'' to
a Hamiltonian $\tilde H$ over a possibly smaller space, as in \Cref{fig:rounding_to_reduction}. In particular, $\tilde
H$ is defined over registers $\tilde{\reg R} = \tilde{\reg R}_{i_1} \otimes \dots \otimes \tilde{\reg R}_{i_\ell}$, with
$\tilde{\reg R}$ is a subspace of $\reg{R}$. As long as we can ensure $\tilde H$ remains commuting and has a zero-energy ground
state if and only if $H$ does, this step will allow us to iteratively \emph{simplify} the Hamiltonian. This idea of
rounding projectors is not new (prior works such as \cite{Bravyi_BV2004_CommutativeVersionKlocal,
Irani_IJ2023_CommutingLocalHamiltonian} can be viewed as implementing a rounding scheme). Our contribution, however, is
to formalize this notion and introduce more general rounding schemes.

\begin{figure}
    \centering
    \begin{tikzpicture}[
        vertexp/.style = {draw, fill, Sepia, circle, minimum width=0.15cm, inner sep=0},
        vertexs/.style = {draw, fill, Sepia, circle, minimum width=0.1cm, inner sep=0},
        my label/.style n args={2}{label={[font=\scriptsize]#1:#2}},
        term/.style = {SpringGreen, line width=0.4mm, fill=SpringGreen, fill opacity=0.3, rounded corners}
    ]
        \begin{pgfonlayer}{fg}
        \foreach \x in {0, 1, 2} {
            \foreach \y in {0, 1} {
                \pgfmathtruncatemacro{\num}{\x+3*\y}
                \node[vertexp, my label={above left}{$\reg R_\num$} ] (u\x\y) at (\x, \y) {};
            }
        }
        \end{pgfonlayer}

        \draw[term] ($ (u00) - (0.25,0.25) $) rectangle ++ (0.5, 1.5);
        \draw[term] ($ (u00) - (0.25,0.25) $) rectangle ++ (1.5, 0.5);
        \draw[term] ($ (u10) - (0.25,0.25) $) rectangle ++ (0.5, 1.5);
        \scoped{
            \path[clip] ($ (u11) - (0.27,0.27) $) -- ($ (u11) - (0.27,-0.27) $) -- ($ (u21) + (0.27,0.27) $) --
            ($ (u21) + (0.27, 0) $) -- (u21.center) -- ($ (u21) - (0.27,0.27) $) -- cycle;
            \draw[term] ($ (u11) - (0.25,0.25) $) rectangle ++ (1.5, 0.5);
        }
        \scoped{
            \clip[overlay] ($ (u20) - (0.27,0.27) $) -- ($ (u21) - (0.27,0.27) $) -- (u21.center) -- ($ (u21) + (0.27, 0)
            $) -- ($ (u20) + (0.27,-0.27) $) -- cycle;
            \draw[term] ($ (u20) - (0.25, 0.25) $) rectangle ++ (0.5, 1.5);
        }

        \draw[line width=1pt, double distance=1pt, arrows = {-Latex[width=0pt 3, length=0pt 3 .5]}] (3,0.5) -- node[midway] (arrowmid) {} (4,0.5);
        \node[my label={below}{$\substack{\text{First}\\\text{Rounding}}$}] at (arrowmid) {};

         \begin{pgfonlayer}{fg}
         \foreach \x in {5, 6, 7} {
             \foreach \y in {0, 1} {
                 \pgfmathtruncatemacro{\dx}{\x-5}
                 \pgfmathtruncatemacro{\num}{\dx+3*\y}
                 \ifthenelse{\dx = 0 \AND \y = 0}{
                     \node[] (v\dx\y) at (\x, \y) {};
                     \node[vertexs, my label={left}{$\tilde{\reg R}_0^a$} ] (av\dx\y) at ($ (\x, \y) - (0.1, -0.1)
                     $){};
                     \node[vertexs, my label={below}{$\tilde{\reg R}_0^b$} ] (bv\dx\y) at ($ (\x, \y) + (0.1,-0.1)
                     $){};
                 }{
                     \node[vertexp, my label={above left}{$\reg R_\num$} ] (v\dx\y) at (\x, \y) {};
                 }
             }
         }
         \end{pgfonlayer}

        \draw[term] ($ (v00) - (0.25, 0.25) $) {[sharp corners] --  ($ (v00) + (0.25, 0.25) $)} -- ($ (v01) +
        (0.25,0.25) $) -- ($ (v01) + (-0.25,0.25) $) [sharp corners] -- cycle;
        \draw[term] ($ (v00) - (0.25, 0.25) $) -- ($ (v10) + (0.25,-0.25) $) -- ($ (v10) + (0.25, 0.25) $) {[sharp corners]  -- ($ (v00) + (0.25,0.25) $) -- cycle};
        \draw[term] ($ (v10) - (0.25,0.25) $) rectangle ++ (0.5, 1.5);
        \scoped{
            \path[clip] ($ (v11) - (0.27,0.27) $) -- ($ (v11) - (0.27,-0.27) $) -- ($ (v21) + (0.27,0.27) $) --
            ($ (v21) + (0.27, 0) $) -- (v21.center) -- ($ (v21) - (0.27,0.27) $) -- cycle;
            \draw[term] ($ (v11) - (0.25,0.25) $) rectangle ++ (1.5, 0.5);
        }
        \scoped{
            \clip[overlay] ($ (v20) - (0.27,0.27) $) -- ($ (v21) - (0.27,0.27) $) -- (v21.center) -- ($ (v21) + (0.27, 0)
            $) -- ($ (v20) + (0.27,-0.27) $) -- cycle;
            \draw[term] ($ (v20) - (0.25, 0.25) $) rectangle ++ (0.5, 1.5);
        }

        \draw[line width=1pt, double distance=1pt, arrows = {-Latex[width=0pt 3, length=0pt 3 .5]}] (8,0.5) --
        node[midway] (arrowmid2) {} (9,0.5);
        \node[my label={below}{$\substack{\text{Second}\\\text{Rounding}}$}] at (arrowmid2) {};

         \begin{pgfonlayer}{fg}
         \foreach \x in {10, 11, 12} {
             \foreach \y in {0, 1} {
                 \pgfmathtruncatemacro{\dx}{\x-10}
                 \pgfmathtruncatemacro{\num}{\dx+3*\y}
                 \ifthenelse{\dx = 0 \AND \y = 0}{
                     \node[] (w\dx\y) at (\x, \y) {};
                     \node[vertexs, my label={left}{$\tilde{\reg R}_0^a$} ] (aw\dx\y) at ($ (\x, \y) - (0.1, -0.1)
                     $){};
                     \node[vertexs, my label={below}{$\tilde{\reg R}_0^b$} ] (bw\dx\y) at ($ (\x, \y) + (0.1,-0.1)
                     $){};
                 }{
                     \ifthenelse{\num = 4}{
                         \node[] (w11) at (11.5,1) {};
                         \node[vertexp, fill=none, Sepia!40] at (\x, \y) {};
                     }{
                         \node[vertexp, my label={above left}{$\reg R_\num$} ] (w\dx\y) at (\x, \y) {};
                     }
                 }
             }
         }
         \end{pgfonlayer}

        \draw[term] ($ (w00) - (0.25, 0.25) $) {[sharp corners] --  ($ (w00) + (0.25, 0.25) $)} -- ($ (w01) +
        (0.25,0.25) $) -- ($ (w01) + (-0.25,0.25) $) [sharp corners] -- cycle;
        \draw[term] ($ (w00) - (0.25, 0.25) $) -- ($ (w10) + (0.25,-0.25) $) -- ($ (w10) + (0.25, 0.25) $) {[sharp corners]  -- ($ (w00) + (0.25,0.25) $) -- cycle};
        \draw[term] ($ (w10) - (0.25,0.25) $) rectangle ++ (0.5, 1);
        \scoped{
            \path[clip] ($ (w11) - (0.27,0.27) $) -- ($ (w11) - (0.27,-0.27) $) -- ($ (w21) + (0.27,0.27) $) --
            ($ (w21) + (0.27, 0) $) -- (w21.center) -- ($ (w21) - (0.27,0.27) $) -- cycle;
            \draw[term] ($ (w11) - (0.25,0.25) $) rectangle ++ (1, 0.5);
        }
        \scoped{
            \clip[overlay] ($ (w20) - (0.27,0.27) $) -- ($ (w21) - (0.27,0.27) $) -- (w21.center) -- ($ (w21) + (0.27, 0)
            $) -- ($ (w20) + (0.27,-0.27) $) -- cycle;
            \draw[term] ($ (w20) - (0.25, 0.25) $) rectangle ++ (0.5, 1.5);
        }

        \draw [line width=0.4mm, decoration={ brace, mirror, raise=0.5cm }, decorate ] ($ (u00) - (0.3,0.5) $) -- node[midway]
        (bracemid) {} ($ (w20) + (0.3,-0.5) $);
        \node[label={[label distance=0.5cm]below:Guided Reduction}] at (bracemid) {};
    \end{tikzpicture}
    \captionsetup{width=0.8\textwidth}
    \caption{
        An example of a guided reduction. On the left is a Hamiltonian over registers $\reg R_1 \otimes \dots \otimes \reg R_5$, with the green regions
        denoting individual Hamiltonian terms. A guided reduction is composed of a series of applications of a rounding schemes, which iteratively
        simplify the Hamiltonian. In the first round, the register $\tilde{\reg R}_0$ is decomposed into registers
        $\tilde{\reg R}_0^a$ and $\tilde{\reg R}_0^b$ (while the non-tilde'd registers are unchanged). In the second
        round, $\tilde{\reg R}_4$ is entirely removed.
    }
    \label{fig:rounding_to_reduction}
\end{figure}

In \cite{Irani_IJ2023_CommutingLocalHamiltonian}, the authors devised a rounding scheme in the case where all but a
single Hamiltonian term commutes with a set of projectors. In our work, we construct a rounding scheme which works even
when there are two non-commuting terms, increasing the generality of this approach.
\begin{theorem}[Rounding pairs of projectors (informal version of \Cref{thm:rounding})]
    Let $H = \sum h$ be a commuting local Hamiltonian and $\pi_1$, $\pi_2$ be a pair of projectors such that for all but one term $h$ that does not commute with $\pi_1$, it is removed by $\pi_2$ (in the sense that $\pi_2 h \pi_2 = 0$), and vice versa.
    Then $H$ can be rounded to a new commuting local Hamiltonian $\tilde H$ that has a zero-energy ground state if and only if $H$ has a $0$ energy ground state.
\end{theorem}

Our main insight behind this theorem is that in the case when only two local projectors survive, we can recover a
commuting local Hamiltonian by projecting onto the Jordan blocks of the pair that are non-zero (see
\Cref{lem:jordans_lemma}). We can view this as an extension of the rounding scheme from
\cite{Irani_IJ2023_CommutingLocalHamiltonian} to pairs of projectors, where we look at blocks of size $2$ instead of
size $1$. The use of Jordan's Lemma requires a much more careful and detailed analysis of the commutative properties of
Jordan blocks.

\paragraph{Characterization of commutation for rank $1$ Hamiltonians}

Our second technical result is a characterization of rank $1$ commuting Hamiltonians.
The Structure Lemma can be viewed as providing a general scaffolding for how two operators can commute.
When both operators are restricted to be rank $1$, we can strengthen the Structure Lemma to show that commutation occurs in one of two specific ways.
\begin{theorem}[Characterization of commuting rank $1$ projectors (informal version of \Cref{thm:rank_1_commutation_main})]
    Let $P$ and $Q$ be rank $1$ projectors so that $P$ acts non-trivially only on registers $\reg A, \reg B$ and $Q$ acts non-trivially only on $\reg B, \reg C$. Further assume that $P$ and $Q$ commute.  Then, restricted to the subspace $\reg B$, either
    \begin{enumerate}
        \item (Singular commutation) both $P$ and $Q$ are projections onto the same state $\psi$, or
        \item (Reducing commutation) $P$ and $Q$ are projections onto orthogonal spaces.
    \end{enumerate}
\end{theorem}

This result allows us to employ a strategy of ``puncturing and repairing'' paths through local Hamiltonians. In
particular, we can partition the local terms adjacent to a qudit into two sets such that terms in each set interact
\emph{only}  on the qudit. Within each set, if we find that the local terms commute in a singular way, we can show that
the qudit is a so-called classical qudit and can be removed. In the case when the local terms commute in a reducing way,
the prover can provide a pair of projector that removes all but two of the terms. The prover's projectors essentially
cuts a small patch into the local Hamiltonian, but leaves terms that no longer commute.  We patch these holes using the
rounding technique described previously, and by repeating this protocol we can cut long strings into geometrically local
commuting Hamiltonians. 

\paragraph{Cubulation and triangulation techniques}

Our final contribution is a more detailed application of the triangulation technique of
\cite{Aharonov_AKV2018_ComplexityTwoDimensional}. In particular, we use a technique we call ``cubulation'', where we
super-impose a 3D cubic lattice on top of a general polyhedral lattice.  We subsequently use our cutting and repairing
technique to remove Hamiltonian terms intersecting both the vertices and edges of the super-imposed cubic lattice,
allowing us to reduce a 3D commuting local Hamiltonian problem to a 2-local problem, which is known to be in
$\mathsf{NP}$.

We note that applying the cubulation technique to a general polyhedral lattice requires dealing with many different cases
related to how the Hamiltonian terms intersect with the edges of the super-imposed lattice.

\paragraph{Verifiers for commuting local Hamiltonians}

In \Cref{sec:verifiers}, we provide a simple model of verifiers that captures the complexity of commuting local Hamiltonians.
As we discuss in the next subsection, we believe that demonstrating hardness for commuting local Hamiltonian problems is one of the most interesting directions of research related to commuting Hamiltonians.
We admit that the model we provide does not make any progress towards this goal, but we hope that providing a tangible ``complexity class'' associated with commuting local Hamiltonians, we will at least provide a language that can be used to describe this problem.

\subsection{Discussion and future work}

In this paper, we present a number of new results related to the geometric commuting local Hamiltonian problem.  
We show that in 2D, verifying the ground energy of rank-1 commuting local Hamiltonians of \emph{any} local dimension are in $\mathsf{NP}$.  
We do this by combining the triangulation technique from \cite{Aharonov_AKV2018_ComplexityTwoDimensional} with a novel commuting Hamiltonian rounding scheme with a characterization of commutation for rank-1 commuting local Hamiltonians to puncture paths in 2D lattices.
The rounding technique we provide works for much more generic conditions than rank-1 commuting local Hamiltonians, and we believe it is of independent interest.  

Secondly, we show that a wide family of 3D commuting local Hamiltonians are in $\mathsf{NP}$.  
We show how to extend the triangulation technique to a cubulation technique.  
Using our puncturing technique, we can remove local terms that lie on either edges or vertices of the cubic lattice, again reducing the problem to the 2-local commuting local Hamiltonian problem.
We note that neither of our reductions are ``constructive,'' in that it is not clear how to recover a succinct description of the commuting local Hamiltonian, even given the proof that it is in $\mathsf{NP}$.

There are several questions that could be consider direct follow ups to our work.  For example, what is the complexity of general rank-1 commuting local Hamiltonians.  Can any tools be developed that allow for analysis of commuting Hamiltonians, even those that are rank-1 that do not have geometric locality?
We note that our techniques ultimately rely heavily on triangulation or cubulation.  
Although we could potentially apply our rounding technique to puncture holes in more complicated commuting local Hamiltonians, without geometric locality it is very difficult to see how puncturing holes or paths through a Hamiltonian could be useful for reducing the problem to the 2-local commuting local Hamiltonian problem.

Another natural question to ask is whether our results, and those of \cite{Irani_IJ2023_CommutingLocalHamiltonian} can be turned into constructive proofs.  
Opposed to this question, the authors wonder if commuting local Hamiltonians could be an instance where state or unitary complexity differ from decision complexity, for example, if it was shown that the complexity of preparing ground states of commuting local Hamiltonians in 2D or 3D is outside of $\mathsf{stateBQP}^{\mathsf{NP}}$.
This would constitute one of the first instances where these theories differ.

Finally, the authors note that almost all results pertaining to commuting local Hamiltonians (this one included) have been focused on showing that increasingly complicated versions of the commuting local Hamiltonian problem are classically verifiable.  
Despite having limited evidence, from the work of \cite{gosset2017qcma}, that commuting local Hamiltonians should be harder than $\mathsf{NP}$, there has been no formal evidence that commuting local Hamiltonians are hard for any class larger than $\mathsf{NP}$.  
The authors believe providing \emph{any} evidence that commuting local Hamiltonians go beyond $\mathsf{NP}$ is a fascinating open question.
A natural class to study is ``next biggest class'', $\mathsf{MA}$, although we think that such a reduction to commuting local Hamiltonian problems might entirely capture the difficulty of this problem.
Specifically, we can think of a $\mathsf{MA}$ protocol as starting from some state $\ket{x||0}$, first measuring some qubits in the $\{\proj{+}, \proj{-}\}$ basis, and then performing some reversible quantum circuit.
Thus, simulating $\mathsf{MA}$ with a commuting local Hamiltonian would probably require dealing with a single non-commuting measurement.  It is known that adding a single non-commuting measurement into a commuting local Hamiltonian yields a class of local Hamiltonians that is complete for $\mathsf{QMA}$.

\ifanon
\else
\paragraph{Acknowledgements}
We thank Sandy Irani and Jiaqing Jiang for helpful discussions related to the Structure Lemma and its limitations. We also thank Chinmay Nirkhe for recommending to us the problem of commuting local Hamiltonians.

YH is supported by the National Science Foundation Graduate Research Fellowship under Grant No. 2140743. Any opinion, findings, and conclusions or recommendations expressed in this material are those of the authors(s) and do not necessarily reflect the views of the National Science Foundation. JB is supported by Henry Yuen's AFORS (award FA9550-21-1-036) and NSF CAREER (award CCF2144219). This work was done in part while the authors were visiting the Simons Institute for the Theory of Computing, supported by NSF QLCI Grant No. 2016245.
\fi

\section{Technical overview}

\subsection{Rounding schemes via the Structure Lemma}
\label{sec:rounding_via_structure_lemma}
As noted in the introduction, the central tool used in the study of the commuting local Hamiltonian problem has been the
Structure Lemma of \cite{Bravyi_BV2004_CommutativeVersionKlocal}. An easy consequence of this lemma is the following
corollary,
\begin{restatable*}[Structure of two commuting operators]{cor}{commutingstructure}
    \label{lem:structure_two_terms}
    Let $\mathcal{A}_h \subseteq \mc L(\reg{R})$ be the $C^*$-algebra induced by a Hermitian $h$ and $\mathcal{A}_{h'}$ be the $C^*$-algebra on $\reg R$ induced by $h'$ that commutes with $h$.  Let $\reg{R}_{i}^{j}$ be the decomposition induced by \cref{lem:structure} applied to $\mathcal{A}_h$.  Then the following holds:
    \begin{align}
        \mathcal{A}_h &= \bigoplus_i \mathcal{L}(\reg{R}_i^{1}) \otimes \id(\reg{R}_i^{2})\\
        \mathcal{A}_{h'} &\subseteq \bigoplus_i \id(\reg{R}_i^{1}) \otimes \mathcal{L}(\reg{R}_i^{2})\,.
    \end{align}
    Crucially, all operators keep the decomposition $\reg{R} = \bigoplus_{i} \reg{R}_i$ invariant.
\end{restatable*}
Informally, this states that by projecting onto a subspace spanned by $\reg R_i$, the operators $h$ and $h'$ can be
thought of as decoupled, with $h$ acting non-trivially only on $\reg R_i^1$, and $h'$ acting non-trivially only on $\reg
R_i^2$.

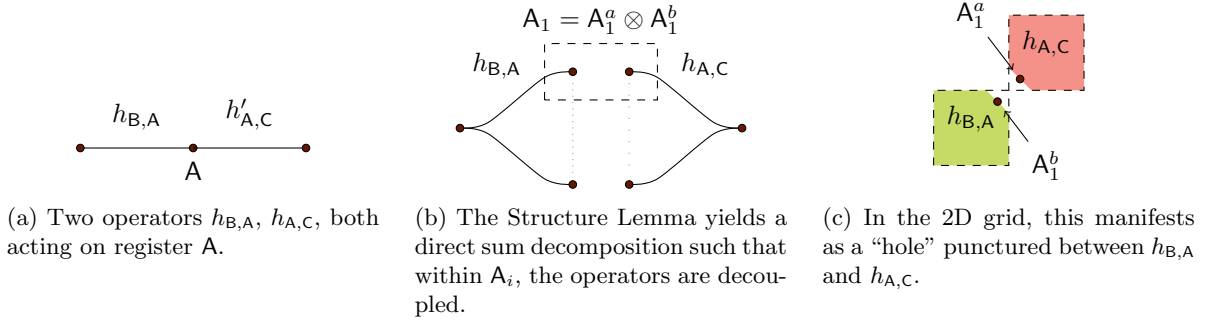
\begin{figure}
    \centering
    \begin{subfigure}[t]{0.3\textwidth}
        \centering
        \begin{tikzpicture}[
            scale=1.5,
            vertexp/.style = {draw, fill=Sepia, circle, minimum width=0.1cm, inner sep=0}
        ]
            \draw (-1,0) node[vertexp] {} -- node[midway, label={above:$h_{\reg B, \reg A}$}] {} (0,0);
            \node[vertexp, label={below:$\reg A$}] at (0,0) {};
            \draw (0,0) -- node[midway, label={above:$h'_{\reg A, \reg C}$}] {} (1,0) node[vertexp] {};
        \end{tikzpicture}
        \caption{Two operators $h_{\reg B, \reg A}$, $h_{\reg A, \reg C}$, both acting on register $\reg A$.}
    \end{subfigure}
    \quad
    \begin{subfigure}[t]{0.3\textwidth}
        \centering
        \begin{tikzpicture}[
            scale=1.5,
            vertexp/.style = {draw, fill=Sepia, circle, minimum width=0.1cm, inner sep=0}
        ]
            \draw plot [smooth] coordinates {(0,0) (0.25,0.05) (0.75,0.45) (1,0.5)};
            \node[vertexp] at (1,0.5) {};
            \draw[dotted,black!40] (1,0.4) -- (1,-0.4);
            \draw plot [smooth] coordinates {(0,0) (0.25,-0.05) (0.75,-0.45) (1,-0.5)};
            \node[vertexp, label={[yshift=0.5cm]above right:$h_{\reg B, \reg A}$}] at (0,0) {};
            \node[vertexp] at (1,-0.5) {};
            \draw[dashed] (0.75, 0.25) rectangle ++(1, 0.5) node[fitting node,label={above:$\reg A_1 = \reg A_1^a \otimes \reg A_1^b$}] {};
            
            \draw plot [smooth] coordinates {(2.5,0) (2.25,0.05) (1.75,0.45) (1.5,0.5)};
            \node[vertexp] at (1.5,0.5) {};
            \draw[loosely dotted,black!40] (1.5,0.4) -- (1.5,-0.4);
            \draw plot [smooth] coordinates {(2.5,0) (2.25,-0.05) (1.75,-0.45) (1.5,-0.5)};
            \node[vertexp] at (1.5,-0.5) {};
            \node[vertexp, label={[yshift=0.5cm]above left:$h_{\reg A, \reg C}$}] at (2.5,0) {};
        \end{tikzpicture}
        \caption{The Structure Lemma yields a direct sum decomposition such that within $\reg A_i$, the operators are decoupled.}
    \end{subfigure}
    \quad
    \begin{subfigure}[t]{0.3\textwidth}
        \centering
        \begin{tikzpicture}[
            vertexp/.style = {draw, fill=Sepia, circle, minimum width=0.1cm, inner sep=0}
        ]
            \draw[draw=none, fill=SpringGreen] (0,0) -- (0,1) -- (0.7,1) -- node[midway] (botmid) {} (1,0.7) -- (1,0) -- cycle;
            \draw[fill=none, dashed] (0,0) rectangle ++(1,1) node[fitting node] {$h_{\reg B, \reg A}$};
            \draw[draw=none, fill=Salmon] (2,2) -- (1,2) -- (1,1.3) -- node[midway] (topmid) {} (1.3,1) -- (2,1) -- cycle;
            \draw[fill=none, dashed] (1,1) rectangle ++(1,1) node[fitting node] {$h_{\reg A, \reg C}$};

            \node (q1) at (0.5,2) {$\reg A_1^a$};
            \node[vertexp] at (topmid) {};
            \node (q2) at (1.5,0) {$\reg A_1^b$};
            \node[vertexp] at (botmid) {};

            \draw[->] (q1) -- (topmid);
            \draw[->] (q2) -- (botmid);
        \end{tikzpicture}
        \caption{In the 2D grid, this manifests as a ``hole'' punctured between $h_{\reg B, \reg A}$ and $h_{\reg A, \reg C}$.}
    \end{subfigure}
    \caption{Puncturing holes via the Structure Lemma.}
    \label{fig:simple_structure_lemma}
\end{figure}

\paragraph{$2$-local setting.}\label{par:2local} In the 2-local setting of
\cite{Bravyi_BV2004_CommutativeVersionKlocal}, every pair of terms $h, h'$ either do not interact, or interact
\emph{only} on a single-qudit register $\reg R$. This means that for each register, a single decomposition $\reg R =
\oplus_i \reg R^i$ can be constructed so that every operators acts invariantly on the corresponding projectors $\pi_i
\dfn (\pi_i)_{\reg{R}_i}$. Concretely, each term $h$ is \emph{block diagonal} with respect to the subspaces $\reg R_i$. This
decomposition naturally lends itself to a rounding scheme, where the projector is chosen to be one of the $\pi_i$'s.
The resulting Hamiltonian is
\[
    \tilde H_i = \sum_{h \in H} \pi_i h \pi_i
\]
Certainly $\lambda_0(\tilde H_i) = 0 \implies \lambda_0(H) = 0$. For the other direction, we use that for any CLH
instance, $\lambda_0(H) = 0 \iff \tr[\prod_j (\Id - h_j)] = \sum_i \tr[\prod_h \pi_i (\Id - h) \pi_i] = 0$. Moreover,
since all $h$ commute with $\pi_i$, each trace in the sum is non-negative. Thus we get the following equivalent
condition to the existence of a ground state:
\begin{equation}
    \label{eq:simplest_trace_eq}    
    \lambda_0(H) = 0 \iff \exists i \text{ s.t. } \tr\Brac{\prod_j (\pi_i - \pi_i h_j \pi_i)} > 0\,.
\end{equation}
The expression on the right is equivalent to existence of a zero-energy eigenstate for the Hamiltonian $\tilde H_i =
\sum_h \pi_i h \pi_i$, defined over registers $(\reg R_1, \dots, \pi_i \reg R_i \pi_i, \dots, \reg R_n)$, where $\pi_i \reg{R}_i \pi_i \leq \reg{R}_i$ is the sub-space generated by projecting vectors of $\reg{R}_i$ onto $\pi_i$. Thus, mapping $H
\rightarrow \tilde H_i$ yields our desired rounding scheme. Crucially, the resulting Hamiltonian is also \emph{simpler}
than the original; the Structure Lemma tells us that $\reg R_i = \otimes_h \reg R_i^{(h)}$ where each $\pi_i h \pi_i$
acts on non-trivially only on $\reg R_i^{(h)}$. This effectively decouples all terms on $\reg R$, as depicted in
\Cref{fig:simple_structure_lemma}.

\paragraph{$k$-local setting.}\label{par:klocal} For higher localities, the situation becomes more complicated. In
general, if one has a 2D CLH instance with a $k$-wise interaction on a register $\reg R$, then commutation between
adjacent terms is not entirely determined by the terms' structure on $\reg R$. For instance, imagine $h = \ketbra 1
\otimes \tilde h_{\reg R}$ and $h' = \ketbra 0 \otimes \tilde h'_{\reg R}$; these terms commute regardless of our choice
of $\tilde h$ and $\tilde h'$! However, a variation of the above argument can be made to work.
\begin{figure}[b]
    \centering
    \begin{tikzpicture}[
        vertexp/.style = {draw, fill=Sepia, circle, minimum width=0.1cm, inner sep=0}
    ]
        \node[draw, rectangle, fill=SpringGreen, minimum width=1cm, minimum height=1cm] at (0,0) {$h_4$};
        \node[draw, rectangle, fill=SpringGreen, minimum width=1cm, minimum height=1cm] at (1,1) {$h_2$};
        \node[draw, rectangle, fill=pink, minimum width=1cm, minimum height=1cm] at (0,1) {$h_1$};
        \node[draw, rectangle, fill=pink, minimum width=1cm, minimum height=1cm] at (1,0) {$h_3$};

        \node[vertexp,label={[label distance=-0.1cm]above left:$\reg R$}] at (0.5,0.5) {};
    \end{tikzpicture}
    \caption{A degree $4$ register $\reg R$.}
    \label{fig:basic_grid}
\end{figure}
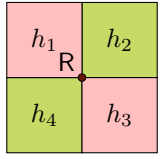
Consider when $k = 4$, as in \Cref{fig:basic_grid}. Then the pairs $h_2$ and $h_4$ induce some
decomposition $\reg R= \oplus_i \reg R_i$, and pairs $h_1$ and $h_3$ induce a decomposition $\reg R = \oplus_i
\tilde{\reg R}_i$. In general these decompositions are not the same. Let the projectors for these subspaces be
$\{\pi_i\}_i$ and $\{\tilde \pi_i\}_i$. Then, rather than \Cref{eq:simplest_trace_eq}, we get
\begin{equation}
    \label{eq:double_trace_eq}    
    \lambda_0(H) = 0 \iff \exists i,j \text{ s.t. } \tr\Brac{\tilde \pi_j (\Id - h_1) \tilde \pi_j\tilde
    \pi_j (\Id - h_3) \tilde \pi_j \pi_i (\Id - h_2) \pi_i  \pi_i (\Id - h_4) \pi_i \prod_{j > 4} (\Id - h_j)} > 0
\end{equation}
As in the $2$-local case, this equivalence only holds when the traces on the RHS are non-negative. Fortunately, the two matrices
\begin{align}
    A &\dfn \tilde \pi_j (\Id - h_1) \tilde \pi_j\tilde \pi_j (\Id - h_3) \tilde \pi_j\\
    B &\dfn \pi_i (\Id - h_2) \pi_i  \pi_i (\Id - h_4) \pi_i \prod_{j > 4} (\Id - h_j)
\end{align}
are both PSD, and $\tr[AB] \geq 0$. Note that this is sensitive to ordering and the same may not hold if we order the
terms as $h_1, h_2, h_3, h_4$. As this equivalence will be heavily used going forward, we formalize it in the following
lemma.
\begin{restatable}[Equivalence under projectors, implicit in
    \cite{Irani_IJ2023_CommutingLocalHamiltonian}]{lemma}{equivprojector}
    \label{lem:equiv_projector}
    Suppose we have a CLH instance $H = \sum_{i \in [m]} h_i$ and two POVM measurements $\Pi_1 = \{\pi_1^1, \dots,
    \pi_{k_1}^1\}$ and $\Pi_2 = \{\pi_1^2, \dots, \pi_{k_2}^2\}$ and
    disjoint sets of Hamiltonian terms $S, T \subseteq [m]$ such that,
    \begin{itemize}
        \item All terms $h_i$ with $i \in [m] \setminus T$ commute with $\Pi_1$.
        \item All terms $h_i$ with $i \in [m] \setminus S$ commute with $\Pi_2$.
        \item All remaining terms $h_i$ with $i \in [m] \setminus (T \cup S)$ commute with both $\Pi_1$ and $\Pi_2$.
    \end{itemize}
    Then,
    \begin{equation}
        \lambda_0(H) = 0 \iff \exists i \in [k_1], j \in [k_2] \text{ such that } \tr\left[\prod_{i \in S}(\pi^1_{i} - \pi^1_i h_1 \pi^1_i)
        \prod_{i \in T} (\pi^2_j - \pi^2_j h_2 \pi^2_j) \overline H_\text{rest}\right]\,,
    \end{equation}
    where $\overline H_\text{rest} \dfn \prod_{i \in [m] \setminus (S \cup T)} (\id - h_i)$. Note that the ordering of the terms on the RHS matters.
\end{restatable}
The proof is deferred to \Cref{sec:misc_lemmas}. \Cref{eq:double_trace_eq} is recovered by taking $S = \{1,2\}$ and $T =
\{3, 4\}$. This lemma therefore yields a statement equivalent to \Cref{eq:simplest_trace_eq}. However, even with this
equivalence it is not clear how to convert the right-hand-side into a CLH instance. Not only do the terms $\tilde \pi_j
h_1 \tilde \pi_j$ and $\pi_i h_2 \pi_i$ not commute, we do not have a consistent subspace onto which can restrict $\reg
R$.

Previous works dealt with this in a couple different ways. In \cite{Schuch_Sch2011_ComplexityCommutingHamiltonians}, any
non-trivial decomposition (i.e. $|\{\reg R_i\}_i| > 1$) on qubits (so dimension 2 registers $\reg R$) requires $\dim(\reg R_i) = 1$. Therefore,
\[
 \tilde \pi_j = \ketbra \psi \quad \text{ and } \quad \pi_i = \ketbra \phi\,.
\]
For $h_2$ and $h_4$ we perform the same restriction as before:
\begin{alignat*}{2}
    h_2 &\rightarrow \ketbra \phi h_2 \ketbra \phi = \ketbra \phi \otimes \tilde h_2 &\qquad& h_1 \rightarrow \ketbra
\psi h_1 \ketbra \psi = \ketbra \psi \otimes \tilde h_1\\
    h_4 &\rightarrow \ketbra \phi h_4 \ketbra \phi = \ketbra \phi \otimes \tilde h_4 && h_3 \rightarrow \ketbra \psi h_3
    \ketbra \psi = \ketbra \psi \otimes \tilde h_3\,.
\end{alignat*}
After this step, $\tilde h_1$ and $\tilde h_3$ may no longer commute. However, notice that every term acts as $\ketbra
\phi$ on $\reg R$ and thus this register (which \cite{Schuch_Sch2011_ComplexityCommutingHamiltonians} calls a \emph{removable} qudit) can traced out from the system. \cite{Schuch_Sch2011_ComplexityCommutingHamiltonians}
shows that ``puncturing'' out all removeable qudits yields a 1-dimensional Hamiltonian. Thus, the author obtains a
guided reduction (where the prover has provided $\tilde \pi_j$ and $\pi_i$) from 2-local qubit CLHs on the 2D lattice
to 1-dimensional (non-commuting) local Hamiltonians. Such Hamiltonians can be easily seen to be contained in $\cNP$.

It is not clear how to extend this argument beyond qudit dimension $d = 2$, as the subspaces from the Structure Lemma
are no longer 1 dimensional and thus the Hamiltonian terms may act non-trivially within each subspace. Still,
\cite{Irani_IJ2023_CommutingLocalHamiltonian} show that in the case of qutrits, Schuch's techniques can be extended. One
of their key insights is identifying \emph{semi-separable} qutrits, which means that there is a non-trivial
decomposition on which \emph{all but one} term acts invariantly. In this case,
\cite{Irani_IJ2023_CommutingLocalHamiltonian} demonstrate a rounding technique which reduces the original Hamiltonian
$H$ into another CLH instance $H'$ with all semi-separable qutrits removed (the authors call this a ``self-reduction'').
Following this step, the Structure Lemma is applied and a careful casework of possible resulting decompositions of $H'$
allow the authors to recover a 1D structure.

\subsection{Our Techniques}
\label{sec:techniques}
%

In this work, we combine a improved rounding scheme with the idea of ``triangulations'' from
\cite{Aharonov_AKV2018_ComplexityTwoDimensional}.

\tikzset{pics/tile1/.style n args={3}{code={
    \draw (#1,#2) -- ($ (2,0) + (#1,#2) $) -- ($ (1,1.72) + (#1,#2) $) -- cycle ;
    \node (c#1#3) at ($ (#1, #2) + (1,0.86)$) {};
}}}
\tikzset{pics/tile2/.style n args={3}{code={
    \draw ($ (2,1.72) + (#1,#2) $) -- ($ (1,0) + (#1,#2) $) -- ($ (0,1.72) + (#1,#2) $) -- cycle;
    \node (c#1#3) at ($ (#1, #2) + (1,0.86)$) {};
}}}
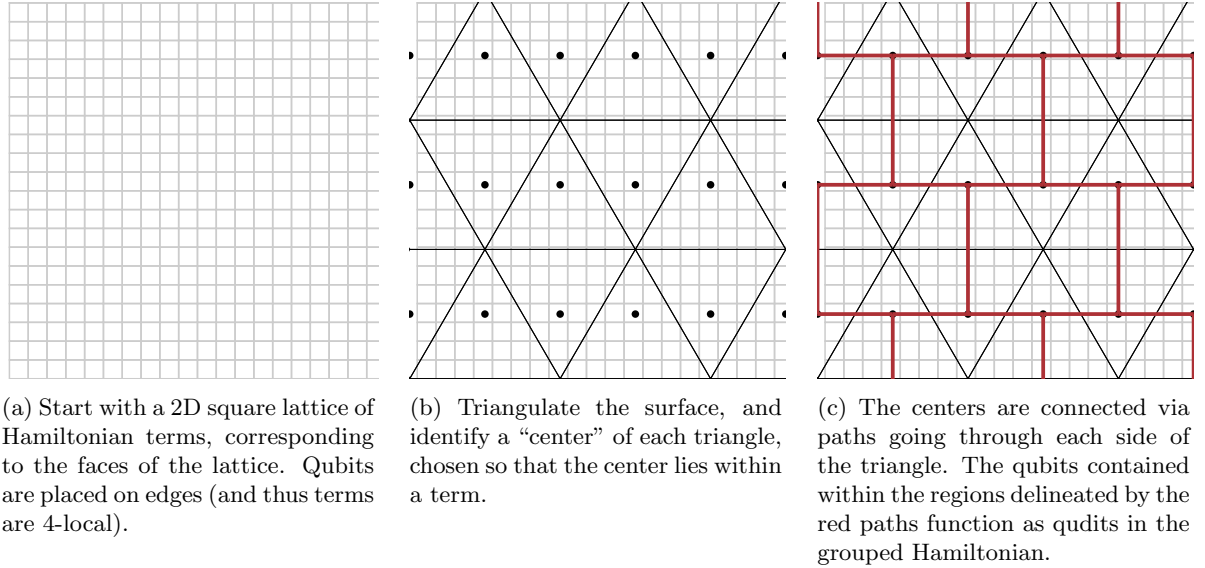
\begin{figure}
    \centering
    \begin{subfigure}[t]{0.3\textwidth}
        \centering
        \begin{tikzpicture}
            \begin{scope}
            \clip (0,0) rectangle ++(5,5);
            \newcommand\delt{0.25}
            \foreach \x in {0,...,29} {
                \foreach \y in {0,...,29} {
                    \draw[black!20,line width=0.2mm] ($ \delt*(\x,\y) + (0.1,0) $) rectangle ++(\delt,\delt);
                }
            }
            \end{scope}
            
        \end{tikzpicture}
        \caption{Start with a 2D square lattice of Hamiltonian terms, corresponding to the faces of the lattice. Qubits are placed on edges (and thus terms are $4$-local).}
        \label{fig:akv_step_0}
    \end{subfigure}
    \quad
    \begin{subfigure}[t]{0.3\textwidth}
        \centering
        \begin{tikzpicture}
            \begin{scope}
            \clip (0,0) rectangle ++(5,5);
            \newcommand\delt{0.25}
            \foreach \x in {0,...,29} {
                \foreach \y in {0,...,29} {
                    \draw[black!20,line width=0.2mm] ($ \delt*(\x,\y) + (0.1,0) $) rectangle ++(\delt,\delt);
                }
            }
            \foreach \x in {-3,-2,0,2,4} {
                \foreach \i in {-2,-1,0,1,2,3} {
                    \pgfmathsetmacro{\y}{\i*1.72}
                    \pgfmathtruncatemacro{\xd}{\x+1}
                    \ifthenelse{\equal{\intcalcMod{\i}{2}}{0}}{
                        \pic {tile1={\x}{\y}{\i}};
                        \pic {tile2={\xd}{\y}{\i}};
                    }{
                        \pic {tile2={\x}{\y}{\i}};
                        \pic {tile1={\xd}{\y}{\i}};
                    }
                } 
            }
            \foreach \x in {-2,-1,...,4,5} {
                \foreach \i in {-1,0,...,2,3} {
                    \node[fill,circle, minimum width=0.1cm,inner sep=0] at (c\x\i) {};
                }
            }
            \end{scope}
        \end{tikzpicture}
        \caption{Triangulate the surface, and identify a ``center'' of each triangle, chosen so that the center lies within a term.}
        \label{fig:akv_step_1}
    \end{subfigure}
    \quad
    \begin{subfigure}[t]{0.3\textwidth}
        \begin{tikzpicture}
            \begin{scope}
            \clip (0,0) rectangle ++(5,5);
            \newcommand\delt{0.25}
            \foreach \x in {0,...,29} {
                \foreach \y in {0,...,29} {
                    \draw[black!20,line width=0.2mm] ($ \delt*(\x,\y) + (0.1,0) $) rectangle ++(\delt,\delt);
                }
            }
            \foreach \x in {-3,-2,0,2,4} {
                \foreach \i in {-2,-1,0,1,2,3} {
                    \pgfmathsetmacro{\y}{\i*1.72}
                    \pgfmathtruncatemacro{\xd}{\x+1}
                    \ifthenelse{\equal{\intcalcMod{\i}{2}}{0}}{
                        \pic {tile1={\x}{\y}{\i}};
                        \pic {tile2={\xd}{\y}{\i}};
                    }{
                        \pic {tile2={\x}{\y}{\i}};
                        \pic {tile1={\xd}{\y}{\i}};
                    }
                } 
            }
            \foreach \x in {-2,-1,...,4,5} {
                \foreach \i in {-1,0,...,2,3} {
                    \node[fill,circle, minimum width=0.1cm,inner sep=0] at (c\x\i) {};
                    \pgfmathtruncatemacro{\xs}{\x-1}
                    \pgfmathtruncatemacro{\is}{\i-1}
                    \draw[Maroon, line width=0.5mm] (c\x\i.center) -- (c\xs\i.center);
                    \ifthenelse{\equal{\intcalcMod{\x+\i}{2}}{0}}{
                            \draw[Maroon, line width=0.5mm] (c\x\i.center) -- (c\x\is.center);
                    }{}
                }
            }
            \end{scope}
        \end{tikzpicture}
        \caption{The centers are connected via paths going through each side of the triangle. The qubits contained within the regions delineated by the red paths function as qudits in the grouped Hamiltonian.}
        \label{fig:akv_step_2}
    \end{subfigure}
    \captionsetup{width=0.8\linewidth}
    \caption{The process of reducing a 2D CLH instance to a 2-local CLH instance. Notice in the final figure that all terms besides those containing the center of each triangle are $2$-local. In \cite{Aharonov_AKV2018_ComplexityTwoDimensional} the center terms are simply removed and corrected later.}
    \label{fig:akv_triangulation}
\end{figure}
\paragraph{Guided reduction via triangulation.} The key insight from \cite{Aharonov_AKV2018_ComplexityTwoDimensional} is
that assuming enough terms are removed from a 2D CLH, the remaining system can be viewed as two local. For now, consider
a 2D grid (where terms are on the faces and qubits are on edges), as all the key ideas can be seen from this restricted
case. First, a triangulation is constructed over the 2D lattice (\Cref{fig:akv_step_1}). This step \emph{does not
depend} on the underlying Hamiltonian, although the \emph{centers} of the triangles should be selected as to reside
within some Hamiltonian term. Then, paths are drawn from the center, through each edge of the triangle, and connecting
an adjacent center. By construction, besides the center of the triangle, all terms are $2$-local.
\cite{Aharonov_AKV2018_ComplexityTwoDimensional} deal with the $3$-local center terms by showing that arbitrary 2D qubit
Hamiltonians are equivalent to a (generalization of) the Toric code. Using properties of the Toric code,
\cite{Aharonov_AKV2018_ComplexityTwoDimensional} argue that the Hamiltonian contains enough ``correctable'' terms so
that the triangle centers can be chosen to coincide with these correctable terms. These terms are removed, then
corrected after preparing the ground state of the $2$-local Hamiltonian.

\paragraph{Our rounding scheme.} In the reduction to $2$-local Hamiltonians of
\cite{Aharonov_AKV2018_ComplexityTwoDimensional}, all that is really needed is a way create holes within each triangle.
They accomplish this primarily through the removal of correctable terms.  At a high level, we use our rounding scheme to
puncture holes in the Hamiltonian in more varied ways, allowing us to apply the techniques from
\cite{Aharonov_AKV2018_ComplexityTwoDimensional} in more general settings than qubit 2D CLH instances.

We accomplish this through analyzing the possible local algebras of rank-1 commuting operators via the Structure Lemma.
In the simplest case, imagine there are only two operators acting non-trivially on a register $\reg R$. This is
essentially the \emph{2-local} setting of \cite{Bravyi_BV2004_CommutativeVersionKlocal}. There a simple consequence of
the Structure Lemma is the following lemma.

\begin{lemma}[Informal version of \Cref{cor:product_structure_2_local}]
    \label{lem:informal_remove_blocks}
    There exists a subspace (sub-register) $\reg R_i = \reg R_i^a \otimes \reg R_i^b \subseteq \reg R$ such that when
    $h$ and $h'$ are restricted to $\reg R_i$, $h$ acts non-trivially only on $\reg R_i^a$ and $h'$ acts non-trivially
    only on $\reg R_i^b$.
\end{lemma}

As long as this subspace has non-zero overlap with the ground space, restricting our Hamiltonian to $\reg R_i$
punctures a ``hole'' in the geometric structure of the Hamiltonian, as in \Cref{fig:simple_structure_lemma}. But like in
the discussion in \Cref{par:klocal}, this argument does not directly apply when interactions are more than $2$-local. To
generalize this argument, we show that the local algebra of rank-1 commuting operators can be classified as
\emph{reducing} or \emph{singular} where intuitively,
\begin{itemize}
    \item If two operators commute in a \emph{reducing} way, then in $\reg R_i$ at least one operator becomes
        the $0$-operator.
    \item If two operators commute in a \emph{singular} way, then $\dim \reg R_i = 1$ and both act as a scalar.
\end{itemize}
For each combination of reducing and singular cases, we show that our rounding scheme yields a Hamiltonian where a hole
has been punctured near $\reg R$ and thus a corner of the co-triangulation can be placed in this hole. A crucial part of
this argument is that rounding schemes yield \emph{another} CLH instance. This fact will be crucial in the 3D setting,
where after an initial application of our rounding scheme, there are still ``obstructions'' in the 3D surface.
However, these obstructions turn out to be $2$-local and, since our intermediate Hamiltonian is commuting, we may apply
the Structure Lemma via \Cref{lem:informal_remove_blocks} to remove the obstructions.

In comparison to previous techniques, 
\begin{enumerate}
    \item The semi-separable self-reduction of \cite{Irani_IJ2023_CommutingLocalHamiltonian} also retains commutativity.
        However, our reduction is more general and (assuming the rank-$1$ constraint) can be applied generically to
        single qudit register in the Hamiltonian.
    \item \cite{Schuch_Sch2011_ComplexityCommutingHamiltonians} reduction is also quite general, however, the final
        Hamiltonian generated is no longer commuting. This makes it hard to imagine how one would generalize their
        techniques to more complex structures, where one might want to perform the reduction \emph{recursively}.
    \item The result of \cite{Aharonov_AKV2018_ComplexityTwoDimensional} is in some sense closest to ours; for any
        triangulation and center, they can find a term (or adjacent term) which is removable. However, their ideas rely
        on the exact characterization of 2D qubit Hamiltonians as related to the defected Toric code.\footnote{The
        defected Toric code is a generalization of the Toric code where real coefficients are permitted in front of each
        term.}
\end{enumerate}

\paragraph{Paper Overview}
The remainder of the paper is laid out as follows.
\begin{itemize}
    \item In \Cref{sec:preliminaries} we introduce some notation and review the basics of the local Hamiltonian problem
        and the Structure Lemma, both of which will play a major part in this paper. Furthermore, we introduce some
        geometric variations of the commuting local Hamiltonian problem.
    \item In \Cref{sec:rounding}, we give the key technical tools we will need in this paper. In \Cref{sec:our_rounding}, we
        describe our \emph{rounding scheme}, which allows us to reduce from a CLH instance to a ``simpler'' CLH
        instance. To apply this rounding scheme to rank-1 CLH instances, we will need our characterization of local
        algebras of rank-1 CLHs as singular and reducing. This is done in \Cref{sec:local_alg_rank1}.
    \item The bulk of our results are contained in \Crefrange{sec:warmup_2d}{sec:proof_3d}.
    \begin{itemize}
        \item In \Cref{sec:warmup_2d}, we prove that \twodclhr{1} is in $\cNP$ (\Cref{thm:2D_rank1_np})
        \item In \Cref{sec:proof_3d}, we prove that \threedclhr{1} is in $\cNP$ (\Cref{thm:3d_theorem}).
    \end{itemize}
\end{itemize}

\section{Preliminaries}
\label{sec:preliminaries}

\subsection{Quantum preliminaries and notation}

A register $\reg{R}$ is a named finite-dimensional complex Hilbert space.  If $\reg{A}$, $\reg{B}$, and $\reg{C}$ are
registers, then $\reg{ABC}$ denotes the tensor product of the associated Hilbert spaces. We write $\mc L(\reg R)$ to
denote the set of all linear transformations over $\reg R$. When $\reg{R}$ appears alongside other registers, for a linear transformation $L \in \mc L(\reg R)$ we
sometimes write $L_{\reg{R}}$ to make explicit that $L$ acts non-trivially on $\reg{R}$ and as identity on other registers. Additionally, we write
$\rho_{\reg{R}}$ to indicate that $\rho$ is a state on $\reg{R}$. Generally, one may think of a register $\reg R$ as
corresponding to a qudit.

We denote the set of positive semidefinite operators on a register $\reg{R}$ by $\pos(\reg{R})$. The set of density
matrices on $\reg R$ is denoted $\states(\reg R)$. For a pure state $\ket\varphi$, we write $\phi$ to denote the density
matrix $\ketbra{\phi}$. We denote the identity operator as $\id$. For an operator $X \in \linear(\reg R)$, we define $\| X
\|_\infty$ to be its operator norm, and $\| X\|_1 = \tr(|X|)$ to denote its trace norm, where $|X| = \sqrt{X^\dagger
X}$. For Hermitian matrices $A,B$, we use the notation $A \succeq B$ to indicate that $A - B$ is positive semi-definite
(PSD). Similarly, $A \preceq B$ indicates that $B - A$ is PSD. We also define the commutator, $[A, B] = AB - BA$.

We will also use Jordan's Lemma, which characterizes the decomposition of a subspace induced by a pair of projectors.
\begin{lemma}[Jordan's Lemma~\cite{jordan1875essai}]
    \label{lem:jordans_lemma}
    For any two Hermitian projectors $\Pi_{P}$ and $\Pi_{Q}$ on a register $\reg{R}$, there exists a orthogonal
    decomposition of $\reg{R} = \bigoplus_{b} \reg{R}_{b}$ (the Jordan decomposition with respect to $\Pi_P, \Pi_Q$)
    into one- and two-dimensional subspaces $\mathcal{B} = \{\reg{S}_b\}_{b}$ (Jordan subspaces, or blocks) where each
    $\reg{R}_b$ is invariant under both $\Pi_P$ and $\Pi_Q$.  Moreover
    \begin{itemize}
        \item In each one-dimensional subspace, $\Pi_P$ and $\Pi_Q$ act as identity or rank-0 projectors.
        \item In each two-dimensional subspace, $\Pi_P$ and $\Pi_Q$ are rank-1 projectors.  In particular, there exist distinct orthogonal bases $\{\ket{p_b^0}, \ket{p_b^1}\}$ and $\{\ket{q_b^0}, \ket{q_b^1}\}$ for $\reg{S}_b$ such that $\Pi_P$ projects onto $\ket{p_b^1}$ and $\Pi_Q$ projects onto $\ket{q_b^1}$. 
    \end{itemize}
\end{lemma}

\subsection{Local Hamiltonian problems}
\begin{definition}[Local Hamiltonian]
   A $k$-local Hamiltonian over registers $\mc R = \{\reg R_1, \dots, \reg R_n\}$ is a Hermitian operator $H \in
   \linear(\reg R_1\dots \reg R_n)$ such that $H$ can be written as $H = \sum_{i=1}^\ell h_i$, where each $h_i$ is PSD
   and $\|h_i\|_\infty \leq 1$. Each $h_i$ acts non-trivially on registers $\mc S \subseteq \mc R$ with $|\mc S| \leq
   k$. We say that a $k$-local Hamiltonian acts on $d$-dimensional qudits if each register $\reg R_i$ corresponds to the
   Hilbert space $\C^{d}$.
\end{definition}

\begin{definition}[Local Hamiltonian problem]
    Given a family $\mc H$ of $k$-local Hamiltonians and parameters $c,s$ with $s - c \geq \tfrac 1 {\cpoly(n)}$, the
    local Hamiltonian problem is a promise problem so that, promised that $H \in \mc H$ has minimum eigenvalue
    $\lambda_0(H) \geq s$ or $\lambda_0(H) \leq c$, to decide which is the case.
\end{definition}

\begin{definition}[Commuting local Hamiltonian problem]
    The commuting local Hamiltonian (CLH) problem is identical to the local Hamiltonian problem, except that the
    Hamiltonian family $\mc H$ is composed of commuting Hamiltonians $H$. A Hamiltonian $H = \sum_i h_i$ is commuting if
    for all $i,j$, $[h_i, h_j] = 0$.
\end{definition}

\begin{remark}[Projective CLH]
    In the context of showing that the local Hamiltonian problem for a family of \emph{commuting} local Hamiltonians is
    in $\cNP$, it's sufficient to restrict to the case where each $h_i$ is a projector. This is because the prover may
    first identify an eigenspace $\Pi_i$ and eigenvalue $\lambda_i$ of each term such that $\sum_i \lambda_i \leq c$,
    the completeness parameter of the CLH problem. The prover then helps verify that $\sum_i (\Id - \Pi_i)$ has a
    zero-energy eigenspace. Going forward, we'll assume that each $h_i$ is a projector.
\end{remark}
This remark has the important consequence that we may always take $c = 0$ and $s = 1$. Next, a particularly important notion in this work is a \emph{rank-constrained} CLH.
\begin{definition}[Rank constrained commuting local Hamiltonian]
    A rank-r constrained, $k$-local CLH, denoted $\text{CLH}^{(r)}$, is a $k$-local CLH where each $h_i$ is a matrix of rank at most $r$.
\end{definition}

We also give some simple definitions related to local Hamiltonians.
\begin{definition}[Degree of a register]
    \label{def:reg_degree}
    Suppose $H = \sum_i h_i$ is a CLH over $n$ qudits, corresponding to registers $\reg R_1, \dots, \reg R_n$ , the \emph{degree} of a register $\reg R_i$, denoted $\text{deg}_H(\reg R_i)$ is the number of terms in $H$ which act non-trivially on $\reg R_i$. If $H$ is clear from context, we write this as simply $\text{deg}(\reg R_i)$.
\end{definition}
Then, the \emph{degree of a Hamiltonian} is defined simply as $\text{deg}(H) = \max_i \text{deg}_H(\reg{R}_i)$. Note that this notion only makes sense once we identify a partitioning of the full space into registers $\reg R_1 \dots \reg R_n$. If unspecified, we associate registers with each qudit of the Hamiltonian.

When we have bounds on the degree, locality, and qudit dimension of a Hamiltonian, one can show that a wide range of Hamiltonians are always satisfiable.
\begin{lemma}[Quantum Lov\'asz Local Lemma (qLLL) \cite{Ambainis_AKS2012_QuantumLovaszLocal}]
\label{lem:qlll}
Let $H = \sum_i h_i$ be a Hamiltonian with locality $k$, degree $g$, qudit dimension $d$, and $\max_i \rank(h_i) \leq r$. Then if $g \leq \tfrac{d^k}{r e}$ then $\lambda_0(H) = 0$.
\end{lemma}
In our work, since we generally take $r = 1$, this implies only instances with large maximum degree are interesting.

\subsection{$C^*$-algebras and the Structure Lemma}

In this section we review the notion of a $C^*$-algebra, the Structure Lemma from \cite{Bravyi_BV2004_CommutativeVersionKlocal} and the connection to commuting local Hamiltonians.  

\begin{definition}[$C^*$-algebra]
    Let $\reg{R}$ be a register, a $C^*$-algebra is any complex algebra $\mathcal{A} \subseteq \linear (\reg{R})$ that is closed under the $\dagger$ operation and includes the identity. 
\end{definition}

\begin{definition}[Commuting algebras]
    Let $\mathcal{A}$ and $\mathcal{A}'$ be two $C^*$-algebras on $\reg{R}$.  We say $\mathcal{A}$ and $\mathcal{A}'$ \emph{commute} if $[h, h'] = 0$ for all $h \in \mathcal{A}$ and $h' \in \mathcal{A}'$.
\end{definition}

The connection between local Hamiltonians and algebras is made through the concept of an ``induced algebra''.

\begin{definition}[Induced algebra]
    Let $h$ be a Hermitian operator acting on $\reg{AB}$, and consider a decomposition of $h$ into
    \begin{equation}
        h = \sum_{i, j} (h_{ij})_{\reg{A}} \otimes (\ketbratwo i j)_{\reg{B}}\,.
    \end{equation}
    Here $\{\ket{i}_{\reg{B}}\}_{i}$ is an orthogonal basis of $\reg{B}$.  Then the induced algebra of $h$ on $\reg{A}$, denoted $\mathcal{A}_{h}^{\reg{A}}$, is defined to be the algebra generated by $\left\langle \{ (h_{ij})_{\reg{A}}\}, \id_{\reg{A}} \right\rangle$.
\end{definition}
The following lemma tells us that the induced algebra is independent of the choice of basis for $\reg{B}$.
\begin{lemma}[Claim B.3 of \cite{Aharonov_AKV2018_ComplexityTwoDimensional}]
    \label{lem:equivalent_algebras}
    Let $h$ be a Hermitian operator and consider two decompositions of $h$
    \begin{equation}
        h = \sum_{i, j} (h_{ij})_{\reg{A}} \otimes (g_{ij})_{\reg{B}} = \sum_{i, j} (\hat{h}_{ij})_{\reg{A}} \otimes (\hat{g}_{ij})_{\reg{B}}\,,
    \end{equation}
    where both the sets $\{g_{ij}\}$ and $\{\hat{g}_{ij}\}$ are linearly independent.  Then the $C^*$-algebra generated by $\{h_{ij}\}_{ij}$ and $\{\hat{h}_{ij}\}_{ij}$ are the same.  
\end{lemma}
The induced algebra gives us a tool which we can use to determine if two Hermitian operators commute via the following lemma.
\begin{lemma}
    \label{lem:algebra_commutation}
    Let $(h_1)_{\reg{AB}}$ and $(h_2)_{\reg{BC}}$ be two Hermitian operators.  Then $[h_1, h_2] = 0$ if and only if $\mathcal{A}^{\reg{B}}_{h_1}$ commutes with $\mathcal{A}^{\reg{B}}_{h_2}$.   
\end{lemma}

Finally, we recall the Structure Lemma from \cite{Bravyi_BV2004_CommutativeVersionKlocal}, which is the main tool prior work used to characterize the commutation between adjacent commuting terms. At a high level, the Structure Lemma says that algebras can be block-diagonalized, and that within each of these blocks, the algebra takes on a tensor product structure. For a proof of the lemma, see \cite[Section~7.3]{Gharibian_GHLS2015_QuantumHamiltonianComplexity}.
\begin{lemma}[The Structure Lemma]\label{lem:structure}
    Let $\mathcal{A} \subseteq \mathcal{L}(\reg{R})$ be a $C^*$-algebra on $\reg{R}$.  Then there exists a direct sum decomposition $\reg{R} = \bigoplus_{i} \reg{R}_i$ and a tensor product structure $\reg{R}_i = \reg{R}_i^1 \otimes \reg{R}_i^{2}$ such that
    \begin{equation}
        \mathcal{A} = \bigoplus_i \mathcal{L}(\reg{R}_i^{1}) \otimes \id_{\reg{R}_i^{2}}\,.
    \end{equation}
\end{lemma}

A corollary of this lemma, and the reason why it is so useful for characterizing the properties of commuting local
Hamiltonians, is that in order to commute with an algebra, another algebra must live entirely within the $\reg{R}_i^2$
subspaces.  Formally, we have the following
\commutingstructure
An easy application of the Structure Lemma can be seen in \Cref{fig:simple_structure_lemma}; there, $2$-local
interactions can be simplified to generate $1$-local terms. Finally, we introduce the notion of a ``classical''
register $\reg C$. Intuitively, this means that under some local unitary $U_{\reg C}$, the register ``looks classical,''
in that every eigenstate is can be written as $\ket x_{\reg C} \otimes \ket \psi$, for a classical string $x$.

\begin{definition}[Classical register]
    For a Hamiltonian $H$, a register $\reg C$ is \emph{classical} if there exists a decomposition of $\reg{C}$ into $1$-dimensional spaces $\reg{R}_1 \oplus \dots \oplus \reg{R}_d$ which is invariant under all terms $h$ of $H$.
\end{definition}

Classical registers will be very useful for us; given any classical register $\reg C$, we can induce a ``hole'' in the
Hamiltonian $H$ on that register. This is because if each Hamiltonian term $h$ acts invariants w.r.t. the decomposition
$(\mc H_{\reg R})_1 \oplus \dots \oplus (\mc H_{\reg R})_d$, this implies that each $h = \sum_i \ketbra i_{\reg R} \otimes h_i$ ,
where $\ketbra i$ is the projector onto the 1-dimensional subspace $(\mc H_{\reg R})_i$, and the entire Hamiltonian can
be written as
\[
    H = \sum_i \ketbra i_{\reg R} \otimes H_i \quad \text{ with } \quad H_i = \sum_h h_i\,,
\]
and thus every eigenstate, and in particular the ground state, is of the form $\ket i_{\mc R} \otimes \ket \psi$.
Therefore, the prover can provide the subspace containing the ground space and the Hamiltonian $\tilde H \dfn H_i$
consists of terms which all act trivially on $\reg R$.

\begin{observation}[Classical restriction]
    \label{obs:classical_restriction}
    Suppose $\reg C$ is a classical qubit for the Hamiltonian $H$. Then there exists a one-dimensional projector
    $\pi_{\reg R}$ such that,
    \begin{equation}
        \lambda_0(H) = 0 \iff \lambda_0(\pi H \pi) = 0\,.
    \end{equation}
    Moreover, $\pi H \pi = \pi \otimes \tilde H$, where $\tilde H$ is a CLH instance where each term acts trivially on
    $\reg C$. Thus the local Hamiltonian of $H$ reduces to solving the local Hamiltonian problem on $\tilde H$, where
    the verifier implicitly keeps track of the 1-dimensional assignment on $\reg C$.
\end{observation}

\subsection{Geometrically constrained commuting local Hamiltonians}
\label{sec:geometrically_contrained}

Now we formally define the notion of 2D and 3D commuting local Hamiltonians in terms of chain complexes.  We refer to a chain complex in 2D as a polygonal complex.

\begin{definition}[Polygonal complex]
    A polygonal complex $\mathcal{K}$ is a collection of points, lines, and polygons such that
    \begin{enumerate}
        \item Every side of a polygon in $\mathcal{K}$ is a line in $\mathcal{K}$, and every endpoint of a line in $\mathcal{K}$ is a point in $\mathcal{K}$. 
        \item For every pair of polygons, their intersection is either the empty set, or a single line in $\mathcal{K}$, along with its endpoints. For every pair of lines in $\mathcal{K}$, their intersection is either the empty set, or it is a single point in $\mathcal{K}$. 
    \end{enumerate}
\end{definition}

In 3D, we refer to a similar chain complex as a polyhedra complex, although it is clear how this would generalize to arbitrarily high dimensions via the concept of a chain complex (which, in reality, we are just defining in geometric terms).

\begin{definition}[Polyhedral complex]
    \label{def:polyhedral_complex}
    A polyhedral complex $\mathcal{K}$ is a collection of points, lines, polygons, and polyhedra such that
    \begin{enumerate}
        \item Every face of a polyhedral in $\mathcal{K}$ is a polygon in $\mathcal{K}$, and every side of a polygon in $\mathcal{K}$ is a line in $\mathcal{K}$, and every endpoint of a line in $\mathcal{K}$ is a point in $\mathcal{K}$.
        \item For every pair of polyhedral, their intersection is either the empty set or a single polygon in $\mathcal{K}$, along with its sides and their endpoints.  For every pair of polygons in $\mathcal{K}$, their intersection is either the empty set, or a single line in $\mathcal{K}$, along with its endpoints.  For every pair of lines in $\mathcal{K}$, their intersection is either the empty set, or a single point in $\mathcal{K}$.  
    \end{enumerate}
\end{definition}

There are many ways to define a commuting local Hamiltonian instance on a chain complex, depending on which terms correspond to Hamiltonian terms and which correspond to qudits.  We present a few of the notable versions of the commuting local Hamiltonian problem that have been studied in the past.
\begin{definition}[2D-CLH]
    \label{def:2d_clh}
    A $(k, d)$-2D-CLH instance is an instance of the commuting $k$-local Hamiltonian problem where the $d$-dimensional qudits can be mapped to vertices and Hamiltonian terms can be mapped to faces in a polygonal complex in such a way that the Hamiltonian term on a face only acts non-trivially on the qudits associated with endpoints of the sides of the face.  
\end{definition}

This is the most general way to define a two dimensional commuting local Hamiltonian problem, but often times more constraints are required to produce non-trivial results.  \cite{Aharonov_AKV2018_ComplexityTwoDimensional} define another way to associate a polygonal complex with a Hamiltonian, which they call 2D-CLH*.  They also show that every 2D-CLH* instance can be transformed into a 2D-CLH instance without changing the local dimension, making this problem strictly easier than the general 2D-CLH problem.\footnote{The authors of \cite{Aharonov_AKV2018_ComplexityTwoDimensional} also show that 2D-CLH* instances can be turned into 2D-CLH instances, but at a cost of expanding the local dimension. Since their result only holds for \emph{qubits}, it cannot be extended to general 2D-CLH instances.}

\begin{definition}[\twodclh]
    \label{def:2dstar_clh}
    A $(k, d)$-\twodclh{} instance is an instance of the commuting $k$-local Hamiltonian problem where the
    $d$-dimensional qudits can be mapped to edges and Hamiltonian terms can be mapped to faces and vertices of a
    polygonal complex in such a way that the Hamiltonian term associated with a vertex acts on qudits associated with the edges adjacent to the vertex, and similarly for Hamiltonian terms on faces. 
\end{definition}

The works of \cite{Schuch_Sch2011_ComplexityCommutingHamiltonians,Jiang_Jia2023_LocalHamiltonianProblem} consider an even more constrained version of the problem in two dimensions, where the polygonal complex is a square lattice.  

\begin{definition}[\twodclhgrid]
    The $d$-\twodclhgrid{} problem is the $(4, d)$-\twodclh{} problem where the polygonal complex is always a square lattice.
\end{definition}

Similar to the definition of 2D-CLH, we can define the 3D-CLH problem as follows.

\begin{definition}[3D-CLH]
    A $(k, d)$-3D-CLH instance is an instance of the commuting $k$-local Hamiltonian problem where $d$-dimensional qudits can be mapped to vertices and Hamiltonian terms can be mapped to volumes in a polyhedral complex such that the Hamiltonian term associated with a volume only acts non-trivially on qudits associated with vertices that lie on the volume. 
\end{definition}

As in the 2D case, we also define a ``starred'' version, where qudits are placed on edges, rather than vertices.
\begin{definition}[\threedclh]
    A $(k, d)$-\threedclh{} instance is an instance of the commuting $k$-local Hamiltonian problem where $d$-dimensional
    qudits can be mapped to \emph{edges} and Hamiltonian terms can be mapped to \emph{volumes} in a polyhedral complex such
    that the Hamiltonian term associated with a volume only acts non-trivially on qudits associated with edges that lie
    on the volume. 
\end{definition}

Finally, we can define a very constrained version of \threedclh{} where the polyhedral complex is constrained to be a
cubic lattice. This case turns out to be trivial due to the quantum Lov\'asz Local Lemma, but is useful for pedagogical
purposes.

\begin{definition}[\threedclhgrid]
    A $d$-\threedclhgrid{} instance is an instance of the commuting $12$-local Hamiltonian problem where the $d$-dimensional qudits can be mapped to edges in a 3D cubic lattice and Hamiltonian terms can be mapped to cubes in the same lattice such that the Hamiltonian term associated with a volume only acts on qudits associated with edges of faces of the volume. 
\end{definition}

\begin{remark}
    For the CLH families \twodclh, \twodclhgrid, \threedclh, and \threedclhgrid, if the parameters $(k,d)$ are left unspecified, we take this to mean there is some $k,d \in \cpoly(n)$ for which every Hamiltonian in the family is a $(k,d)$-CLH instance.
\end{remark}

Finally, we remark that for our results, we make some mild assumptions on the 3D geometric structure.
\begin{assumption}[Uniformly high degree]
\label{as:high_degree}
    Given a Hamiltonian $H$ over registers $\reg R_1, \dots \reg R_n$, we assume that $\min_{i \in [n]} \deg_H(\reg R_i) \geq 4$.
\end{assumption}
This assumption is necessary in order to apply our rounding scheme. This assumption is somewhat justified by
\Cref{lem:qlll}; if the degree of \emph{all} registers was strictly less than $4$, then the resulting (rank-1)
Hamiltonian would be trivial. However, even a single high degree term makes qLLL inapplicable and we need this assumption to hold for all registers.

Lastly the following assumption enforces a nice structure on the 3D surface.
\begin{restatable}[Generalized cubic lattice]{assumption}{generalizedlattice}
    \label{as:generalized_cubic_lattice}
    We maintain some connection with the simpler cubic lattice case. This connection is through two restrictions on the lattice.
    \begin{enumerate}
        \item If there are two Hamiltonian terms $h_1, h_2$ sharing a face, then there is no term which shares a face
            with \emph{both} $h_1$ and $h_2$. Intuitively, this prevents ``Jenga-like'' structures, which will make it hard to draw an analogue to the tunnel picture (\Cref{fig:tube}) from the simpler case.
        \item Each Hamiltonian term should have at least $5$ faces. This condition will ensure that when we punctures holes for the vertex, there are enough faces so that each boundary path (e.g. the lines in \Cref{fig:vertex_cube_slice_edges}) can exit through a distinct face.
    \end{enumerate}
\end{restatable}

\section{Rounding commuting local Hamiltonians}
\label{sec:rounding}
In this section, we describe the rounding schemes used in our paper. As stated previously, these rounding schemes serve
as a single step in a guided reduction; the existence of a rounding scheme asserts that there exists a projector $\pi$
(or a set of projectors $\{\pi_i\}_i$) which can be provided by the prover to help the verifier simplify the
Hamiltonian. At a high level, our rounding schemes emerge when \Cref{lem:equiv_projector} not only provides an equivalent
condition to the existence of a ground state, but the resulting trace expression can be ``rounded'' back to an
equivalent (and hopefully simplified) CLH instance.

Formally, a rounding scheme for Hamiltonians is defined as follows:
\begin{definition}[Rounding scheme for commuting local Hamiltonians]
    \label{def:rounding_scheme}
    Let $H = (h_1)_{\reg{A_1}\reg{B}} + \ldots + (h_k)_{\reg{A}_k\reg{B}}$ be a CLH instance defined over registers $\reg A_1, \dots, \reg A_k, \reg B$. The registers $\reg A_i, \reg A_j$ are allowed to overlap, in the sense that $\reg A_i \cap \reg A_j \neq \emptyset$. Let $\Pi = \{(\pi_{i})_{\reg{B}}\}_{i}$ be a set of projectors acting only
    on $\reg{B}$. A rounding scheme is an efficient classical algorithm that takes as input a description of $h_1,
    \ldots, h_k$, and $\{\pi_i\}_i$, and outputs a CLH instance $\Tilde{H}_\Pi$ such that $H$ has a $0$-energy ground
    state if and only if $\Tilde{H}_\Pi$ does.
\end{definition} 

\begin{figure}[!ht]
\centering
\begin{subfigure}[t]{0.4\textwidth}
    \centering
    \begin{tikzpicture}[scale=1]
        \foreach \x in {0,1,2,3} {
            \foreach \y in {0,1,2,3} {
                \pgfmathtruncatemacro{\xynum}{\y*4+\x}
                \pgfmathtruncatemacro{\xdist}{2*abs(\x-1.5)}
                \pgfmathtruncatemacro{\ydist}{2*abs(\y-1.5)}
                \ifthenelse{\NOT\(\xdist > 1 \OR \ydist > 1\)}{
                    \draw[fill=CornflowerBlue!50] (\x,\y) rectangle ++ (1,1) node[midway] {$h_{\xynum}$};
                }{
                    \draw[fill=CornflowerBlue!20] (\x,\y) rectangle ++ (1,1) node[midway] {$h_{\xynum}$};
                }
            }
        }
        \node[label={[label distance=-0.35cm]below left:$\reg R$}] at (2,2) {\textbullet};
    \end{tikzpicture}
    \caption{Terms surrounding register $\reg R$.}
\end{subfigure}
\quad
\begin{subfigure}[t]{0.4\textwidth}
    \centering
    \begin{tikzpicture}
        \foreach \x in {0,1,2,3} {
            \foreach \y in {0,1,2,3} {
                \pgfmathtruncatemacro{\xynum}{\y*4+\x}
                \pgfmathtruncatemacro{\xdist}{2*abs(\x-1.5)}
                \pgfmathtruncatemacro{\ydist}{2*abs(\y-1.5)}
                \ifthenelse{\NOT\(\xdist > 1 \OR \ydist > 1\)}{
                }{
                    \draw[fill=CornflowerBlue!20] (\x,\y) rectangle ++ (1,1) node[midway] {$h_{\xynum}$};
                }
            }
        }
        \begin{scope}[xshift=1cm, yshift=1cm]
            \foreach \x in {0,1} {
                \foreach \y in {0,1} {
                    \draw[dashed,black!40] (\x,\y) rectangle ++(1,1);
                }
            } 
            \draw[fill=CornflowerBlue!50] (0,0) -- (1,0) -- (1,0.7) -- (0.7,1) -- (0,1) -- cycle;
            \draw[fill=CornflowerBlue!50] (2,2) -- (2,1) -- (1.3,1) -- (1,1.3) -- (1,2) -- cycle;
            \draw[fill=CornflowerBlue!50] (0,1) -- (0.7,1) -- (1,1.3) -- (1,2) -- (0,2) --cycle;
            \draw[fill=CornflowerBlue!50] (1,0) -- (1,0.7) -- (1.3,1) -- (2,1) -- (2,0) -- cycle;
            \path (0,1) rectangle ++ (1,1) node[midway] {$h_9$};
            \path (1,0) rectangle ++ (1,1) node[midway] {$h_6$}; 
            \path (1,1) rectangle ++ (1,1) node[midway] {$h_{10}$}; 
            \path (0,0) rectangle ++ (1,1) node[midway] {$h_5$}; 
        \end{scope}
    \end{tikzpicture}
    \caption{When register $\reg R$ is classical, there is a choice of projector such that the terms in the resulting
    Hamiltonian act \emph{trivially} on $\reg R$.}
    \label{fig:classical_restriction}
\end{subfigure}
\captionsetup{width=0.8\textwidth}
\caption{Result of \Cref{cor:classical_guided_reduction}}
\label{fig:types_of_restrictions}
\end{figure}

Although in the above definition, we write $\tilde H_\Pi$ to emphasize that the resulting projector is constructed using
$\Pi$, we'll usually drop the $\Pi$ subscript for ease of notation, and refer to the resulting Hamiltonian as $\tilde
H$. This definition may some a bit opaque in that it does not specify \emph{how} $\tilde H$ should be constructed.
Naively, one might assume we sequentially apply each $\pi \in \Pi$ to $H$; in 2-local setting described in
\Cref{sec:rounding_via_structure_lemma}, this is exactly what we did. In general, however, these projectors may not
commute with all terms and more complicated rounding schemes are often necessary.

In addition to 2-local rounding, we have seen another example of a rounding scheme. \Cref{obs:classical_restriction}
yields the following,
\begin{cor}[Rephrasing of \Cref{obs:classical_restriction}]
    \label{cor:classical_guided_reduction}
    Let $H = \sum_h h$ be a CLH instance over $(\reg R_1, \dots, \reg R_n)$ with classical register $\reg C = \reg R_i$. Let
    $\{\pi_i = \ketbra{\psi}_i\}_i$ be the projectors on to the
    corresponding 1-dimensional subspaces. Then there is a rounding scheme for $H$ and $\{\pi_i\}_i$. Moreover, the
    rounding scheme yields a simplified CLH instance $\tilde H = \sum_h \tilde h$ on $(\reg R_1, \dots, \reg R_{i-1},
    \reg R_{i+1}, \dots, \reg R_n)$ such that,
    \begin{itemize}
        \item If $h$ acts trivially on $\reg C$, then $\tilde h = h$.
        \item Otherwise, $\tilde h = \left(\id \otimes \bra{\psi_i}_{\reg{C}}\right) h \left(\id \otimes \ket{\psi_i}_{\reg{C}}\right)$.
    \end{itemize}
\end{cor}
The result of this rounding is depicted in \Cref{fig:types_of_restrictions}.

In this section, we will first show how the key technical tools in previous papers (e.g.
\cite{Bravyi_BV2004_CommutativeVersionKlocal, Irani_IJ2023_CommutingLocalHamiltonian}) can be phrased in terms of a
rounding scheme. Then, we will give analyze the local algebra of commuting, rank-1 operators, then show how this yields
new rounding scheme for rank-1 CLH instances. By showing that these rounding schemes significantly simplify the
Hamiltonian, applying them iteratively (with corresponding projectors supplied by the prover) will yield a \emph{guided
reduction} from 2D and 3D rank-1 CLH instances to 2-local CLH instances, known to be contain in $\cNP$.

\subsection{Previous works recasted as rounding schemes}
\paragraph{Rounding scheme of Bravyi-Vyalyi} A central result of \cite{Bravyi_BV2004_CommutativeVersionKlocal} is that
\emph{2-local} CLH instances are contained in $\cNP$. At the heart of this result is the following rounding scheme.
Consider a Hamiltonian $H$ defined over registers $\{\reg R_1, \dots, \reg R_n\}$. Let $\reg B = \reg R_k$ to be any qudit
register, and let $S_{\reg B}$ be the set of Hamiltonian terms acting non-trivially on $\reg B$. Fix some $h \in S_{\reg{B}}$ and consider the pair of commuting operators $h$ and $\overline{h} = \sum_{h' \neq h \in S_{\reg{B}}}$.  Applying the Structure Lemma yields a decomposition into a direct sum of tensor product spaces in which $\overline{h}$ and $h$ act on different particles.  Recursively applying this argument to $\overline{h}$ yields a decomposition
\begin{equation}
    \reg B = \bigoplus_i \reg B_i = \bigoplus_i \bigotimes_{h \in S_{\reg B}} \reg B_i^h\,,
\end{equation}
where the algebra of any term satisfies
\begin{equation}
    \mc A_h^{\reg B} \subseteq \bigoplus_i \id \otimes \dots \otimes \mc L(\reg B_i^h) \otimes \dots \otimes
    \id\,.
\end{equation}
In particular, $h$ acts non-trivially only on the subspaces $\{\reg B_i^h\}_i$ and is block-diagonal with
respect to the subspaces $\{\reg B_i\}_i$. The rounding scheme is then to set each $\pi_i$ to be the projector
onto $\reg B_i$, yielding the Hamiltonians $H_i = \pi_i H \pi_i = \sum_j \pi_i h_j \pi_i$. We see that this is
a valid rounding scheme as
\begin{itemize}
    \item Each pairs of terms $h_j, h_k \in H_i$ is commuting since both are block-diagonal across the subspace
        corresponding to $\pi_i$ (either because both projectors act trivially on $\reg B$, or if they do not act trivially then because $\pi_i$ corresponds to the direct sum induced by the Structure Lemma).
    \item In the statement of \Cref{lem:equiv_projector} set $S = \{h\}_{h \in H}$, $\mc M = \{\pi_i\}_i$, $T =
        \emptyset$, and $\tilde M = \{\id\}$. Then,
        \begin{equation}
            \lambda(H) = 0 \iff \exists i \text{ s.t. } \tr\Brac{\prod_j (\pi_i - \pi_i h_j \pi_i)} > 0\,.
        \end{equation}
        But the RHS is equivalent to the existence of a ground state of the Hamiltonian $\tilde H = H_i = \sum_j \pi_i
        h_j \pi_i$, over the reduced space $\{\reg R_1, \dots, \pi_i \reg R_k \pi_i, \dots, \reg R_n\}$. Thus, there
        exists a $H_i$ such that $\lambda_0(H_i) = 0 \iff \lambda(H) = 0$.
\end{itemize}
 This implies the following rounding scheme:
\begin{lemma}[Rounding scheme for 2-local Hamiltonians]
    \label{lem:2_local_rounding}
    Suppose there exists a register $\reg B$ such that all terms acting non-trivially on $\reg B$ interact on no other
    registers (i.e. their corresponding $\reg A_i$'s are orthogonal). Then there exists a rounding scheme for $H$.
\end{lemma}
Moreover, this rounding scheme significantly simplifies the Hamiltonian.
\begin{cor}[Product structure from 2-local rounding]
    \label{cor:product_structure_2_local}
    Let $H$ be a Hamiltonian with a register $\reg B$ satisfying the requirements of \Cref{lem:2_local_rounding}. After
    rounding to generate $\tilde H$, a register $\reg B$ gets mapped to $\tilde{\reg B} \subseteq \reg B$. Moreover,
    $\tilde{\reg B}$ can be represented as a tensor product of registers $\tilde{\reg B} = \otimes_i {\tilde{\reg B}}_i$
    such that each term of $\tilde H$ either act trivially act on $\tilde{\reg B}$, or act on a unique sub-register
    ${\tilde{\reg B}}_i$.
\end{cor}
For two terms, this is demonstrated in \Cref{fig:simple_structure_lemma}. If $H$ is $2$-local, each register $\reg R_i$
satisfies \Cref{lem:2_local_rounding}. Thus, after applying the rounding scheme to each register,
\Cref{cor:product_structure_2_local} implies that the resulting Hamiltonian is made 1-local, and the ground state can be
verified in classical polynomial time. Since the prover must provide the projector required for each step of the
rounding scheme, this yields containment in $\cNP$. In fact, this implies a \emph{guided reduction} for 2-local
Hamiltonians:

\begin{cor}[Guided reduction for 2-local Hamiltonians (restatement of \cite{Bravyi_BV2004_CommutativeVersionKlocal}]
    There exists a guided reduction from 2-local Hamiltonians to 1-local Hamiltonians.
\end{cor}

\paragraph{Rounding scheme of Irani-Jiang}
In \cite{Irani_IJ2023_CommutingLocalHamiltonian}, a rounding scheme is introduced to remove what the authors call
``semi-separable qudits,'' which are qudits for which there exists a non-trivial decomposition, induced by the Structure
Lemma, such that all but \emph{a single} term acts invariantly with respect to this decomposition. In a sense, this is a
weakening of the assumption of \Cref{lem:2_local_rounding}, where we required a decomposition under which all terms
acted invariantly. Let the projectors corresponding to this decomposition be $\{\pi_i\}_i$. Then,
\cite{Irani_IJ2023_CommutingLocalHamiltonian} show,
\begin{theorem}[Rounding scheme for \twodclhgrid{} \cite{Irani_IJ2023_CommutingLocalHamiltonian}]
    \label{thm:2d_grid_rounding_scheme}
    Let $H$ an instance of \twodclhgrid{} and $\{\pi_i\}_{i \in \ell}$ (with $\ell > 1$) be a set of projectors on a register $\reg B$ such that all
    but one term commute with each projector. Then there is a rounding scheme for $H$ and $\{\pi_i\}_i$, yielding
    another \twodclhgrid{} instance $\tilde H = \pi_i H \pi_i$.
\end{theorem}
Since $\ell > 1$, the result of this rounding scheme is a reduction in the local dimension of $\reg B$. Thus, by
iteratively applying this rounding scheme, we can remove all semi-separable qudits from $H$, yielding a guided reduction
from general \twodclhgrid{} instances to \twodclhgrid-without-semi-separable-qudits instances.
\begin{cor}
    There is a guided reduction from \twodclhgrid{} to \twodclhgrid-without-semi-separable-qudits, which iteratively
    applies \Cref{thm:2d_grid_rounding_scheme}.
\end{cor}
To obtain a containment in $\cNP$, the authors need to consider qutrits, rather than general qudit instances. They show
that for \emph{qutrit} Hamiltonians with no semi-separable qutrits, their single-register algebras can be well
characterized and a careful case analysis shows that the resulting Hamiltonian can be made to have a 1D structure and
thus solveable in $\cNP$.

\subsection{An improved rounding scheme}
\label{sec:our_rounding}
In this section, we demonstrate a rounding scheme whenever there exists projectors $\pi_1, \pi_2$, each of which commute
with all but one term.  We note that while the setting in which we apply our rounding scheme is incomparable to that of
\cite{Irani_IJ2023_CommutingLocalHamiltonian}, the improved rounding scheme we supply applies in all the situations
where the rounding scheme of \cite{Irani_IJ2023_CommutingLocalHamiltonian} would apply, taking one of the projectors to
be $\id$, the identity operator. In \Cref{sec:warmup_2d}, using tools developed in \Cref{sec:local_alg_rank1}, we will
see how this can be applied to the rank-1 setting to yield a guided reduction for \twodclhrgrid{1} and \twodclhr{1}. In
its most general form, however, \Cref{thm:rounding} works for general CLH instances.
\begin{definition} We say that a matrix $P$ \emph{survives} a projector $\pi$ if $\pi P \pi \neq 0$.
\end{definition}
\begin{theorem}[Rank-1 rounding scheme]
    \label{thm:rounding}
    Let $H$ be a sum of commuting local projectors and let $\pi_1$, $\pi_2$ be a pair of projectors such that for both
    $\pi_1$ and $\pi_2$, there is at most a single term that both does not commute with it and survives the other
    projector; call this term $\pi_1$'s non-commuting operator $P$ and $\pi_2$'s operator $Q$. Then there is a rounding
    scheme for $H$ and $\{\pi_1, \pi_2\}$. Moreover, in the rounded Hamiltonian $\tilde H$, $P$ and $Q$ and replaced by
    a \emph{single} merged term $R$, spanning the supports of both $P$ and $Q$.
\end{theorem}
Intuitively, this means that if both $\pi_1$ and $\pi_2$ are over a single register $\reg B$, applying \emph{both} $\pi_1$ and $\pi_2$ kills off all but two terms which act non-trivially on $\reg B$.
\begin{proof}

    Let $P$ and $Q$ be as defined in the thoerem statement. Let $H_{\mathrm{rest}}$ be the remaining terms of $H$ that
    survive both $\pi_1$ and $\pi_2$ (and thus commute with both). For clarity, we will re-label $\pi_2 = \pi_P$ and $\pi_1
    = \pi_Q$ so it is clear which operators commute. We also define $\overline{H}_{\mathrm{rest}} = \prod_{h \in
    H_{\mathrm{rest}}}(\id - h)$.  First, from \Cref{lem:equiv_projector} we have that,
    \begin{align}
        \lambda_0(H) = 0 \iff \tr[(\pi_P - \pi_P P \pi_P)(\pi_Q - \pi_Q Q \pi_Q) \overline{H}_{\mathrm{rest}}] > 0\,. 
    \end{align}

    Now, define $\Tilde{P} = \pi_P - \pi_P P \pi_P$ and $\Tilde{Q} = \pi_Q - \pi_Q Q \pi_Q$.  Note that because $P$ and
    $\pi_P$ commute, $\Tilde{P}$ is a projector, and similarly for $\Tilde{Q}$.  By Jordan's lemma
    (\Cref{lem:jordans_lemma}), we can express $\Tilde{P}\Tilde{Q}$ as
    \begin{equation}
        \Tilde{P}\Tilde{Q} = \sum_{b} \eta_b \ket{p_b}\!\bra{q_b}\,,
    \end{equation}
    where $b$ indexes the Jordan blocks on $(\reg{A}_Q \cup \reg{A}_P)\otimes \reg{B}$.  Here, since $\reg{A}_P$ and $\reg{A}_Q$ might overlap, we write the union to denote the Hilbert space spanned by both. Formally, each block $B \in \mc B$ is spanned by $\ket{p_b}$ and $\ket{q_b}$; $B$ is 1-dimensional if $\ket{p_b} = \ket{q_b}$ and 2-dimensional otherwise. Let $\reg{B}'$ be the following subspace:
    \begin{equation}
        \reg{B}' = \mathrm{span}\{\ket{p_b}, \ket{q_b}: \eta_b > 0\}\,,
    \end{equation}
    and $\Delta$ be the projection onto $\reg{B}'$. Further define $\Pi^{P}_{\reg{B}'} = \sum_{b : \eta_b > 0} \proj{p_b}$ and $\Pi^{Q}_{\reg{B}'} = \sum_{b : \eta_b > 0} \proj{q_b}$.  The rounded commuting local projector instance that our algorithm will output is $\Tilde{H} = (\id_{\reg{B'}} - \Delta) + H_{\mathrm{rest}}$.  We prove that the Hamiltonian satisfies the following two properties:
    \begin{enumerate}
        \item (Commutation): $[\Delta, \overline{H}_{\mathrm{rest}}] = 0$.
        \item (Completeness and soundness): $\tr[\Delta \overline{H}_{\mathrm{rest}}] > 0 \iff \tr[(\pi_P - P)(\pi_Q - Q) \overline{H}_{\mathrm{rest}}] > 0$.
    \end{enumerate}
    These two properties guarantee that the new Hamiltonian we output is a commuting projector instance, and that it is a valid rounding scheme. 
    Let's begin by showing commutation.
    \begin{claim}\label{claim:commutation}
        $\Delta$ and $H_{\mathrm{rest}}$ commute.
    \end{claim}
    \begin{proof}
    Within each Jordan block of dimension $2$, let $\ket{\Tilde{q}_b} = \tfrac{(\id - \proj{p_b})\ket{q_b})}{\|(\id - \proj{p_b})\ket{q_b})\|_2}$ so that $\ket{\Tilde q_b} \perp \ket{p_b}$. Define $\Pi^{P}_{\reg{B}'} = \sum_{b: \eta_b > 0} \proj{p_b}$ and $\Pi^{Q}_{\reg{B}'} = \sum_{b: \eta_b > 0} \proj{\Tilde{q}_b}$. Orthogonality of each $\ket{\Tilde q_b}$ and $\ket{p_b}$ implies $\Delta = \Pi_{\reg B'} = \Pi^P_{\reg B'} + \Pi^Q_{\reg B'}$. Then to prove the claim, it suffices to show that for all $h \in H_{\mathrm{rest}}$, $h$ commutes with both $\Pi^{P}_{\mathsf{B}'}$ and $\Pi^{\Tilde{Q}}_{\mathsf{B}'}$.  

    By assumption, $h \in H_{\mathrm{rest}}$ commutes with $\pi_{P}$ and $\pi_{Q}$, and because we started with a commuting local Hamiltonian instance, $h$ commutes with $P$ and $Q$.  Therefore, $h$ commutes with any polynomial in $\Tilde{P}$ and $\Tilde{Q}$.  Specifically, consider $\Tilde{P}\Tilde{Q}\Tilde{P}$.  We can write this operator in terms of its Jordan blocks as:
    \begin{equation}
        \Tilde{P}\Tilde{Q}\Tilde{P} = \sum_{b} \eta_b \proj{p_b}\,.
    \end{equation}
    Via \Cref{lem:rounding_preserves_commutation} we conclude that $h$ commutes with $\textsf{round}(\Tilde{P}\Tilde{Q}\Tilde{P})$, which is exactly $\Pi^{P}_{\mathsf{B}'}$.  

    Now we show that $h$ commutes with $\Pi^{\Tilde{Q}}_{\mathsf{B}'}$.  We can always write $\ket{q_b} = \sqrt{\eta_b}\ket{p_b} + \sqrt{1 - \eta_b} \ket{\Tilde{q}_b}$.  Because $\Tilde{Q}\Tilde{P}\Tilde{Q}$ commutes with every $h$, we also have that for all polynomials $p : \mathbb C \rightarrow \mathbb C$ acting individually on each eigenvalue,
    \begin{equation}
        \left[h, p(\Tilde{Q}\Tilde{P}\Tilde{Q})\right] = 0\,.
    \end{equation}
    This essentially follows from \Cref{lem:rounding_preserves_commutation}. Letting $p(x) = \frac{1}{\sqrt{x}}$, we evaluate $c(\Tilde{Q}\Tilde{P}\Tilde{Q})$ as follows:
    \begin{align*}
        \sum_{b} \sqrt{\eta_b} \proj{q_b} &= \sum_{b} \frac{1}{\sqrt{\eta_b}} \left(\eta_b \proj{p_b} + \sqrt{\eta_b(1 -
        \eta_b)}(\ket{p_b}\!\bra{\Tilde{q}_b} + \ket{\Tilde{q_b}}\!\bra{p_b}) + (1 -
    \eta_b)\proj{\Tilde{q_b}}\right)\numberthis\\
        &= \sum_{b} \sqrt{\eta_b} \proj{p_b} + \sqrt{1 - \eta_b} (\ket{p_b}\!\bra{\Tilde{q}_b} + \ket{\Tilde{q_b}}\!\bra{p_b}) + \frac{1 - \eta_b}{\sqrt{\eta_b}} \proj{\Tilde{q_b}}\,.
    \end{align*}
    Then we can evaluate the following function of $\Tilde{Q}$ and $\Tilde{P}$:
    \begin{align}
        \label{eq:qtilde_vs_qpq}
        p(\Tilde{Q}\Tilde{P}\Tilde{Q}) - \Tilde{Q}\Tilde{P} - \Tilde{P}\Tilde{Q} + \sqrt{\Tilde{P}\Tilde{Q}\Tilde{P}} &= \sum_{b \st \eta_b \neq 0} \frac{1 - \eta_b}{\sqrt{\eta_b}} \proj{\Tilde{q_b}}\,.
    \end{align}
    Applying the rounding map to the above expression yields exactly $\Pi^{\tilde Q}_{\reg B'}$, and, therefore, all $h \in \overline H_{\mathrm{rest}}$ commutes with $\Pi^{\Tilde{Q}}_{\mathsf{B}'}$. Since $\Delta$ is the sum of $\Pi^{\Tilde{Q}}_{\mathsf{B}'}$ and $\Pi^{P}_{\mathsf{B}'}$, we have that $\Delta$ commutes with $\overline H_{\mathrm{rest}}$ as desired. This completes the proof of \Cref{claim:commutation}.
    \end{proof}

    \begin{claim}\label{claim:ground_energy}
        $\tr[\Delta \overline H_{\mathrm{rest}}] > 0$ if and only if $\tr[(\pi_P - \pi_P P \pi_P )(\pi_Q - \pi_Q Q \pi_Q)\overline H_{\mathrm{rest}}] > 0$.
    \end{claim}
    \begin{proof}
    First assume that $\tr[\Tilde{P}\Tilde{Q} \overline H_{\mathrm{rest}}] > 0$.  We know that $\Tilde{P}$ is a projector, so we have that $\Tilde{P} = \Tilde{P}^2$.  Using the cyclic property of the trace and the fact that $\Tilde{P}$ commutes with $\overline H_{\mathrm{rest}}$, we have that this condition is equivalent to 
    \begin{equation}
        \tr[\Tilde{P}\Tilde{Q}\Tilde{P}\overline H_{\mathrm{rest}}] >0\,.
    \end{equation}
    Then because $\Delta \succeq \Tilde{P}\Tilde{Q}\Tilde{P}$ we have that $\tr[\Delta \overline H_{\mathrm{rest}}] \geq \tr[\Tilde{P}\Tilde{Q}\Tilde{P} \overline H_{\text{rest}}] > 0$.  

    Now assume that $\tr[\Tilde{P}\Tilde{Q}\overline H_{\mathrm{rest}}] = 0$.  We will show that both $\tr[\Pi^{Q}_{\reg{B}'} \overline H_{\mathrm{rest}}] = \tr[\Pi^{P}_{\reg{B}'}\overline H_{\mathrm{rest}}] = 0$.  Let $\eta_{\min}$ be the smallest non-zero eigenvalue of $\Tilde{P}\Tilde{Q}\Tilde{P}$ so that $\Pi^{P}_{\reg{B}'} \preceq \frac{1}{\eta_{\min}}\Tilde{P}\Tilde{Q}\Tilde{P}$. It immediately follows that $\tr[\Pi^{P}_{\reg{B}'}\overline H_{\mathrm{rest}}] \leq \frac{1}{\eta_{\min}} \tr[\Tilde{P}\Tilde{Q}\Tilde{P} H_{\mathrm{rest}}] = 0$.  Similarly, because $\Tilde{Q}$ is a projector, we have the following:
    \begin{align*}
        \tr\left[\sum_{b \st \eta_b \neq 0}\proj{q_b} \overline H_{\mathrm{rest}}\right] &\leq \frac{1}{\eta_{\min}}
        \tr[\Tilde{Q}\Tilde{P}\Tilde{Q}\overline H_{\mathrm{rest}}]\numberthis\\
        &= \frac{1}{\eta_{\min}} \tr[\Tilde{P}\Tilde{Q}\overline H_{\mathrm{rest}}]\\
        &= 0\,.
    \end{align*}
    Here the second line uses the definition of $\Tilde{Q}\Tilde{P}\Tilde{Q}$, the third line uses the fact that $\Tilde{Q}$ commutes with $\overline H_{\mathrm{rest}}$, the cyclic property of the trace, and that $\Tilde{Q}$ is a projector. As in \Cref{claim:ground_energy} we really want a bound on $\tr[\sum_b \ketbra{\tilde q_b} \overline H_\text{rest}]$. Again we use \Cref{eq:qtilde_vs_qpq} to write,
    \begin{equation*}
        \tr\Brac{\sum_{b \st \eta_b \neq 0} \frac{1-\eta_b}{\sqrt{\eta_b}} \ketbra{\tilde q_b} \overline H_{\mathrm{rest}}} = \tr[c(\tilde Q
        \tilde P \tilde Q) \overline H_{\mathrm{rest}}] - \tr[\tilde P \tilde Q \overline H_{\mathrm{rest}}] - \tr[\tilde Q
        \tilde P \overline H_{\mathrm{rest}}] + \tr[\sqrt{\tilde Q \tilde P \tilde Q} \overline{H_\text{rest}}]\,,
    \end{equation*}
    and the above argument shows that each expression on the RHS is $0$. Finally, we let $\eta_{\min}' = \min_{b \st \eta_b \neq 0} \tfrac{1 - \eta_b}{\sqrt{\eta_b}}$ and bound,
    \[
        \tr[\Pi^{\Tilde{Q}}_{\reg{B}'} \overline H_{\mathrm{rest}}] \leq \frac 1 {\eta_{\min}'} \tr\Brac{\sum_{b \st \eta_b \neq 0} \frac{1-\eta_b}{\sqrt{\eta_b}} \ketbra{\tilde q_b} \overline H_{\mathrm{rest}}} = 0\,,
    \] 
    Again noting that $\Delta = \Pi^{P}_{\reg{B}'} + \Pi^{\tilde Q}_{\reg B'}$, we conclude that $\tr[\Delta \overline H_\text{rest}] = 0$, completing the proof of \Cref{claim:ground_energy}.
    \end{proof}

With the proofs of the two claims, the only remaining fact to check is that $\id_{\reg{B}'} - \Delta$ is a projector.  However since $\Delta$ is a sum of orthogonal projectors and all of the vectors are contained entirely in $\reg{B}'$, it must be a projector.  Thus, $H_{\mathrm{rest}} + (\id_{\reg{B}'} - \Delta)$ is a rounding scheme for $H$ and $\{\pi_1, \pi_2\}$.  
\end{proof}

\section{Tools for rank-1 commuting operators}
\label{sec:local_alg_rank1}
In this section we develop a characterization of the local algebras of rank-1 commuting operators. In particular, we
show that rank $1$ commuting projectors commute in one of two ways, either in a so called \emph{singular}, or
\emph{reducing} way, which we will define below.

\begin{definition}[Reducing commutation for rank $1$ projectors]
    \label{def:rank_1_reducing}
    Two rank $1$ projectors $P_{\reg{AB}}$ and $Q_{\reg{BC}}$ commute in a reducing way if there exists a projector $\Pi$ such that 
    \begin{align*}
        \Pi P \Pi &= P \text{ and }\\
        \Pi Q \Pi &= 0\,.\numberthis
    \end{align*}
\end{definition}

\begin{definition}[Singular commutation for rank $1$ projectors]
    \label{def:rank_1_singular}
    Two rank $1$ projectors $P_{\reg{AB}}$ and $Q_{\reg{BC}}$ commute in a singular way if the following holds.
    \begin{align*}
        P &= \proj{\psi}_{\reg{B}} \otimes \wt{P}_{\reg{A}}\text{ and }\\
        Q &= \proj{\psi}_{\reg{B}} \otimes \wt{Q}_{\reg{C}}\,.\numberthis
    \end{align*}
\end{definition}

Our goal for this section will be to prove the following lemma, although in reality we will prove a more general lemma than we need for the rest of the paper.
\begin{theorem}[Commutation of rank $1$ projectors]
    \label{thm:rank_1_commutation_main}
    Let $P$ and $Q$ be two rank $1$ projectors. Then $P$ and $Q$ either commute in a singular (\Cref{def:rank_1_singular}) or reducing way (\Cref{def:rank_1_reducing}). 
\end{theorem}

For the proof of this lemma, we will consider restricted algebras, where the restricting projector must also commute with the algebra. Formally,

\begin{definition}[Subspace restrictions of an algebra]
    \label{def:subspace_restriction}
    Given an algebra $\mathcal{A}$ and subspace $\Pi$ that commutes with all operators of $\mathcal{A}$, we define the restriction of $\mathcal{A}$ onto $\Pi$, $(\mathcal{A})|_{\Pi}$, to be the algebra generated by $\Pi h \Pi$, $h \in \mathcal{A}$.  We sometimes abuse notation and write $(\mathcal{A})|_{\reg{R}}$ to mean that $\Pi$ is the projection onto $\reg{R}$.
\end{definition}
\begin{remark}
    Note that if a projector commutes with $\mathcal{A}$, every element of the algebra on a register can be written as a sum of two operators, one in $\Pi$ and one in $\id - \Pi$. 
\end{remark}

The condition that $\Pi$ commutes with $\mathcal{A}$ implies that the restriction onto $\Pi$ yields another algebra (i.e. the resulting set is closed under sums and products).  In the next section, we show a special case to the Structure Lemma (\Cref{lem:structure}) in the case when one of the operators is rank $1$.  Intuitively, rank $1$ operators ``fully span'' their support, meaning that any term commuting with a rank $1$ projector must be trivial in their overlap.  

\begin{lemma}[Commutation of rank $1$ projectors]
\label{thm:rank_1_commutation}
    Let $P_{\reg{AB}}$ be a rank $1$ projector and $Q_{\reg{BC}}$ be another projector that commutes with $P$.  Let
    $\mathcal{A}_{P}$ and $\mathcal{A}_{Q}$ be the induced algebras by $P$ and $Q$ on $\reg{B}$ respectively.  Then
    there exists a decomposition of $\reg{B} = \reg{B}_P \oplus \reg B_Q$ such that 
    \begin{enumerate}
        \item $\mathcal{A}_{P} = \linear(\reg{B}_P)$, i.e. $P$ acts as the full algebra on $\reg B_P$ and as $0$ on $\reg B_Q$, and 
        \item $(\mathcal{A}_{Q})|_{\reg{B}_P}$ is trivial, and $Q$ acts non-trivially only on $\reg B_Q$.
    \end{enumerate}
\end{lemma}
\begin{proof}
    Let $P = \proj{\psi}$ for some state $\ket{\psi}$.  We can write the Schmidt decomposition of $\ket{\psi}$ as
    \begin{equation}
        \ket{\psi} = \sum_{i} \alpha_i \ket{\psi_i^1}_{\reg{A}} \ket{\psi_i^2}_{\reg{B}}\,.
    \end{equation}
    Then we can write $P$ as follows
    \begin{equation}
        P = \sum_{i, j} \alpha_i \alpha_j^{\dagger} \ket{\psi_i^1}\!\bra{\psi_j^1} \otimes \ket{\psi_i^2}\!\bra{\psi_j^2}\,.
    \end{equation}
    Since the local algebra is independent of the choice of decomposition (\Cref{lem:equivalent_algebras}), the algebra $\mathcal{A}_{P}$ is equal to $\linear(\mathrm{span}(\{\ket{\psi_i^2}\}_i))$. Defining $\reg B_P := \text{span}(\{\ket{\psi_i^2}\}_i)$ shows the first part of the claim.

    Next we characterize $(\mc A_Q)|_{\reg B_P}$. By assumption, $\mathcal{A}_Q$ commutes with $\mathcal{A}_P$. Since the projector onto $\reg B_P$, $\Pi_{\reg B_P}$ is in $\mathcal{A}_{P}$, $\mathcal{A}_Q$ commutes with $\Pi_{\reg{B}_P}$. This implies that any element of $\mathcal A_Q$ is block diagonal with respect to the subspaces $\reg B_P$ and $\reg B^\perp_P$, and we can consider the restricted algebra $(\mc A_Q)|_{\reg B_P}$. Now, it remains to be seen that the restriction of $\mathcal{A}_{Q}$ onto $\reg{B}_{P}$ commutes with $\mathcal{A}_P$ if $\mathcal{A}_Q$ commutes with $\mathcal{A}_P$.  As the only algebra that commutes with the full algebra on a subspace is the trivial algebra, this will complete the theorem.  

    Fix an element $h$ of $(\mathcal{A}_Q)|_{\reg{B}_P}$. By definition $h = \Pi_{\reg{B}_{P}} h' \Pi_{\reg{B}_{P}}$ where $h' \in \mc A_Q$. Additionally, every element of $\mathcal{A}_{P}$ is in the $+1$-eigenspace of $\Pi_{\reg{B}_{P}}$ (and therefore commutes with the projector). Then we have the following for all operators $g \in \mathcal{A}_{P}$:
    \begin{align*}
        [h, g] &= hg - gh\numberthis\\ 
        &= \Pi_{\reg{B}_{P}} h' \Pi_{\reg{B}_{P}} g - g  \Pi_{\reg{B}_{P}} h' \Pi_{\reg{B}_{P}}\\
        &= \Pi_{\reg{B}_{P}} h' g\Pi_{\reg{B}_{P}} - \Pi_{\reg{B}_{P}} g h' \Pi_{\reg{B}_{P}}\\
        &= \Pi_{\reg{B}_{P}} [h', g] \Pi_{\reg{B}_{P}}\\
        &= 0\,.
    \end{align*}
    Here we use the fact that $g$ commutes with the projector, and $h'$ commutes with $g$ by the assumption that $\mathcal{A}_Q$ commutes with $\mathcal{A}_P$.  Therefore, the theorem is proved.
\end{proof}
Combining \Cref{thm:rank_1_commutation} with the Structure Lemma (\Cref{lem:structure}), we get the following corollary about the way that rank $1$ projectors commute with neighboring terms.
\begin{cor}
    \label{cor:one_rank_1}
    Let $P_{\reg{AB}}$ be a rank $1$ projector and $Q$ be another projector that commutes with it and overlaps on
    $\reg{B}$.  Then the direct sum decomposition from \Cref{lem:structure} consists of two subspaces, $\reg{R}_1 \oplus
    \reg{R}_2$ with $\reg{R}_1 = (\reg{B}_P \otimes \reg{B}_Q)$, where $\reg{B}_P$ is the subspace implied by
    \Cref{thm:rank_1_commutation}. Moreover, the following is true.
    \begin{enumerate}
        \item Within $\reg{R}_1$, $Q$ looks like $\id_{\reg{B}_P} \otimes \widetilde{Q}$ where $\widetilde{Q}$ is supported on $\reg B_Q$.
        \item Within $\reg{R}_{2}$, $P$ is $0$. 
    \end{enumerate}
\end{cor}
\begin{proof}
    We first apply the Structure Lemma, which implies that we can decompose the space $\reg{B}$ into $\reg{R}_1 \oplus
    \reg{R}_2$. By \Cref{thm:rank_1_commutation} item (1), $P_{\reg{AB}}$ acts non-trivially only one one of $\reg R_1$
    or $\reg R_2$; assume without loss of generality it is $\reg R_1$. Next, \Cref{thm:rank_1_commutation} further
    implies that the algebra generated by $P_{\reg{AB}}$ on $\reg B$ is the full algebra on a subspace $\reg B_P \subseteq
    \reg R_1$. Take $\reg{B}_Q$ to be the subspace such that $\reg{R}_1 = \reg B_P \otimes \reg B_Q$; such a subspace
    exists by the Structure Lemma. Furthermore, the algebra of $Q$ on $\reg B_P$ is the identity, and thus $Q|_{\reg
    R_1} = \id \otimes \wt{Q}$. Finally, since $P_{\reg{AB}}$ acts non-trivially on $\reg R_1$, it must act as $0$ on
    $\reg R_2$ (as otherwise it would have rank $> 1$). 
\end{proof}

\begin{proof}[Proof of \Cref{thm:rank_1_commutation_main}]
    We apply \Cref{cor:one_rank_1} to the case of two rank-$1$ projectors. If $\tilde Q$ in the above corollary is
    non-zero, then $\rank(Q) \geq \dim(\reg B_P)$; this implies $\dim(\reg B_P) = 1$ and $\reg B_P = \text{span}(\ketbra
    \psi)$ for some state $\ket \psi$. We conclude that $Q = \ketbra \psi \otimes \tilde Q$ and similarly for $P$, as it
    is the full algebra on $\reg B_P$. This is exactly the singular case (\Cref{def:rank_1_singular}).

    Otherwise, if $\tilde Q = 0$ then we may set $\Pi$ to be the projection onto $\reg R_1$ to match the reducing case
    (\Cref{def:rank_1_reducing}).
\end{proof}

\begin{remark}[On rank larger than $1$]
    One may wonder whether these techniques extend to the rank $> 1$ setting. When attempting such a generalization, it
    turns out that the \emph{singular} case becomes challenging. However, if we were able to assume that $Q$ is of a lower rank that
    $\dim(\reg{B}_P)$, we see from the above that the singular case can not happen. Given this (admittedly specific)
    constraint, we can generalize our proof to higher ranks.
\end{remark}

\paragraph{Connection to rounding schemes} As a quick example, let's see how to apply \Cref{thm:rank_1_commutation_main}
in the degree-4, rank-1 case. Consider a qudit register $\reg R$, acted on non-trivially by $P_1, P_2$ and $Q_1, Q_2$,
where the pairs $P_1$ and $P_2$, and $Q_1$ and $Q_2$) only interact on $\reg R$ (as in \Cref{fig:q_surrounding}). By
\Cref{thm:rank_1_commutation_main}, each pair commute in a \emph{reducing} or \emph{reducing} way. We notice that if
both pairs commute in a reducing way, then the corresponding projectors $\pi_P$ and $\pi_Q$ satisfy the requirements of
\Cref{thm:rounding}. This yields a new Hamiltonian where,
\begin{enumerate}
    \item One of $P_1, P_2$ is removed, WLOG let it be $P_2$.
    \item One of $Q_1, Q_2$ is removed, WLOG let it be $Q_2$.
    \item The remaining terms $P_1, P_1$ are combined into a single term $h_\text{merge}$ which commutes with all other
        Hamiltonian  terms.
\end{enumerate}
This is depicted in \Cref{fig:merge_case}. Thus, we get the following guided reduction which punctures a single hole in
the Hamiltonian.
\begin{cor}[Guided reduction for rank-1 \twodclhgrid{}]
    \label{cor:reducing_guided_reduction}
    Suppose for a \twodclhgrid{} instance $H$ there exists a qudit register $\reg R$ on which the diagonal terms commute
    in a reducing way. Then, there exists a guided reduction to a Hamiltonian $\tilde H$ where all terms acting
    trivially on $\reg R$ are unchanged, and there exists only a single term $h_\text{merge}$ acting non-trivially on
    $\reg R$, spanning the supports of two terms which were adjacent and acted non-trivially on $\reg R$ in $H$.
\end{cor}
\noindent In the next section, we'll scale this reduction up by considering the other \emph{singular} commuting terms as
well.

\section{\texorpdfstring{\twodclhr{1} is in $\cNP$}{Rank 1 2D CLH is in NP}}
\label{sec:warmup_2d}
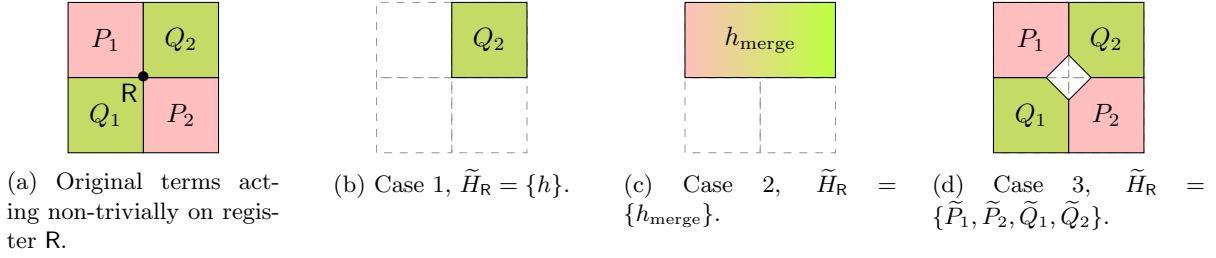
\begin{figure}[!ht]
\centering
\begin{subfigure}[t]{0.22\textwidth}
    \centering
    \begin{tikzpicture}[scale=1]
        \draw[fill=pink] (1,2) rectangle ++ (1,1) node[midway] {$P_1$};
        \draw[fill=pink] (2,1) rectangle ++ (1,1) node[midway] {$P_2$}; 
        \draw[fill=SpringGreen] (2,2) rectangle ++ (1,1) node[midway] {$Q_2$}; 
        \draw[fill=SpringGreen] (1,1) rectangle ++ (1,1) node[midway] {$Q_1$}; 
    
        \node[label={[label distance=-0.35cm]below left:$\reg R$}] at (2,2) {\textbullet};
    \end{tikzpicture}
    \caption{Original terms acting non-trivially on register $\reg R$.}
\end{subfigure}
\quad
\begin{subfigure}[t]{0.22\textwidth}
    \centering
    \begin{tikzpicture}
        \foreach \x in {0,1} {
            \foreach \y in {0,1} {
                \draw[dashed,black!40] (\x,\y) rectangle ++(1,1);
            }
        } 
        \draw[fill=SpringGreen] (1,1) rectangle ++ (1,1) node[midway] {$Q_2$};
    \end{tikzpicture}
    \caption{Case 1, $\tilde H_{\reg R} = \{h\}$.}
\end{subfigure}
\quad
\begin{subfigure}[t]{0.22\textwidth}
    \centering
    \begin{tikzpicture}
        \foreach \x in {0,1} {
            \foreach \y in {0,1} {
                \draw[dashed,black!40] (\x,\y) rectangle ++(1,1);
            }
        } 
        \draw[left color=pink, right color=SpringGreen] (0,1) rectangle ++ (2,1) node[midway] {$h_\text{merge}$};
    \end{tikzpicture}
    \caption{Case 2, $\tilde H_{\reg R}$ $=$ $\{h_\text{merge}\}$.}
    \label{fig:merge_case}
\end{subfigure}
\quad
\begin{subfigure}[t]{0.22\textwidth}
    \centering
    \begin{tikzpicture}
        \foreach \x in {0,1} {
            \foreach \y in {0,1} {
                \draw[dashed,black!40] (\x,\y) rectangle ++(1,1);
            }
        } 
        \draw[fill=SpringGreen] (0,0) -- (1,0) -- (1,0.7) -- (0.7,1) -- (0,1) -- cycle;
        \draw[fill=SpringGreen] (2,2) -- (2,1) -- (1.3,1) -- (1,1.3) -- (1,2) -- cycle;
        \draw[fill=pink] (0,1) -- (0.7,1) -- (1,1.3) -- (1,2) -- (0,2) --cycle;
        \draw[fill=pink] (1,0) -- (1,0.7) -- (1.3,1) -- (2,1) -- (2,0) -- cycle;
        \path (0,1) rectangle ++ (1,1) node[midway] {$P_1$};
        \path (1,0) rectangle ++ (1,1) node[midway] {$P_2$}; 
        \path (1,1) rectangle ++ (1,1) node[midway] {$Q_2$}; 
        \path (0,0) rectangle ++ (1,1) node[midway] {$Q_1$}; 
    \end{tikzpicture}
    \caption{Case 3, $\tilde H_{\reg R}$ $=$ $\{\tilde P_1, \tilde P_2, \tilde Q_1, \tilde Q_2\}$.}
\end{subfigure}
\captionsetup{width=0.8\textwidth}
\caption{The original Hamiltonian and the possible resulting Hamiltonians from the rounding scheme of \Cref{lem:deg_4_puncturing}.}
\label{fig:q_surrounding}
\end{figure}
To begin, we can show that rank-1 2D-CLH instances are in $\mathsf{NP}$ for all local dimensions $d$. As noted before, our proof is through a \emph{guided reduction} (\Cref{def:guided_reduction}).
\begin{theorem}\label{thm:2D_rank1_np}
    There is a guided reduction from rank-1 \twodclhr{1} to 2-local CLH.
\end{theorem}
Then, since any 2-local CLH is in $\cNP$ by \cite{Bravyi_BV2004_CommutativeVersionKlocal}, we get the following simple corollary.
\begin{cor}
    \label{cor:2d_rank1_np}
    The local Hamiltonian problem for \twodclhr{1} is in $\cNP$.
\end{cor}

The proof of \Cref{thm:2D_rank1_np} is inspired by \cite{Aharonov_AKV2018_ComplexityTwoDimensional}. However, our final
result is incomparable as we handle the non-starred version of the problem for any local dimension (see
\Cref{def:2d_clh} versus \Cref{def:2dstar_clh}), but \cite{Aharonov_AKV2018_ComplexityTwoDimensional} is able to handle
Hamiltonian terms of any rank. As in \cite{Aharonov_AKV2018_ComplexityTwoDimensional}, we will show that the rank-1
condition allows us to create many holes in the 2D grid, so that after grouping qudit registers of dimension $d$ into
new registers of dimension at most $\O(d)$, the resulting Hamiltonian becomes 2-local. However, rather than creating
holes by reduction to the Toric code, we use the rounding scheme developed in \Cref{thm:rounding}.

Rounding will allow us to prove following key lemma, which describes how to simplify a set of terms around a single
register. For now we give a version of the lemma that only works for degree-$4$ registers; however, in
\Cref{sec:extending} we'll show that a simple corollary of this lemma is a similar result for registers with higher
degree.
\begin{restatable}[Puncturing a $4$-local register]{lemma}{degfourpuncturing}
    \label{lem:deg_4_puncturing}
    Let $H$ be a Hamiltonian, and let $\reg R$ be a register acted on non-trivially $H_{\reg R} = \{P_1, P_2, Q_1,
    Q_2\}$. So, $H = \sum_{h \in H_{\reg R}} h + \sum_{h \in H_\text{rest}} h$, where $H_\text{rest}$ are terms acting
    trivially on $\reg R$. Furthermore, suppose that $P_1$ and $P_2$ intersect only on $\reg R$, and same for $Q_1,
    Q_2$. Then for any ``way'' the pair $P_1$ and $P_2$ and pair $Q_1$ and $Q_2$ commute (as in
    \Cref{thm:rank_1_commutation_main}), there a rounding scheme yielding a set of at most $3$ Hamiltonian terms $\tilde
    H_{\reg R}$ such that the Hamiltonian
    \begin{equation}
        \tilde H = \sum_{h \in \tilde H_{\reg R}} h + \sum_{h \in H_\text{rest}} h \quad \text{and} \quad
        \lambda_0(\tilde H) = 0 \iff \lambda_0(H) = 0
    \end{equation}
    Furthermore, the set $\tilde H_{\reg R}$ falls into one of three cases. See \Cref{fig:q_surrounding} for matching
    figures.
    \begin{enumerate}
        \item $\tilde H_{\reg R} = \{h\}$, with $h \in H_{\reg R}$. \label{item:puncture_case1}
        \item $\tilde H_{\reg R} = \{h_\text{merge}\}$ with $h_\text{merge}$ having support equal to a union of $P_i$,
            $Q_j$ for some $i,j \in [2]$. \label{item:puncture_case2}
        \item $\tilde H_{\reg R} = \{\tilde P_1, \tilde P_2, \tilde Q_1, \tilde Q_2\}$ where $\reg R$ is removed from
            the support of each operator.\label{item:puncture_case3}
    \end{enumerate}
\end{restatable}

\begin{restatable}[Degree $k > 4$ puncturing]{cor}{localpuncturing}
    \label{cor:local_puncturing}
    \Cref{lem:deg_4_puncturing} holds whenever register $\reg R$ has degree $k > 4$. In this case, either $\reg R$ can
    be removed from the Hamiltonian, or all but $3$ terms can be made to act trivially on $\reg R$ introducing a hole
    in the 2D surface.
\end{restatable}
With this corollary, \Cref{thm:2D_rank1_np} follows identically as in \cite[Section
7]{Aharonov_AKV2018_ComplexityTwoDimensional}.
\begin{proof}[Proof of \Cref{thm:2D_rank1_np}]
    Fix some triangulation $\mc T$ of $H$. For each triangle $T\in \mc T$, identify a qudit register $\reg R$ such that
    each neighbor of $\reg R$ is contained entirely in $T$. For each $\reg R$, we apply \Cref{cor:local_puncturing}, and
    the prover provides the projectors required by the rounding scheme. In each case of \Cref{lem:deg_4_puncturing}, we
    see that we puncture a hole adjacent to $\reg R$. Thus, as in \cite{Aharonov_AKV2018_ComplexityTwoDimensional}, we
    place the corners of the co-triangulation at these holes, and draw paths from each hole to the centers of the edges
    of the surrounding triangles to construct the grouped register ${\reg R}_T$ (representing a higher-dimensional
    qudit). Given this grouping, terms either cross exactly one edge (and thus are $2$-local) or act only on $\reg R_T$
    (and are $1$-local). This yields a guided reduction from \twodclhgrid{} to $2$-local CLH.
\end{proof}

In the remainder of this section, we'll prove \Cref{lem:deg_4_puncturing}, deferring the proof of
\Cref{cor:local_puncturing} to \Cref{sec:extending}. 

\begin{proof}[Proof of \Cref{lem:deg_4_puncturing}]
    As in \Cref{fig:q_surrounding}, consider the pairs $P_1, P_2$ and $Q_1, Q_2$, each of which intersect only on
    register $\reg R$. Then, the possible cases from \Cref{cor:one_rank_1}
    are,\textcolor{white}{\labeltext{\textcircled{\textbullet}}{text:cases}}
    \begin{itemize}
        \item \textbf{(Case A)} \emph{Both} decompositions are reducing.
        \item \textbf{(Case B)} \emph{Exactly one} decomposition is reducing.
        \item \textbf{(Case C)} \emph{Neither} decompositions are reducing. This implies $(P_1)|_{\reg R} = (P_2)|_{\reg
            R} = \ketbra \psi$ and $({\reg R}_1)|_{\reg R} = ({\reg R}_2)|_{\reg R} = \ketbra \phi$.
    \end{itemize}
    In each of these cases, we'll show that we obtain one of the three cases from the lemma statement. Since $P_1, P_2$
    commute and intersect only on $\reg R$, the Structure Lemma induces a direct sum decomposition of $\mc H_{\reg R} =
    \oplus_i (\mc H_{\reg R})_i$. We denote the projector onto one of these subspaces provided by the prover as $\pi_P$.
    Similarly, we obtain a projector $\pi_Q$, from the decomposition induced by $Q_1$ and $Q_2$.

    \vspace{1em}
    \noindent \textbf{Case A (Both reducing)}.
    In the case when both are reducing, we claim that we can apply \Cref{thm:rounding}.  In particular, by the
    definition of reducing, one of two $Q$ terms does not survive $\pi_Q$, leaving only a single term that does not
    commute with $\pi_P$, and similarly only a single $P$ term survives $\pi_P$, leaving a single $P$ term which does
    not commute with $\pi_Q$.  We let $\widetilde{H}_{\reg R}$ be the Hamiltonian produced by \Cref{thm:rounding}. The
    theorem guarantees that $\tilde{H}_{\reg R}$ satisfies \Cref{item:puncture_case2}.
    
    \vspace{1em}
    \noindent \textbf{Case B (One reducing)}. Without loss of generality, say $P_1$, $P_2$ commute in a singular manner
    (so $P_1 = \tilde P_1 \otimes \ketbra \psi$ and $P_2 = \tilde P_2 \otimes \ketbra \psi$) and $Q_1, Q_2$ commute in a
    reducing manner. If $\pi_P$ is onto $\mc H_\text{rest}$, then we again left with a single term, leaving us in
    \Cref{item:puncture_case1}. If $\pi_Q$ is onto $\mc H_\text{rest}$, then all remaining terms act as $\ketbra \psi$
    on $\reg R$, and $\reg R$ is a classical register. Thus, by \Cref{obs:classical_restriction} we can obtain an
    equivalent Hamiltonian where each term acts trivially on $\reg R$ and we obtain \Cref{item:puncture_case3}).
    Otherwise, we assume $\pi_P = \ketbra \psi$ and $\pi_Q$ preserves $Q_1$. Thus (by \Cref{lem:equiv_projector}), we it
    suffices to analyze
    \begin{equation}
        \tr\Brac{(\psi - \tilde P_1 \otimes \psi)(\psi - \tilde P_2 \otimes \psi)(\pi_Q - Q_1) \prod_{h \in S} (\Id
        -h)}\,.
    \end{equation}
    Now, pull out a $\psi$ from either side of the two $P$-type terms, and conjugate $\pi_Q - Q_1 \rightarrow \alpha
    \cdot \psi - \psi Q_1 \psi$ (where $\alpha = \tr[\pi_Q \psi]$). Since $\psi$ and $\psi Q_1 \psi$ are simultaneously
    diagonalizable, we can use the rounding technique from \cite{Irani_IJ2023_CommutingLocalHamiltonian} to round
    $\alpha \psi - \psi Q_1 \psi \rightarrow \psi - \mathcal R(\psi Q_1 \psi)$
    \footnote{
        More precisely, we note that $\alpha \cdot \psi \succeq \psi Q_q \psi$. Then, pull the positive constant
        $\alpha$ out of the trace. Then, within each non-zero eigenspace of $\psi Q_1 \psi$, $\psi$ has eigenvalue $1$.
        Finally, the rounding map $\mc R$ takes each non-zero eigenvalue to $0$, which by
        \cite{Irani_IJ2023_CommutingLocalHamiltonian} yields an equivalent trace expression.
    } 
    which yields a projector commuting with $P_1, P_2$, as well as all remaining terms $h \in S$. Thus, the register
    $\reg R$ again is classical, and we obtain \Cref{item:puncture_case3}.
    
    \vspace{1em}
    \noindent \textbf{Case C (Both singular)}. First, if one of $\pi_P$ or $\pi_Q$ project onto $\mc H_\text{rest}$, then the $q$ becomes
    classical. Otherwise we can write,
    \begin{align*}
        P_1 =\ketbra \psi \otimes \tilde P_1  \qquad & \qquad P_2 =\ketbra \psi \otimes \tilde P_2\numberthis\\
        Q_1 =\ketbra \phi \otimes \tilde Q_1  \qquad & \qquad Q_2 =\ketbra \phi \otimes \tilde Q_2
    \end{align*}
    with $\pi_P = \ketbra \psi$ and $\pi_Q = \ketbra \phi$. By commutation of $P_1, P_2$ and $Q_1, Q_2$, we have that
    $[\tilde P_1, \tilde P_2] = [\tilde Q_1, \tilde Q_2] = 0$. Similarly, since each $h \in S$ acts trivially on $\reg
    R$, their commutation with $P_1, P_2, Q_1, Q_2$ is equivalent to their commutation with the tilde-d terms. Once
    again we can apply $\pi_P$ and $\pi_Q$ to the $P$ and $Q$-type terms respectively via \Cref{lem:equiv_projector},
    and obtain,
    \begin{align*}
        &\tr\Brac{(\psi - \psi \otimes \tilde P_1 )(\psi -\psi \otimes \tilde P_2 )(\phi - \phi \otimes \tilde Q_1)(\phi -
        \phi \otimes \tilde Q_2)\prod_{h \in S} (\id -h)}\numberthis\\
        &= \tr\Brac{\phi_{\reg R} \psi_{\reg R} (\id - \id \otimes \tilde P_1 ) \Paren{\id_{\reg R} \otimes (\id  - \tilde P_2
            )}\psi_{\reg R} \phi_{\reg R} (\id - \id
        \otimes \tilde Q_1) \Paren{\id_{\reg R} \otimes (\id - \tilde Q_2)} \prod_{h \in S} (\Id -h)}\\
        &= \braket{\psi}{\phi}^2\ \tr\Brac{\bra{\psi}_{\reg R}(\id - \id \otimes \tilde P_1) \ket{\psi}_{\reg R} \cdot (\id - \tilde P_2)
        \bra{\phi}_{\reg R} (\id - \id \otimes \tilde Q_1) \ket{\phi_{\reg R}} \cdot (\id - \tilde Q_2) \prod_{h \in S}
    (\id - h)}\\
        &= \braket{\psi}{\phi}^2\ \tr\Brac{(\id - \tilde P_1)(\id - \tilde P_2)(\id - \tilde Q_1) (\id - \tilde Q_2) \prod_{h
    \in S}(\id - h)}
    \end{align*}
    where in the second line, we have pulled out the $\psi, \phi$ projectors, and used the cyclic property of the trace to
    pull one $\phi$ to the LHS. To obtain the third line, we take the trace with respect to $\reg R$. If the original
    Hamiltonian had a non-non-empty ground space, then there is a choice of $\psi, \phi$ such that $\braket \psi \phi^2
    > 0$. Besides the constant factor, checking whether the trace is greater than $0$ is equivalent to the condition that
    \begin{equation}
        \lambda_0(\tilde H) = 0 \quad \text{where}\quad \tilde H = \tilde P_1 + \tilde P_2 + \tilde Q_1 + \tilde Q_2 + \sum_{h
        \in S} h\,,
    \end{equation}
    and $\tilde H$ is a Hamiltonian which does not act on $\reg R$. Moreover, this Hamiltonian is commuting, by our
    observation earlier. Thus, we have rounded to a CLH instance satisfying \Cref{item:puncture_case3}.
\end{proof}

\subsection{\texorpdfstring{Extending to degree $k > 4$}{Extending to degree larger than 4}}
\label{sec:extending}
We now show how \Cref{cor:local_puncturing} follows from \Cref{lem:deg_4_puncturing}.
\localpuncturing*
Recall that the degree of a register is the number of terms acting non-trivially on it. We can extend the above proof to
work for registers with degree larger than $4$. Since we are working with planar, 2D Hamiltonians, we may index the
terms for a single register $\reg R$ as $H_1, \dots, H_k$, such that for each $H_i$, $H_i$ shares two registers with
$H_{i-1}$ and $H_{i+1}$, and only the register $\reg R$ for $H_j$, $|i - j | > 1$\footnote{Arithmetic here is performed
modulo $k$, so that $k+1 \equiv 1$}. Define two sets, $\mc P$ and $\mc Q$ such that,
\begin{align*}
\mc Q &= \{H_{2i-1} \st i \in \floor{k/2}\}\numberthis\\
\mc P &= \{H_{2i} \st i \in \floor{k/2}\}
\end{align*}
i.e. $\mc P$ contains the even indexed $H_i$'s, and $\mc Q$ contains the odds. If $k$ is even then each pair $P_1, P_2
\in \mc P$ and $Q_1, Q_2 \in \mc Q$ share only a single register. On the other hand, if $k$ is odd, then $Q_1$ and $Q_{k}$
intersect on 2 registers. In this case, we redefine $Q_1 \dfn Q_1 + Q_k$, yielding a single rank $2$ term. 
Since \Cref{cor:one_rank_1} works as long as all but a single term is rank-$1$, we may assume that $k$ is even.

Since each $P_i, P_j \in \mc P$ only intersect on a single register, we can analyze their local algebra on $\reg R$
using the Structure Lemma. By iteratively applying \Cref{thm:rank_1_commutation}, we see that each $P_i$ has an induced
algebra $\mc A_{P_i}$ which is full on some subspace. Moreover, these subspaces are orthogonal unless they are
1-dimensional. Group the operators $P_i \in \mc P$ into sets $S^{\mc P}_1, \dots, S^{\mc P}_\ell$ such that every
$S^{\mc P}_i$ has a corresponding register $\reg{R}_i$ with $\mc A_{P_i} = \mc L(\reg R_i)$ for each $P_i \in S^{\mc
P}_i$. These subspaces further satisfy the following properties
\begin{enumerate}
    \item The $\reg{R}_i$'s partition the Hilbert space of register $\reg R$, i.e. $\reg{R} = \oplus_{i=1}^\ell \reg{R}_i$.
    \item If $|S^{\mc P}_i| > 1$, then $\dim(\reg{R}_i) = 1$.
\end{enumerate}
The second conditions follows by applying \Cref{cor:one_rank_1} to the elements that are the full algebra on the same
subspace. Similarly, we may define sets $\mc S^{\mc Q}_i$.  Here we point out that if we have a rank-2 term from $k$
being odd, the second condition still holds.  In particular, if there is another rank $1$ projector in
$S^{\mathcal{P}}_i$, then both terms in the sum $Q_1 + Q_k$ must individually commute with it, and thus the dimension of
$\reg{R}^q_i$ must be $1$ by \Cref{thm:rank_1_commutation}.

The condition that our Hamiltonian as a $0$-energy ground space is equivalent to the condition that,
\begin{equation}
    \tr\Brac{\prod_{P \in \mc P} (\Id-P) \prod_{Q \in \mc Q} (\Id - Q) \prod_{h \in S} (\Id - h)} > 0\,.
\end{equation}
Write $\Pi^{\mc P}_a$ and $\Pi^{\mc Q}_b$ as the projectors onto $\mc H^{\mc P}_a$ and $\mc H^{\mc Q}_b$
respectively. By \Cref{lem:equiv_projector}, we can further write the equivalent statement,
\begin{equation}
\label{eq:general_projected}
\exists a,b \text{ s.t. }\tr\Brac{\prod_{P \in \mc P} (\Pi^{\mc P}_b-\Pi^{\mc
P}_b P \Pi^{\mc P}_b) \prod_{Q \in \mc Q} (\Pi^{\mc Q}_a-\Pi^{\mc Q}_a Q \Pi^{\mc Q}_a)\prod_{h \in S} (\Id - h)} > 0
\end{equation}
This choice of $a,b$ zeroes out all terms except those in $S^{\mc P}_a$ and
$S^{\mc Q}_b$. At this point, we can case on the size of each set, resembling the cases \textbf{A}, \textbf{B}, and
\textbf{C} considered in the previous section.
\begin{itemize}
    \item $|S^{\mc P}_a| = |S^{\mc Q}_b| = 1$. In this case, we're left with a single $Q$-type and $P$-type term, and we
        can applying the analysis of \textbf{Case A}.
    \item $|S^{\mc P}_a| = 1$ and $|S^{\mc Q}_b| > 1$. This implies that each $Q \in S^{\mc Q}_b$ is of the form $\phi
        \otimes \tilde Q$, for a fixed state $\phi$. This corresponds to \textbf{Case B}, yielding a classical qudit.
        The case when $|S^{\mc P}_a| > 1$ and $|S^{\mc Q}_b|  = 1$ is identical.
    \item $|S^{\mc P}_a|, |S^{\mc Q}_b| > 1$. Then each non-zero $P = \psi \otimes \tilde P$ and $Q = \phi \otimes \tilde
        Q$. This corresponds to case \textbf{Case C}, and we can conjugate each $Q$ term with $\psi$. This yields a
        classical qudit.
\end{itemize}

\section{Commuting Hamiltonians in three dimensions}
\label{sec:proof_3d}
\NewDocumentCommand{\DrawCubes}{O{} m m m m m m}{%
    \def\XCoord{#2}
    \def\YCoord{#3}
    \def\ZCoord{#4}
    \def\XWidth{#5}
    \def\YWidth{#6}
    \def\ZWidth{#7}
    \draw[canvas is xy plane at z=\ZCoord, #1] (\XCoord,\YCoord) rectangle ++ (\XWidth,\YWidth) ;
    \draw[canvas is zy plane at x=\XCoord, #1] (\ZCoord,\YCoord) rectangle ++ (\ZWidth, \YWidth) ;
    \draw[canvas is xz plane at y=\YCoord, #1] (\XCoord,\ZCoord) rectangle ++ (\XWidth, \ZWidth) ;
    \draw[canvas is zy plane at x=\XCoord+\XWidth, #1] (\ZCoord,\YCoord) rectangle ++ (\ZWidth, \YWidth) ;
    \draw[canvas is xz plane at y=\YCoord+\YWidth, #1] (\XCoord,\ZCoord) rectangle ++ (\XWidth, \ZWidth) ;
    \draw[canvas is xy plane at z=\ZCoord+\ZWidth, #1] (\XCoord,\YCoord) rectangle ++ (\XWidth, \YWidth) ;
}%

In this section, we show the rank-1 assumption allows us to place commuting Local Hamiltonians over a 3D complex in
$\cNP$. We recall the following important assumption,
\generalizedlattice*
 Formally, we state our main result, and its immediate corollary, as follows.
 \begin{theorem}[Guided reduction for \threedclhr{1}]
    \label{thm:3d_theorem}
    Let $H$ be an instance of \threedclhr{1} with degree $g$ (\Cref{def:reg_degree}) on $d$-dimensional registers.
    Suppose we also have a bound $k$ on the locality of any term. Then,
    \begin{itemize}
        \item If $g \leq \tfrac 1 e {d^k}$, then by \Cref{lem:qlll} the instance $H$ is trivially satisfiable.
        \item Otherwise, assume $g > \tfrac 1 e {d^k} \geq 6$. If $H$ satisfies \Cref{as:high_degree} (large minimum
            degree) and \Cref{as:generalized_cubic_lattice}, then there is a guided reduction from $H$ to a $2$-local
            commuting local Hamiltonian. 
    \end{itemize}
\end{theorem}

\begin{cor}[Local Hamiltonian problem on \threedclhr{1} in $\cNP$]
    For non-trivial \threedclhr{1} instances meeting the requirements of \Cref{thm:3d_theorem}, the local Hamiltonian
    problem on these instances is in $\cNP$.
\end{cor}

Note that the \emph{cubic lattice} implies a degree bound of $g = 4$ and locality $k = 8$, for which the first case of
\Cref{thm:3d_theorem} applies (and thus the instance is trivially satisfiable). However, most of the ideas for the
general case are extensions of ideas from the cubic lattice, so we will first describe our proof on this more restricted
setting.

\begin{figure}[h]
\centering
\begin{tikzpicture}[scale=0.75]
    \foreach \z in {-2,-1,...,3} {
        \foreach \x in {0,...,6} {   
            \foreach \y in {0,...,6} {   
                \pgfmathsetmacro{\c}{40-(3-\z)*5}
                \DrawCubes[black!\c,dashed]{\x}{\y}{\z}{1}{1}{1}
            }
        }
    }
    \foreach \z in {-3,0} {
        \foreach \x in {0,3} {   
            \foreach \y in {0,3} {   
                \pgfmathsetmacro{\xd}{\x+.5}
                \pgfmathsetmacro{\yd}{\y+.5}
                \pgfmathsetmacro{\zd}{\z+.5}
                \DrawCubes[CornflowerBlue, line width=0.5mm]{\xd}{\yd}{\zd}{3}{3}{3}
            }
        }
    }
    \DrawCubes[draw=none, fill=SpringGreen,fill opacity=0.5]{0}{0}{3}{7}{1}{1}
    \draw[CornflowerBlue, line width=0.5mm, canvas is xy plane at z=3.5] (0.5,1) -- (0.5,2);
    \draw[CornflowerBlue, line width=0.5mm, canvas is xy plane at z=3.5] (3.5,1) -- (3.5,2);
    \draw[CornflowerBlue, line width=0.5mm, canvas is xy plane at z=3.5] (6.5,1) -- (6.5,2);
\end{tikzpicture}
\captionsetup{width=.8\linewidth}
\caption{
    An example of a cubulation. The blue edges represent the cubulation, where the qubits (on edges) inside each blue
    cube are grouped into a single qudit. The green shading emphasizes some terms intersecting the cubulation.
}
\label{fig:overall_cubulation}
\end{figure}
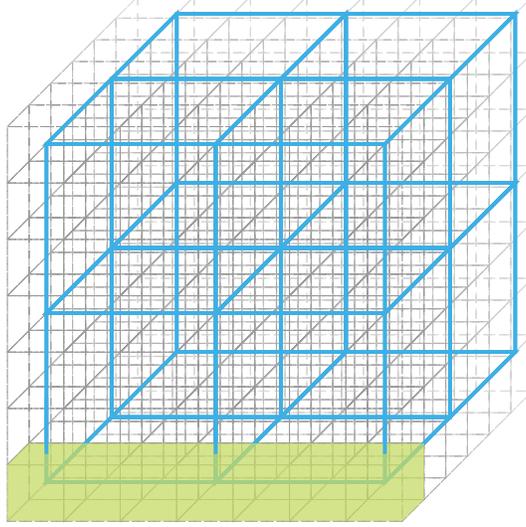

Our strategy in this section is to induce a ``cubulation'' in the 3D space, where a lattice of cubes is superimposed
over the Hamiltonian terms, as in \Cref{fig:overall_cubulation}. We refer to the set of larger cubes as $\mc C$. Given a
cubulation $\mc C$, we can classify terms in the original Hamiltonian as into 3 ``types:'' those intersecting a face of
the cubulation, intersecting an edge, or intersecting a vertex. Those intersected a face are automatically 2-local. In
the 2D setting terms intersecting edges were also 2-local. In the 3D case, these are 4-local. Therefore, we will need to
use our rounding scheme from \Cref{cor:local_puncturing} to create ``tunnels'' so that the edges can pass through holes
in the Hamiltonian. This still is not quite sufficient, as there can blockages in the generated tunnels, see
\Cref{fig:blocked_tunnel}. To address this issue, we use the fact that the result of rounding schemes are also valid CLH
instances and that these blockages are essentially 2-local interactions. Therefore, we will be able to apply the 2-local
rounding scheme in the form of \Cref{cor:product_structure_2_local} to fix this case.

\newsavebox{\imagebox}
  
\tikzset{pics/cubes/.style n args={3}{code={
    \node[canvas is xy plane at z=0] (-center) at (#1, #2) {};
    \foreach \z in {-1,...,1} {
        \foreach \x in {0,1} {   
            \foreach \y in {0,1} {   
                \pgfmathtruncatemacro{\xn}{\x + #1}%
                \pgfmathtruncatemacro{\yn}{\y + #2}%
                \DrawCubes[black!20,dashed]{\xn}{\yn}{\z}{1}{1}{1}
            }
        }
    }
}}}

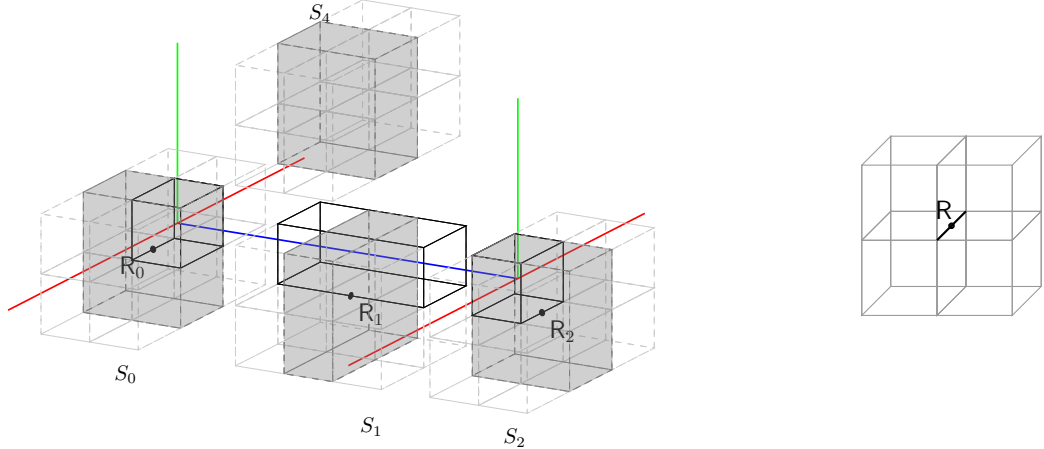
\begin{figure}
\centering

\savebox{\imagebox}{
    \scalebox{0.8}{
    \begin{tikzpicture}[
        grayfill/.style={fill,gray,fill opacity=0.2},
    ]
        \begin{scope}[rotate around y=-20]
        \draw[canvas is zx plane at y=1.5,blue,thick] (0.5,1.5) -- (0.5,8.5);
        \draw[canvas is zy plane at x=1.5,green,thick] (0.5,1.5) -- (0.5,4.5);
        \draw[canvas is zy plane at x=8.5,green,thick] (0.5,1.5) -- (0.5,4.5);
        \draw[canvas is zx plane at y=1.5,red,thick] (4.5,1.5) -- (-2.5,1.5);
        \draw[canvas is zx plane at y=1.5,red,thick] (4.5,8.5) -- (-2.5,8.5);
        
        \draw pic (first) {cubes={0}{0}{0}};
        \draw[canvas is xy plane at z=0.5, fill] (1,1) circle (1.5pt) node[below left] {\large $\reg R_0$};
        \node[below= of first-center] {$S_0$};
        \DrawCubes[black]{1}{1}{0}{1}{1}{1}
        \DrawCubes[grayfill,dashed]{0}{0}{0}{2}{2}{1}

        \foreach \z in {0,1} {
            \foreach \x in {0,...,2} {   
                \foreach \y in {0,1} {   
                    \pgfmathtruncatemacro{\xn}{\x + 4}%
                    \pgfmathtruncatemacro{\yn}{\y + 0}%
                    \DrawCubes[black!20,dashed]{\xn}{\yn}{\z}{1}{1}{1}
                }
            }
        }
        \draw[canvas is zy plane at x=5.5, fill] (1,1) circle (1.5pt) node[below right] (c) {\large $\reg R_1$};
        \node[below=1.5 of c] {$S_1$};
        \DrawCubes[black]{4}{1}{0}{3}{1}{1}
        \DrawCubes[grayfill,dashed]{5}{0}{2}{1}{2}{-2}
        
        \draw pic (third) {cubes={8}{0}{0}};
        \node[below= of third-center] {$S_2$};
        \draw[canvas is xy plane at z=0.5, fill] (9,1) circle (1.5pt) node[below right] {\large $\reg R_2$};
        \DrawCubes[black]{8}{1}{0}{1}{1}{1}
        \DrawCubes[grayfill,dashed]{8}{0}{0}{2}{2}{1}
        
        \draw pic (first) {cubes={4}{3}{0}};
        \node[above=2 of first-center] {$S_4$};
        \DrawCubes[grayfill,dashed]{4}{3}{0}{2}{2}{1}
        \end{scope}
    \end{tikzpicture}
    }
}%
\begin{subfigure}[t]{0.7\textwidth}
\centering\usebox{\imagebox}
\caption{
    The light-gray dashed lines indicate the individual Hamiltonian terms. The black boxes depict the Hamiltonian terms
    intersecting the boundaries of the ``cubulation'' (boundaries depicted by the red, green, and blue lines). The
    shaded in terms indicates a ``slice'' (denoted $S_i$) containing a term intersecting a boundary of the cubulation.
    Notice $S_4$ only intersects the plane defined by the green and blue lines.
}
\label{fig:edge_of_cubulation}
\end{subfigure}
\qquad
\begin{subfigure}[t]{0.2\textwidth}
    \centering\raisebox{\dimexpr.5\ht\imagebox-.5\height}{
    \newcommand{\dimension}{1}
        \begin{tikzpicture}
        \foreach \x in {0,...,\dimension} {   
            \foreach \y in {0,...,\dimension} {   
                \draw[canvas is xy plane at z=0, black!40] (\x,\y) rectangle ++ (1,1) ;
                \draw[canvas is xy plane at z=1, black!40] (\x,\y) rectangle ++ (1,1) ;
                \draw[canvas is zy plane at x=\x, black!40] (0,\y) rectangle ++ (1,1) ;
                \draw[canvas is zy plane at x=\x+1, black!40] (0,\y) rectangle ++ (1,1) ;
            }
        }
        
        \draw[canvas is zx plane at y=1, thick] (0,1) -- (1,1) node[midway, style={circle, inner
            sep={0.75\pgflinewidth}, draw}, font=\scriptsize, label={[label distance=-0.1cm,xshift=0.1cm]above
        left:$\reg R$}] {};
        
        \end{tikzpicture}
    }
    \caption{A single slice centered around a register $\reg R$.}
    \label{fig:basic_cube}
\end{subfigure}
\caption{A local view at a cubulation.}
\label{fig:cubulation}
\end{figure}

\subsection{3D cubic lattice}
\label{sec:cubic_lattice}

In the section, we consider the special case of \threedclhrgrid{1}. As mentioned above, we superimpose a cubulation $\mc
C$ over the 3D space, as in \Cref{fig:overall_cubulation}. Given a cubulation $\mc C$, we can classify the Hamiltonian
terms into 3 types. We use \Cref{fig:edge_of_cubulation} as a reference.
\begin{enumerate}
    \item Terms intersecting a \emph{face} of the cubulation (e.g. contained in $S_4$). \label{item:face_adj}
    \item Terms intersecting an \emph{edge} of the cubulation (e.g. terms contained in $S_1$). \label{item:edge_adj}
    \item Terms intersecting a \emph{vertex} of the cubulation (e.g. terms contained in $S_0$ and $S_2$). \label{item:vertex_adj}
\end{enumerate}
Type \ref{item:face_adj} terms are easy; these are automatically $2$-local as faces are shared by at most two adjacent
$C, C' \in \mc C$. Type \ref{item:edge_adj} and \ref{item:vertex_adj} are harder. For these remaining terms, the idea is
to apply our rounding technique from the 2D case to a ``slice'' of terms, i.e. \Cref{fig:basic_cube}. The key
observations is that from the point of a qudit register $\reg R$ on the center edge, the local geometry ``looks like''
the 2D grid. As a result, we are in a setting where we can apply \Cref{lem:deg_4_puncturing}, restated here for
convenience.
\degfourpuncturing*
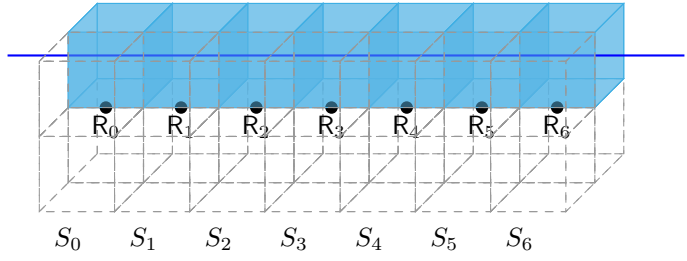
\begin{figure}
\centering
\begin{tikzpicture}[
        wallshadeb/.style={fill,CornflowerBlue!90,fill opacity=0.5},
    ]
    \foreach \x in {0,...,6} {   
        \foreach \y in {0,1} {   
            \foreach \z in {0,1} {
                \ifthenelse{\y = 1 \AND \z = 0}{
                    \DrawCubes[wallshadeb,draw=CornflowerBlue]{\x}{\y}{\z}{1}{1}{1}
                }{
                    \DrawCubes[black!40,dashed]{\x}{\y}{\z}{1}{1}{1}
                }
            }
        }
        \draw[canvas is xy plane at z=1] (\x,1) -- +(1,0) node[midway, circle, fill, scale=0.5,label={[label
            distance=-0.1cm]below:$\reg R_\x$}] {};
        \pgfmathtruncatemacro{\xn}{\x + 0.5}%
        \node[canvas is xy plane at z = 1] at (\xn,-0.75) {$S_\x$};
    }
    \draw[canvas is zx plane at y=1.5,blue,thick] (0.5,-1) -- (0.5,8);
    \foreach \x in {0,...,6} {
        \draw[canvas is xy plane at z=1, wallshadeb] (\x,1) rectangle ++(1,1);
        \DrawCubes[black!40,dashed]{\x}{1}{1}{1}{1}{1}
    }
\end{tikzpicture}
\captionsetup{width=.8\linewidth}
\caption{The set of ``slices'' $S_i$ around an edge of the cubulation (in blue). Each slice is centered around the
    register $\reg R_i$. The Hamiltonian term containing the edge is colored.}
\label{fig:tube}
\end{figure}

\paragraph{Edge-type terms.} Consider the set of terms $S_1$; as indicated in \Cref{fig:edge_of_cubulation}, we
pick the slice so that it is perpendicular to the (blue) edge. These slices look like \Cref{fig:tube}. For each slice we
will apply \Cref{lem:deg_4_puncturing}. As long as the reduced slices permit a (possibly curving) boundary to pass
through empty areas of the geometry, we ensure that only $2$-local terms remain, as the remaining terms pass through
\emph{faces} of the cubulation (and thus are $2$-local). One can check that in most cases, two adjacent slices to create
such a hole:
\begin{itemize}
    \item Type \ref{item:puncture_case1} leaves a hole when adjacent to \emph{any} other type.
    \item Type \ref{item:puncture_case3} removes the center register and thus always leaves a hole.
\end{itemize}
The trouble is when there are two adjacent Type \ref{item:puncture_case2} slices, and specifically, when the terms line
up as in $h_0$ and $h_1$ in \Cref{fig:blocked_tunnel} (it's easy to see that any other orientation also leaves a hole).
However, we now use a crucial aspect of \Cref{lem:deg_4_puncturing}, which is that the merged terms commute with all
terms outside of the slice. Under this type of blockage, the (green and blue) merged terms are the \emph{only} terms
acting non-trivially on their shared edge and thus can be viewed as 2-local operators acting on their intersection.
Moreover, all other terms act trivially there. Thus, we can directly apply \Cref{cor:product_structure_2_local} which
provides a rounding scheme creating a hole between $h_1$ and $h_0$.

\tikzset{pics/cornershade/.style={code={
    \draw[canvas is xy plane at z=1, #1] (-1,0) -- (0,1) -- (-1,1) -- cycle;
    \draw[canvas is xy plane at z=0, #1] (-1,0) -- (0,1) -- (-1,1) -- cycle;
    \draw[canvas is zx plane at y=1, #1] (0,-1) -- (1,-1) -- (1,0) -- (0,0) -- cycle;
    \draw[canvas is zy plane at x=-1, #1] (0,0) -- (0,1) -- (1,1) -- (1,0) -- cycle;
    \begin{scope}[rotate around z=-45]
        \pgfmathsetmacro\xd{sqrt(2)/2} 
        \draw[canvas is zy plane at x=-\xd, #1] (1,-\xd) -- (1,\xd) -- (0,\xd) -- (0,-\xd) -- cycle;
    \end{scope}
}}}
\begin{figure}
    \centering
    \begin{subfigure}[t]{0.4\textwidth}
    \centering
    \begin{tikzpicture}[
        wallshadep/.style={fill, pink,fill opacity=0.5},
        wallshadeg/.style={fill,SpringGreen,fill opacity=0.5},
        wallshadeb/.style={fill,CornflowerBlue!90,fill opacity=0.5},
    ]

    \foreach \z in {-2,-1,0} {
    \foreach \x in {-1,0} {   
        \foreach \y in {-1,0} {   
            \DrawCubes[black!30]{\x}{\y}{\z}{1}{1}{1}
        }
    }}
    \DrawCubes[wallshadeb]{-1}{-1}{-2}{2}{1}{1}
    \draw[canvas is zx plane at y=0, thick] (-2,0) -- (-1,0) node[midway, style={circle, inner sep={0.75\pgflinewidth},
        draw}, font=\scriptsize, label={[label distance=-0.1cm,xshift=0.1cm]above left:$\reg R_0$}] {};

    \DrawCubes[wallshadeg]{-1}{-1}{-1}{1}{2}{1}
    \draw[canvas is zx plane at y=0, thick] (-1,0) -- (0,0) node[midway, style={circle, inner sep={0.75\pgflinewidth},
        draw}, font=\scriptsize, label={[label distance=-0.1cm,xshift=0.1cm]above left:$\reg R_1$}] {};

    \foreach \angle in {0,90,180} {
        \begin{scope}[rotate around z=\angle]
            \pic {cornershade={wallshadep}};
        \end{scope} 
    }
    \begin{scope}[rotate around z=-70]
        \draw[canvas is zy plane at x=0,line width=0.5mm] plot [smooth] coordinates {(3,0.5) (-4,0.5)};
    \end{scope}

    \begin{scope}[rotate around z=270]
        \pic {cornershade={wallshadep}};
    \end{scope} 
    
    \draw[canvas is zx plane at y=0, thick] (0,0) -- (1,0) node[midway, style={circle, inner sep={0.75\pgflinewidth},
        draw}, font=\scriptsize, label={[label distance=-0.1cm,xshift=0.1cm]above left:$\reg R_2$}] {};
    
    \end{tikzpicture}
    \caption{
        The slice for $\reg R_2$ falls into \Cref{item:puncture_case3}, and the slices for $\reg R_1$ and $\reg R_0$ fall into
        \Cref{item:puncture_case2}. The orientation of the merged blocks permits the boundary to pass through.
    }
    \end{subfigure}
    \qquad
    \begin{subfigure}[t]{0.4\textwidth}
    \centering
    \begin{tikzpicture}[
        wallshadep/.style={fill, pink,fill opacity=0.5},
        wallshadeg/.style={fill,SpringGreen,fill opacity=0.5},
        wallshadeb/.style={fill,CornflowerBlue!90,fill opacity=0.5},
    ]

    \foreach \z in {-2,-1,0} {
    \foreach \x in {-1,0} {   
        \foreach \y in {-1,0} {   
            \DrawCubes[black!30]{\x}{\y}{\z}{1}{1}{1}
        }
    }}
    \DrawCubes[wallshadeb]{0}{-1}{-2}{1}{2}{1}
    \draw[canvas is zx plane at y=0, thick] (-2,0) -- (-1,0) node[midway, style={circle, inner sep={0.75\pgflinewidth},
        draw}, font=\scriptsize, label={[label distance=-0.1cm,xshift=0.1cm]above left:$\reg R_0$}] {};

    \DrawCubes[wallshadeg]{-1}{-1}{-1}{1}{2}{1}
    \draw[canvas is zx plane at y=0, thick] (-1,0) -- (0,0) node[midway, style={circle, inner sep={0.75\pgflinewidth},
        draw}, font=\scriptsize, label={[label distance=-0.1cm,xshift=0.1cm]above left:$\reg R_1$}] {};

    \draw[canvas is xy plane at z=-0.5, SpringGreen, line width=0.4mm] (-1.25,1) -- (-2,1) -- (-2,-1) node[midway,label={[label distance=-0.1cm]left:$h_1$}] {} -- (-1.25,-1);

    \draw[canvas is xy plane at z=-1.5, CornflowerBlue, line width=0.4mm] (1.25,1) -- (1.5,1) -- (1.5,-1) node[midway,label={[label distance=-0.1cm]right:$h_0$}] {} -- (1.25,-1);

    \foreach \angle in {0,90,...,270} {
        \begin{scope}[rotate around z=\angle]
            \pic {cornershade={wallshadep}};
        \end{scope} 
    }
    
    \draw[canvas is zx plane at y=0, thick] (0,0) -- (1,0) node[midway, style={circle, inner sep={0.75\pgflinewidth},
        draw}, font=\scriptsize, label={[label distance=-0.1cm,xshift=0.1cm]above left:$\reg R_2$}] {};
    
    \end{tikzpicture}
    \caption{The orientation of the merged terms does not leave a hole for the boundary to pass through.}
    \label{fig:blocked_tunnel}
    \end{subfigure}
    \caption{Given a set of terms with an cubulation edge (in bold black) passing through, we can apply \Cref{lem:deg_4_puncturing} to attempt to create a hole. Depending on the orientation, this process may fail.}
\end{figure}
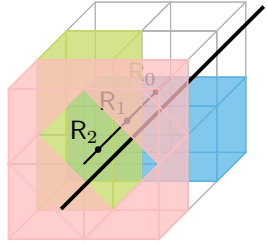
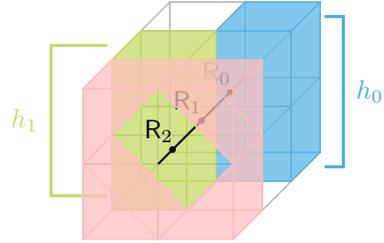

\begin{figure}
    \centering
    \begin{subfigure}[t]{0.3\textwidth}
        \centering
        \begin{tikzpicture}[rotate around x=5,rotate around y=5, rotate around z=-2]
            \foreach \x in {-1,0} {
                \foreach \y in {-1,0} {
                    \DrawCubes[dashed,black!40]{\x}{\y}{0}{1}{1}{1}
                }
            }
            \node[canvas is xy plane at z=0.5] at (-1.25, 1.25) {$S_0$};
    
            \draw[densely dotted, canvas is xy plane at z=0.5,line width=0.4mm,ForestGreen] (0.5,-2.5) -- (0.5,2.5);
            \draw[densely dotted, canvas is zx plane at y=0.5,line width=0.4mm, RoyalBlue] (0.5,-1.5) -- (0.5,2.5);
    
            \DrawCubes[black!90,fill,fill opacity=.3]{-1}{-1}{0}{2}{1}{1}
            \node[canvas is xy plane at z=0.5] at (-1.5,-0.5) {$h$};
            
            \draw[densely dotted, canvas is zx plane at y=0.5,line width=0.4mm, Maroon] (-1.5,0.5) -- (2.5,0.5);
            \draw[canvas is zx plane at y=1, Maroon, line width=0.6mm] (1,0) -- (2,0);
            \draw[canvas is zx plane at y=0, Maroon, line width=0.6mm] (0,0) -- (-1,0);
            \draw[canvas is zy plane at x=1, OliveGreen, line width=0.6mm] (1,0) -- (1,-1);
            \draw[canvas is zy plane at x=0, OliveGreen, line width=0.6mm] (0,1) -- (0,2);
            \draw[canvas is xy plane at z=0, RoyalBlue, line width=0.6mm] (-1,0) -- (-2,0);
            \draw[canvas is xy plane at z=0, RoyalBlue, line width=0.6mm] (1,0) -- (2,0);

        \end{tikzpicture}
        \caption{A slice $S_0$ containing the vertex term, which yields the merged term $h$. The dotted blue, red, and green lines indicate the boundaries of the cubulation. The solid lines indicate the edges on which \Cref{lem:deg_4_puncturing} will be applied.}
        \label{fig:vertex_cube_slice_edges}
    \end{subfigure}
    \quad
    \begin{subfigure}[t]{0.3\textwidth}
        \centering
        \begin{tikzpicture}
            \DrawCubes[dashed,fill=Maroon,Maroon,fill opacity=0.3]{-1}{-1}{-1}{2}{2}{1};
            \foreach \x in {-1,0} {
                \foreach \y in {-1,0} {
                    \DrawCubes[dashed,black!40]{\x}{\y}{0}{1}{1}{1}
                }
            }
    
            \DrawCubes[black!90,fill,fill opacity=.3]{-1}{-1}{0}{2}{1}{1}
            
            \draw[canvas is zx plane at y=1, Maroon, line width=0.6mm] (1,0) -- (2,0);
            \draw[canvas is zx plane at y=0, Maroon, line width=0.6mm] (0,0) -- (-1,0);
            \DrawCubes[dashed,fill=Maroon,Maroon, fill opacity=0.3]{-1}{0}{1}{2}{2}{1};

        \end{tikzpicture}
        \caption{An example of the slices on which we'll apply \Cref{lem:deg_4_puncturing}.}
    \end{subfigure}
    \quad
    \begin{subfigure}[t]{0.3\textwidth}
        \centering
        \begin{tikzpicture}[rotate around x=5,rotate around y=5, rotate around z=-2]
            \foreach \x in {-1,0} {
                \foreach \y in {-1,0} {
                    \DrawCubes[dashed,black!40]{\x}{\y}{0}{1}{1}{1}
                }
            }
    
            \draw[densely dotted, canvas is zx plane at y=0.5,line width=0.4mm, RoyalBlue] (0.5,-1.5) -- (0.5,2.5);

            \draw[canvas is xy plane at z=0, RoyalBlue, line width=0.6mm] (-1,0) -- (-2,0);
            \draw[canvas is xy plane at z=0, RoyalBlue, line width=0.6mm] (1,0) -- (2,0);
            \DrawCubes[RoyalBlue,fill=none]{-2}{-1}{-1}{1}{2}{2};
            \DrawCubes[RoyalBlue!90,fill,fill opacity=.3]{-2}{0}{-1}{1}{1}{2};
            
            \DrawCubes[black!90,fill,fill opacity=.3]{-1}{-1}{0}{2}{1}{1}
            \node[canvas is xy plane at z=0.5] at (-1.5,-0.5) {$h$};
            
            \DrawCubes[RoyalBlue,fill=none]{1}{-1}{-1}{1}{2}{2};
            \DrawCubes[RoyalBlue!90,fill,fill opacity=.3]{1}{-1}{0}{1}{2}{1};
            \draw[canvas is zx plane at y = 0,line width=0.5mm] (0,-1) -- node[midway, label={[label
                distance=-0.3cm]above left:$\reg R_0$}] {} (1,-1);
            \draw[canvas is zx plane at y = 0,line width=0.5mm] (0,1) -- node[midway, label={[label
                distance=-0.3cm]below right:$\reg R_1$}] {} (1,1);
            \draw[canvas is zy plane at x = 1,line width=0.5mm] (0,0) -- node[midway, label={[label
                distance=-0.2cm]right:$\reg R_2$}] {} (0,1);
            \node[canvas is xy plane at z= 0,RoyalBlue] at (2.2, -0.15) {$\reg R_3$};
            
            \node[canvas is xy plane at z=1] at (0, -1.25) {$S_0$};
            \node[canvas is xy plane at z=-1] at (2.25, 0) {$S_1$};
        \end{tikzpicture}
        \caption{An example of possible ``blockages.'' The merged blue terms on the left and right block the blue boundary from pass through an empty region.}
        \label{fig:vertex_blockages}
    \end{subfigure}
    \caption{The process of constructing tubes of slices for a vertex of the cubulation.}
    \label{fig:vertex_cube}
    
\end{figure}
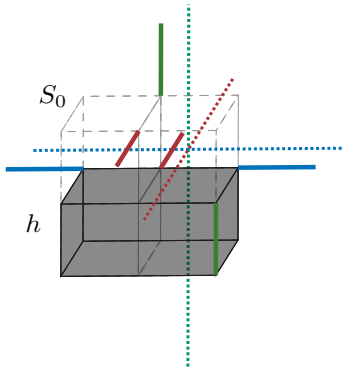
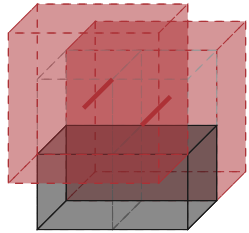
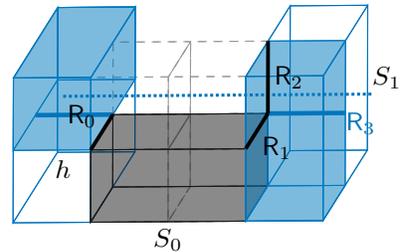
\paragraph{Vertex-type terms.}
\label{par:vertex_type}
As before, we'll apply \Cref{lem:deg_4_puncturing} to a slice with the term containing the vertex (slice $S_0$ in
\Cref{fig:vertex_cube}. The corner is then placed in the hole created by the lemma. The next step is to apply puncturing
again to surround qudit registers to create holes for the cubulation boundaries. In \Cref{fig:vertex_cube_slice_edges},
we draw the (dotted) boundaries, and describe the edges (registers) on which \Cref{lem:deg_4_puncturing} will be applied
to create holes. Two things to call out,
\begin{enumerate}
    \item The edges are chosen so that slices intersect on at most one term. This prevents ``merged'' terms from
        \Cref{lem:deg_4_puncturing} impacting the other edges. For instance, although the bottom green register intersects
        $h$ (and possibly the merged term of the right blue boundary), its placement ensures the green register remains
        degree-$4$.
    \item For the north/south green boundary, we do not pick the edge pointing straight down, as the merged term $h$
        blocks our path. Thus, we first go ``forwards,'' then eventually turn back down. \label{item:turning}
\end{enumerate}

For slices which are adjacent to two removed terms from $S_0$ (e.g. the top green, and both red boundaries), no special
care is needed; we can apply \Cref{cor:product_structure_2_local} to puncture holes if obstructions arise. However, there
are also slices which are perpendicular to $S_0$, see, e.g. \Cref{fig:vertex_blockages}. For these, it's not immediately
clear the 2-local rounding scheme applies. Nonetheless, we can show that as long as a slice is adjacent is to \emph{at
least one} removed term of a previous slice, we can make a continuous hole. This will allow us to puncture holes for the
blue and bottom green edges. For reference, the names in the following lemma are chosen to line up with the right volume
in \Cref{fig:vertex_blockages}.

\begin{lemma}[General blockages]
    \label{lem:general_blockages}
    Suppose there exists a slice $S_1$ which is adjacent (via a face) to a removed term $h_\text{removed}$. After
    applying \Cref{lem:deg_4_puncturing} (or \Cref{cor:local_puncturing}) to $S_1$, we either immediately have punctured
    a continuous hole, or there is a further projection that can be applied to an edge adjacent to $h_\text{removed}$
    and $S_1$ which punctures a hole from $h_\text{removed}$ through $S_1$.
\end{lemma}
\begin{proof}
    Let $f$ be a face of the removed term $h_\text{removed}$, from which we'll attempt to pass through $S_1$. Let $h$ be
    the (original) term of $S_1$ adjacent to $f$. Assume we have applied \Cref{cor:local_puncturing} on $\reg R_3$, the
    central register of $S_1$. If $h$ is removed via \Cref{cor:local_puncturing}, then we are done. Otherwise, we make
    the two geometric observations:
    \begin{itemize}
        \item There are at most two edges on $f$ ($\reg R_1, \reg R_2$) which are adjacent to $\reg R_3$. This is
            because any third edge would subdivide $f$.
        \item For any edge $\reg R$ on $f$, there are at most $2$ terms on which $h$ shares more than just the edge
            $\reg R$. This is because any edge induces a slice, sandwiching $h$ by two other terms.
    \end{itemize} 

    Since $h_\text{removed}$ is adjacent to $h$ via $f$ (and thus via $\reg R_1$, $\reg R_2$), and there exists two
    distinct terms $h_1, h_2$ which are adjacent to $h$ \emph{within} the slice $S_1$, the second item implies that
    \begin{itemize}
        \item Across $\reg R_1$, $h$ is adjacent to \emph{only} $h_\text{removed}$ and some $h_1 \in S_1$ on more than just
            $\reg R_1$.
        \item Across $\reg R_2$, $h$ is adjacent to \emph{only} $h_\text{removed}$ and some $h_2 \in S_2 \setminus \{h_1\}$
            on more than just $\reg R_2$.
    \end{itemize}
    But \Cref{cor:local_puncturing} merges $h$ with at most one adjacent term in $S_1$, implying either $h_1$ or $h_2$
    are removed. Without loss of generality, assume $h_2$ is removed. This implies that every other term acting
    non-trivially on $\reg R_2$ intersects $h$ \emph{only} on $\reg R_2$ and thus across $\reg R_2$, all terms can be
    treated as 2-local. As before, we apply \Cref{cor:product_structure_2_local} to create a hole through this edge. 
\end{proof}

Beyond the obstruction depicted in \Cref{fig:vertex_blockages}, this also permits us to ``turn,'' as required by the
green boundary in \Cref{fig:vertex_cube_slice_edges}.

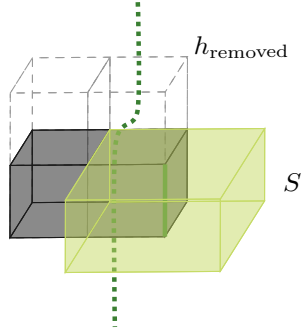
\begin{figure}[H]
    \centering
    \begin{tikzpicture}[rotate around x=5,rotate around y=5, rotate around z=-2,]
        \foreach \x in {-1,0} {
            \foreach \y in {-1,0} {
                \DrawCubes[dashed,black!40]{\x}{\y}{0}{1}{1}{1}
            }
        }

        \DrawCubes[black!90,fill,fill opacity=.3]{-1}{-1}{0}{2}{1}{1}
        
        \draw[canvas is zy plane at x=1, OliveGreen, line width=0.6mm] (1,0) -- (1,-1);
        \draw[canvas is zy plane at x=0.5, OliveGreen, dotted, line width=0.6mm] plot [smooth] coordinates {(0.5, 2) (0.5, 0.5) (1.5, 0.5)
        (1.5, -2)};
        \DrawCubes[SpringGreen,fill,fill opacity = 0.3]{0}{-1}{0}{2}{1}{2};
        \node[canvas is xy plane at z=0.5] at (2.5, -0.5) {$S$};
        \node[canvas is xy plane at z = 0.5, anchor=south west] at (1.1,1.1) {$h_\text{removed}$};
    \end{tikzpicture}
    \captionsetup{width=0.8\textwidth}
    \caption{The green slice and ``turning'' green boundary. One can check that applying \Cref{lem:general_blockages}
    with the above choice of $h_\text{removed}$ and $S_1 \dfn S$, we puncture a hole through $S$.}
\end{figure}

\subsection{General 3D lattices}
We now turn to the second case in \Cref{thm:3d_theorem}. Although similar ideas from the cubic lattice case work here,
there are several components which need to be generalized. First, the idea of a tube of slices is not well-defined. One
idea is to choose choose some path $P = (e_1, e_2, \dots)$ and attempt to puncture a hole along this edge (e.g. by
applying \Cref{lem:deg_4_puncturing} do each edge $e_i$). This is has the undesirable property that the resulting
``cubulation'' depends heavily on the individual Hamiltonian terms, unlike in the proof of
\cite{Aharonov_AE2011_ComplexityCommutingLocal}, where the triangulation can be made independent of the Hamiltonian. To
emulate this idea, we partition the 3D space in a cubic grid, large enough that within each element $C$ of the grid $\mc
C$, there is a term (volume) $h_C$ such that all neighboring volumes are fully contained within $C$. We'll use this
center term $h_C$ to define the co-cubulation. At a high level, this involves selecting a center of grid element $C$,
and then drawing paths from the center to the geometric centers of each face of $C$. This is essentially the
co-triangulation idea of \cite{Aharonov_AE2011_ComplexityCommutingLocal}. We now show how to pick the centers and paths
so that the Hamiltonian can be made $2$-local. We recall the key assumption we make in this section for convenience:

\generalizedlattice*

\paragraph{Puncturing edges} As in \Cref{sec:cubic_lattice}, we'd like to define a notion of a ``tube'' of slices. In
the cubic lattice case, such a tube is essentially defined by a sequence of faces, $F_1, F_2, \dots$, such that there is
a term $h$ sharing adjacent faces $F_i, F_{i+1}$. Then, the slice $S_i$ will be defined by an edge
(register) $\reg R_i$ connecting $F_i$ and $F_{i+1}$. The terms of the slice are those volumes which act non-trivially on
$\reg R_i$. This generalizes the notion of tubes, and we can apply \Cref{cor:local_puncturing} to each $\reg R_i$. There
are several complications which arise in the general 3D case.
\begin{enumerate}
    \item Two faces $F_i, F_{i+1}$ share an edge.
    \item A volume of $S_i$ is not directly adjacent to a volume in $S_{i+1}$, as in  \Cref{fig:slice_no_shared_face}
    \item There are no adjacent removed terms between slices. This generalizes \Cref{fig:blocked_tunnel} and is depicted
        in \Cref{fig:generalized_blocked_tunnel}.
\end{enumerate}

Item 1 is easily dealt with; any two non-equal faces sharing a volume must be have \emph{some} edge separating
them so we can apply \Cref{cor:local_puncturing} to this edge instead.

For Item 2, we may no longer have a direct path between the two slices, e.g. if the pink terms in
\Cref{fig:slice_no_shared_face} are removed. However, in this case, any term in the blue region does not intersect $h$
or $h'$ on a face, as this would violate the first item of \Cref{as:generalized_cubic_lattice}. Since $h$ and $h'$ both
commute with any term in the blue region, there exists a projector on $\reg R_3$ which decouples these two terms and induces
a hole from face $F_{k-1}$ to $F_{k+1}$. 

For Item 3, \Cref{cor:local_puncturing} ensures that (the regions corresponding to) $3$ terms are non-trivial. Since the
local degree is at least $6$, this implies that the $3$ non-trivial terms can only create a blockage when they do not
share any faces. But then we can directly apply \Cref{cor:product_structure_2_local} to create a hole.
\tikzset{pics/facevolume/.style={code={
    \node[canvas is zy plane at x = -1, faces] at (0,0) {};
    \node[canvas is zy plane at x = 0, faces] at (0,0) {};
    \draw[canvas is xy plane at z = -0.75, fill opacity= 0.5, fill=SpringGreen] (-1,-0.75) -- (0,-0.75) -- (0, 0.75) -- (-1,0.75) -- cycle; 
    \draw[canvas is xy plane at z = 0.75,text opacity=1, fill opacity= 0.5, fill=SpringGreen] (-1,-0.75) rectangle ++(1,1.5) node[midway,scale=1.25,yshift=-0.1cm] {#1} ; 
}}}

\tikzset{pics/wedge/.style={code={
    \draw[canvas is xy plane at z = -0.75, volfaces] (-1, 0.75) -- (0,0.75) -- (-1,1.5) -- cycle;
    \draw[canvas is zy plane at x = -1, volfaces] (0.75,0.75) -- (0.75,1.5) -- (-0.75,1.5) -- (-0.75, 0.75) -- cycle;
    \draw[canvas is xy plane at z = 0.75, volfaces] (-1, 0.75) -- (0,0.75) -- (-1,1.5) -- cycle;
    \begin{scope}[rotate around z=55, transform canvas={yshift=0.75cm}]
        \draw[canvas is zy plane at x = 0, volfaces] (0.75, 0) -- (0.75, 1.25) -- (-0.75,1.25) -- (-0.75,0) -- cycle;
    \end{scope}
}}}

\tikzset{pics/page/.style={code={
    \pgfmathsetmacro{\x}{cos(60)}
    \pgfmathsetmacro{\y}{sin(60)}
    \draw[dashed] (0,0,0) -- (0,1,0) -- (\x,\y,0) -- cycle;
    \draw[fill opacity=0.5, fill=SpringGreen] (0,0,0) -- (0,0,1) -- (0,1,1) -- (0,1,0) -- cycle;
    \draw[fill opacity=0.5, fill=SpringGreen] (0,0,0) -- (0,0,1) -- (\x,\y,1) -- (\x,\y,0) -- cycle;
    \draw[dashed] (0,1,0) -- (0,1,1) -- (\x,\y,1) -- (\x,\y,0) -- cycle;
    \begin{pgfonlayer}{fg}
        \draw[dashed] (0,0,1) -- (0,1,1) -- (\x,\y,1) -- cycle;
    \end{pgfonlayer}
}}}

\begin{figure}
    \centering
    \savebox{\imagebox}{
        \scalebox{0.8}{
            \begin{tikzpicture}[
                    faces/.style={draw, rectangle, minimum width=1.5cm, minimum height=1.5cm, fill opacity=0.75,fill=SpringGreen},    
                    volfaces/.style={fill opacity=0.75,fill=pink},
                    topface/.style={fill opacity=0.75,fill=CornflowerBlue},
                    rotate around x=5
                ]
        
                \node[canvas is zy plane at x = -3, faces, label={[label distance=0.5cm]below:$F_1$}] at (0,0) {};
                \draw[canvas is zx plane at y = 0.25, dotted, line width=0.75mm] (0.75, -2.25) -- (0.75, -1.25);
                \draw[canvas is zx plane at y = 0.25, dotted, line width=0.75mm] (0.75, 1.75) -- (0.75, 2.75);
                \node[canvas is zy plane at x = 3, faces, label={[label distance=0.5cm]below:$F_\ell$}] at (0,0) {};

                \pic {facevolume={$h$}} ; 
                \begin{scope}[transform canvas={xshift=1cm}]
                    \pic {facevolume={$h'$}};
                \end{scope}
                \draw[canvas is xy plane at z = 0.75,decorate,decoration={brace,mirror,raise=3pt}] (-0.9,0.75) --
                    node[below=3pt] {\small $\reg R_0$} (-0.1,0.75);
                \draw[canvas is xy plane at z = 0.75,decorate,decoration={brace,mirror,raise=3pt}] (0.1,0.75) --
                    node[below=3pt] {\small $\reg R_1$} (0.9,0.75);

                \pic {wedge};
                \begin{scope}[rotate around y=180]
                    \pic {wedge};
                \end{scope}
                \node[canvas is xy plane at z=0.75] at (-1, -1) {$F_{k-1}$};
                \node[canvas is xy plane at z=0.75] at (0, -1) {$F_{k}$};
                \node[canvas is xy plane at z=0.75] at (1, -1) {$F_{k+1}$};

                \begin{pgfonlayer}{bg}
                \end{pgfonlayer}
                \draw[canvas is xy plane at z=-0.75,topface] (-1,1.5) -- (0,0.75) -- (1,1.5) -- cycle;
                \draw[canvas is xy plane at z=0.75,topface] (-1,1.5) -- (0,0.75) -- (1,1.5) -- cycle;
                \draw[canvas is zx plane at y=1.5,topface] (0.75, -1) -- (0.75,1) -- (-0.75,1) -- (-0.75,-1) -- cycle;

                \draw[canvas is zy plane at x=0,line width=0.75mm] (0.75,0.75) -- node[midway] (center) {} (-0.75,0.75);
                \node[canvas is xy plane at z=0] (label) at (-1,2.5) {$\reg R_2$};
                \draw[->, line width=0.5mm] (label) -- (center);
                
            \end{tikzpicture}
        }
    }%
    \begin{subfigure}[t]{0.3\textwidth}
        \centering\usebox{\imagebox}
        \caption{
            Two adjacent terms sharing a face. In this example, two of the surrounding (pink) terms do not also share a
            face. However, by treating the terms in the blue region, together with the right pink term, as one larger
            term, we can recover the ``cubic'' structure of \Cref{sec:cubic_lattice}.
        }
        \label{fig:slice_no_shared_face}
    \end{subfigure}
    \quad
    \begin{subfigure}[t]{0.3\textwidth}
        \centering
        \begin{tikzpicture}[
            volfaces/.style={fill opacity=0.5,fill=pink},
        ]
            \DrawCubes[fill=SpringGreen, opacity=0.75]{-1}{0}{0}{1}{1}{1};
            \DrawCubes[fill=SpringGreen, opacity=0.75]{0}{-1}{0}{1}{1}{1};
            \foreach \angle in {0,30,60} {
                \begin{scope}[rotate around z=-\angle]
                    \pic {page};
                \end{scope}
            }
            \pgfmathsetmacro{\x}{cos(60)}
            \pgfmathsetmacro{\y}{sin(60)}
            \draw[line width=0.5mm, canvas is xy plane at z = 0.5] plot [smooth] coordinates {(-2,0.5) (0.2, 0.5) (0.8,0.25) (2, 0.25)};
            \begin{scope}[rotate around z=-60]
                \draw[fill opacity=0.5, fill=SpringGreen] (0,1,0) -- (0,1,1) -- (\x,\y,1) -- (\x,\y,0) -- cycle;
            \end{scope}
        \end{tikzpicture}
        \caption{A sequence of faces where adjacent faces share an edge.}
        \label{fig:shared_edge}
    \end{subfigure}
    \quad
    \begin{subfigure}[t]{0.3\textwidth}
        \centering
        \begin{tikzpicture}[
          x={(1cm,0.4cm)},
          y={(8mm, -3mm)},
          z={(0cm,1cm)},
          line cap=round,
          line join=round,
          radius=1,
          delta angle=-180,
          front/.style={canvas is yz plane at x=5.5},
          back/.style={canvas is yz plane at x=6.5},
          mid/.style={canvas is yz plane at x=6},
        ]
            \draw[back, gray] (125:1) arc[start angle=125, delta angle=180];

            \draw {
                [back]
                  (125:1) arc[start angle=125]
            }{
                [front]
                -- (125+180:1) arc[end angle=125]
            } --cycle;

            \def\sone{252}
            \draw[fill=CornflowerBlue,fill opacity=.8] {
                [back]
                (108:1) -- (0,0) -- (0:1) arc[start angle=0, delta angle=-55]
            }{
                [mid]
                -- (-55:1) arc[start angle=-55, delta angle=-197]
            } --cycle;
            \draw {
                [back]
                (0:1)
            } {
                [mid]
                -- (0:1) -- (0,0) -- (108:1)
            } {
                [back]
                -- (108:1) -- (0,0)
            } {
                [front]
                -- (0,0)
            };

            \draw[fill=SpringGreen,fill opacity=.8] {
              [mid]
              (125:1) arc[start angle=125, delta angle=-125]
            }{
              [front]
              -- (0:1) -- (0,0) -- (\sone:1) arc[start angle=\sone, delta angle=-127]
            } --cycle;
            \draw[fill=SpringGreen,fill opacity=.8] {
                [mid]
                (0,0) -- (-108:1)
            } {
                [front]
                -- (-108:1) -- (0,0)
            } -- cycle;

            \foreach \r in {0,36,...,144} {
                \begin{scope}[rotate around x=\r]
                    \draw[dashed, gray] {
                        [back,] (-1, 0) -- (1,0)
                    } {
                        [front]
                        -- (1,0) -- (-1,0)
                    } -- cycle;
                \end{scope}
            }

            \draw[front]      (125:1) arc[start angle=125];
        \end{tikzpicture}
        \caption{
            One might imagine that we could have completely disjoint removed terms, such that we cannot even apply
            \Cref{cor:product_structure_2_local}. However, this is avoided by the number of terms guaranteed to be
            trivial by \Cref{cor:local_puncturing}.
        }
        \label{fig:generalized_blocked_tunnel}
    \end{subfigure}
    \caption{Complications in general 3D geometries.}
\end{figure}

\paragraph{Puncturing vertices} Choose an edge $\reg R$ (i.e. a qudit register) of the term $h_C$, and consider the set
of terms acting non-trivially on $\reg R$. Since we are in the high degree setting, these form a slice $S_0$ with degree $>
4$, and we can apply \Cref{cor:local_puncturing}. This ensures that there are at least $2$ face-adjacent terms which are
now trivial in the resulting Hamiltonian and we can place the corner of the co-cubulation in this hole. We now need to
route 6 paths through the surrounding faces. Suppose there are $k$ removed terms, and each term has at least $f$ faces.
Since each term is adjacent to two other terms in the slice, $2$ faces from each term are internal. Thus, the number of
faces surrounding the hole is at least,
\begin{equation}
    N_\text{faces} \geq (f-2) \cdot k
\end{equation}
By Item 2 of \Cref{as:generalized_cubic_lattice}, each term has at least $5$ faces, and $k$ is at least $2$ so
$N_\text{faces} \geq 6$, which is exactly the number of distinct edges exiting the corner. We now can make an argument
similar to \Cref{par:vertex_type}. First, we pick the edges $\reg R_1, \dots, \reg R_6)$ so that each pair $\reg R_i
\neq \reg R_j$ are not adjacent via a single edge. It's possible that $\reg R_i, \reg R_j$ share a single face (and thus
one volume), but any other volume would need to span two edges to intersect both both registers, which violates
convexity. One can check that these edges can be chosen in this way, by imagining flattening out the $N_\text{faces}$
exposed faces of $S_0$ and pick $6$ vertices to serve as the origins of each edge on the resulting 2D surface.

Once this is done, as in the cubic case, any edge adjacent to two removed terms can be easily dealt with. For edges
adjacent to only single term, one can check that the observations made in the proof of \Cref{lem:general_blockages}
still hold in the general 3D lattice case and therefore holes can be made through these terms as well.

\printbibliography

\appendix 
\section{Verifiers for commuting local Hamiltonians}
\label{sec:verifiers}

Much of the work surrounding commuting local Hamiltonians has been directed towards showing ever increasing classes of them are classically verifiable.
However, showing that commuting local Hamiltonians contain problems that are not in $\mathsf{NP}$ has proven much harder; to this day it is not even known if commuting local Hamiltonians can capture the complexity of problems in $\mathsf{MA}$.
In this section, we do not make any formal progress towards showing the hardness of commuting local Hamiltonians, but try to provide a conceptual framework that might be useful for studying commuting local Hamiltonians in the future.  
We provide a class of quantum verifiers for whom the problems that they can verify exactly captures the complexity of commuting local Hamiltonians, although the definition of the verifier is mostly specifically designed to capture commuting local Hamiltonians.  
Consider the following complexity class

\begin{definition}[$\mathsf{QIMA}_{k}$]
    An $k$-local instant verifier $V$ is a time-uniform polynomial-time algorithm on two registers $\reg{AB}$ that consists of $m = \poly(n)$ many $k+1$-local $\reg{B}$-controlled gates $G_1, \ldots, G_{m}$ that commute with each other.  $V$ acts on a witness $\ket{\psi}_{\reg{A}}$, and an auxillary register $\reg{B}$ initialized in the state $\ket{+}^{\otimes n_+}$, and $V$ measures $\proj{+}^{\otimes n_{+}} \otimes \id$ for some $l$ to accept. 
    
    A promise problem $L$ belongs to $\mathsf{QIMA}_{k}$ if there exists a $k$-local instant verifier such that for all $x \in L$, there exists a witness $\ket{\psi_x}$ such that $V_x(\ket{\psi_s})$ accepts with probability $1$, and for all $x \not\in L$, for all $\ket{\psi}$, $V_x(\ket{\psi})$ accepts with probability at most $1/2$.  
\end{definition}

\begin{lemma}
    The $k$-local commuting local projectors problem is in $\mathsf{QIMA}_{k}$.
\end{lemma}
\begin{proof}
    Consider the instant verifier that performs the following unitary
    \begin{equation*}
        U_i = \id - 2\Pi_i\,,
    \end{equation*}
    controlled on the $i$'th $\ket{+}$ ancilla register.  
    Then the verifier measures all $m$ ancilla registers.  
    If $\ket{\psi}$ is a joint $0$-energy state of all verifiers, then for all $i$, $U_i \ket{\psi} = \ket{\psi}$, and otherwise for every state $\ket{\psi}$ there exists an $i$ such that $\mathrm{Re}(\bra{\psi} U_i \ket{\psi}) < c$ for some $c$ inverse-polynomial away from $1$.  Thus, if the commuting projectors have a $0$ energy ground state, the ground state is accepted by the verifier with probability $1$, and otherwise every state is rejected by the verifier with probability at least inverse polynomial probability.  Applying parallel repetition allows us to amplify the advantage without changing the locality of the unitaries.  
\end{proof}

\begin{lemma}
    For all $k$, the $k$-local commuting local projectors problem is hard for $\mathsf{QIMA}_{k}$.  
\end{lemma}
\begin{proof}
    Let $G_1, \ldots, G_m$ be the gates of a $k$-local instant verifier $V$.  Then consider the following commuting local Hamiltonian:
    \begin{equation*}
        h_i = \id - \frac{G_i + G_i^{\dagger}}{2}\,.
    \end{equation*}
    We note that every every $h_i$ commutes, and if $\ket{\psi}$ (the witness) is a $+1$-eigenstate of $G_i$, it is a $0$-energy eigenstate of $h_i$.  Note that not all of the $h_i$ are projectors, however using the reduction from \cite{Irani_IJ2023_CommutingLocalHamiltonian}, the prover can send a (classical) $\mathsf{NP}$ witness containing $\lambda_i$.  
    Finally note that every $h_i$ is Hermitian and commutes with every other $h_i$ if the $G_i$ commute with each other.  
\end{proof}

We can define the complexity class $\mathsf{QIMA}_{\log}$ to be the complexity class defined in a similar fashion where $k$ is allowed to be a function of the input size, and is $\log(n)$.
Unfortunately very little is known about reductions between commuting local Hamiltonian problems, for example it is not known different localities are equally hard, or if reducing the rank of local projectors changes the difficulty of the problem.  
We think these are interesting open questions for future work, and we hope that by defining some complexity class that captures the difficulty of commuting local Hamiltonians, there will be more of a formal framework to study the hardness of these problems.
\section{Lemmas for commuting local Hamiltonians}
\label{sec:misc_lemmas}
\begin{lemma}[Rounding preserves commutation]
    \label{lem:rounding_preserves_commutation}
    Suppose $A$ and $B$ are two commuting, $N \times N$ Hermitian matrices. Suppose $B$ has eigendecomposition $B = \sum_i \lambda^A_i \Pi_i$, with $\lambda^A_i \neq \lambda^A_j$ for $i\neq j$. Then, the \emph{rounded} matrix $\textsf{round}(B) = \sum_{i \st \lambda^A_i > 0} \Pi_i$ commutes with $A$.
\end{lemma}
\begin{proof}
    Since $A$ and $B$ commute, $A$ acts invariantly on each eigenspace $\Pi_i$ of $B$ and thus commutes with $\Pi_i$ individual. This easily implies that $A$ commutes with $\textsf{round}(B)$.
\end{proof}

\equivprojector*
\begin{proof}
    Since $H$ is commuting,
    \[
        \lambda_0(H) = 0 \iff \tr\Brac{\prod_i (\id - h_i)} > 0\,.
    \]
    This fact can be easily seen by considering a simultaneously diagonalizing basis for all $h_i$'s. For the RHS, we have
    \begin{align}
        \tr\Brac{\prod_i (\id - h_i)} = \sum_{i,j} \tr\Brac{\pi^1_i(\id - h_1)\pi^1_i \pi^2_j(\id - h_2) \pi^2_j
        \overline H_\text{rest}}\,,
    \end{align}
    where we used that $\pi^1_i$ commutes with $h_1$ and thus $\pi^1_i h_1 \pi^1_k = 0$ for $i \neq k$ and similarly for
    $h_2$. Additionally, we write $\overline H_\text{rest} = \prod_{i > 2} (\id - h_i)$. Suppose each summand on the RHS
    is non-negative. If the LHS is $0$, this immediately means all terms on the right are $0$. On the other hand, if the
    LHS is positive, then there must be some $i,j$ such that 
    \[
    \tr\Brac{\pi^1_i(\id - h_1)\pi^1_i \pi^2_j(\id - h_2) \pi^2_j \overline H_\text{rest}} > 0\,.
    \]
    This proves the claim. It remains to show each summand is positive.

    Consider arbitrary $i,j$ and write $\pi_1 = \pi^1_i$ and $\pi_2 = \pi^2_j$. Since $\pi_2$ commutes with $h_2$, we know that these operators are diagonal in the same basis. In this basis, write $\pi_2 = \sum_i \ketbra i$ and $h_2 = \sum_i \lambda_i \ketbra i$. Thus,
    \[
        \pi_2 - \pi_2 h_2 \pi_2 = \sum_i (1 - \lambda_i) \ketbra i
    \] 
    But since $\|h_2\|_\infty \leq 1$, each coefficient is non-negative and this matrix is PSD. Next, by assumption $h_i$ with $i > 2$ commute with both $\pi_2$ and $h_2$; thus, $[\pi_2 - \pi_2 h_2 \pi_2, \overline H_\text{rest}] = 0$. The product of two commuting PSD operators is also PSD and thus $(\pi_2 - \pi_2 h_2 \pi_2) \overline H_\text{rest} \succeq 0$. Finally, we have by the same argument that $\pi_1 - \pi_1 h_1 \pi_1 \succeq 0$. Thus
    \[
        (\pi_1 - \pi_1 h_1 \pi_1)\Paren{(\pi_2 - \pi_2 h_2 \pi_2) \overline H_\text{rest}}
    \]
    is the product of two PSD operators and this implies the trace is non-negative.
\end{proof}

\end{document}